\documentclass{aa}  

\usepackage{natbib}
\usepackage{graphicx}
\usepackage{txfonts}
\usepackage{hyperref}
\usepackage{color}
\usepackage{upquote} 
\usepackage{xspace}
\usepackage{ulem}  
\usepackage{scalerel}
\usepackage{booktabs}
\usepackage{longtable}
\usepackage[dvipsnames,usenames,table]{xcolor}
\newcommand{\orcit}[1]{\protect\href{https://orcid.org/#1}{\protect\includegraphics[width=8pt]{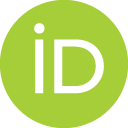}}}

\usepackage{multirow}
\usepackage{subfig}

\def\Gaia{\textit{Gaia}\xspace}
\def\gmag{\ensuremath{G}\xspace}
\def\gbp{\ensuremath{G_{\rm BP}}\xspace}
\def\grp{\ensuremath{G_{\rm RP}}\xspace}
\newcommand{\bpminrp}{\ensuremath{G_\mathrm{BP}-G_\mathrm{RP}}\xspace}

\newcommand{\qrg}{QR_5(G)}

\newcommand{\mediandeltawl}{\langle\Delta\lambda\rangle_{\rm RP}}

\newcommand{\ks}{K_{\rm s}}
\newcommand{\wjk}{W_{\scaleto{\rm J,K_s}{4.5pt}}}

\newcommand{\tableclas}{\texttt{vari\_classifier\_result}\xspace}
\newcommand{\tablesos}{\texttt{vari\_long\_period\_variable}\xspace}

\begin{document} 

  \title{\Gaia Data Release 3: The second \Gaia catalogue of long-period variable candidates}
  \subtitle{}
  \titlerunning{\Gaia DR3 long-period variable candidates}
  
  \author{
 T.~Lebzelter\orcit{0000-0002-0702-7551}\inst{\ref{inst1}\fnmsep\thanks{Corresponding~author: T. Lebzelter (\href{mailto:thomas.lebzelter@univie.ac.at}{\tt thomas.lebzelter@univie.ac.at})}}
\and N.~Mowlavi\orcit{0000-0003-1578-6993}\inst{\ref{inst2},\ref{inst3}\fnmsep\thanks{Corresponding author: N. Mowlavi (\href{mailto:nami.mowlavi@unige.ch}{\tt nami.mowlavi@unige.ch})}}
\and I.~Lecoeur-Taibi\orcit{0000-0003-0029-8575}\inst{\ref{inst3}}
\and M.~Trabucchi\orcit{0000-0002-1429-2388}\inst{\ref{inst4},\ref{inst2}\fnmsep\thanks{Corresponding author: M. Trabucchi (\href{mailto:michele.trabucchi@unipd.it}{\tt michele.trabucchi@unipd.it})}}
\and M.~Audard\orcit{0000-0003-4721-034X}\inst{\ref{inst2},\ref{inst3}}
\and P.~Garc\'{i}a-Lario\orcit{0000-0003-4039-8212}\inst{\ref{inst:0013}}
\and P.~Gavras\orcit{0000-0002-4383-4836}\inst{\ref{inst:0022}}
\and B.~Holl\orcit{0000-0001-6220-3266}\inst{\ref{inst2},\ref{inst3}}
\and G.~Jevardat~de~Fombelle\inst{\ref{inst3}}
\and K.~Nienartowicz\orcit{0000-0001-5415-0547}\inst{\ref{inst6}}
\and L.~Rimoldini\orcit{0000-0002-0306-585X}\inst{\ref{inst3}}
\and L.~Eyer\orcit{0000-0002-0182-8040}\inst{\ref{inst2},\ref{inst3}}
}
         
  \authorrunning{Lebzelter et al.}
  
  \institute{University of Vienna, Department of Astrophysics, Tuerkenschanz\-strasse 17, A1180 Vienna, Austria\label{inst1}
             \and
             Department of Astronomy, University of Geneva, Ch. Pegasi 51, CH-1290 Versoix, Switzerland\label{inst2}
             \and
             Department of Astronomy, University of Geneva, Ch. d'Ecogia 16, CH-1290 Versoix, Switzerland\label{inst3}
             \and
             Dipartimento di Fisica e Astronomia, Universit\`a di Padova, Vicolo dell'Osservatorio 2, I-35122 Padova, Italy\label{inst4}
             \and
             SixSq, Rue du Bois-du-Lan 8, CH-1217 Geneva, Switzerland\label{inst6}
             \and European Space Agency (ESA), European Space Astronomy Centre (ESAC), Camino bajo del Castillo, s/n, Urbanizacion Villafranca del Castillo, Villanueva de la Ca\~{n}ada, 28692 Madrid, Spain\label{inst:0013}
             \and RHEA for European Space Agency (ESA), Camino bajo del Castillo, s/n, Urbanizacion Villafranca del Castillo, Villanueva de la Ca\~{n}ada, 28692 Madrid, Spain\label{inst:0022}
            }

\date{June 2022}

\abstract{
The third Gaia Data Release covers 34 months of data and includes the second \Gaia catalogue of long-period variables (LPVs), with \gmag variability amplitudes larger than 0.1~mag (5-95\% quantile range).
}{
The paper describes the production and content of the second$^{}$ \Gaia catalogue of LPVs and the methods we used to compute the published variability parameters and identify C-star candidates.
}{
We applied various filtering criteria to minimise contamination from variable star types other than LPVs. The period and amplitude of the detected variability were derived from model fits to the \gmag-band light curve wherever possible. C stars were identified using their molecular signature in the low-resolution $RP$ spectra.
}{
The catalogue contains 1\,720\,558 LPV candidates, including 392\,240 stars with published periods (ranging from 35 to $\sim$1000 days) and 546\,468 stars classified as C-star candidates.
Comparison with literature data (OGLE and ASAS-SN) leads to an estimated completeness of 80\%.
The recovery rate is about 90\% for the most regular stars (typically miras) and 60\% for SRVs and irregular stars.
At the same time, the number of known LPVs is increased by a factor of 6 with respect to literature data for amplitudes larger than 0.1 mag in \gmag, and the contamination is estimated to be below 2\%.
Our C-star classification, based on solid theoretical arguments, is consistent with spectroscopically identified C stars in the literature.
Caution must be taken in crowded regions, however, where the signal-ro-noise ratio of the RP spectra can become very low, or if the source is reddened by some kind of extinction.
The quality and potential of the catalogue are illustrated by presenting and discussing LPVs in the solar neighbourhood, in globular clusters, and in galaxies of the Local Group.
}{
This is the largest all-sky LPVs catalogue to date. The photometric depth reaches \gmag = 20 mag. This is a unique dataset for research into the late stages of stellar evolution.}

\keywords{stars: variables - stars: AGB and post-AGB - stars: carbon - galaxies: stellar content - catalogs - methods: data analysis}

\maketitle

\section{Introduction}
\label{sec:Introduction}

Due to their large variability amplitudes, in particular in the visual range, long-period variables (LPVs) have been known and studied for a long time.
They represent late evolutionary stages of low- and intermediate-mass stars and thus indicate phases in the life of a star that are of high astrophysical interest.
Nucleosynthesis of carbon, oxygen, and s-process elements, dredge-up events, and high mass-loss rates occurring in these stars are relevant for understanding stellar and galactic evolution.

In addition to the most prominent members of this type, the miras, many more objects are classified as LPVs today. They range from small-amplitude red giants (SARG or OSARG) near the tip of the red giant branch (RGB) up to variable red supergiants.
Studying this wide class of variables has profited much from large monitoring surveys.
Due to their high intrinsic brightness, LPVs can be detected in an extensive volume of space.
As a consequence, large catalogues of LPVs have been produced as by-products of various sky surveys. 
One of the most influential catalogues in this context was produced by the Optical Gravitational Lensing Experiment (OGLE) team \citep{soszynski_etal_2009_ogle3lmc,soszynski_etal_2011_ogle3smc,soszynski_etal_2013_ogle3bulge}. 
This database revealed several parallel pulsation sequences on the upper giant branch, which turned out to be a key for relating pulsation properties with mass and evolutionary stage of an LPV \citep[][]{Wood2015}.
At the same time, the variability of LPVs plays a key role in producing significant mass loss because periodic levitation of the atmosphere enhances this process 
\citep{hoefner_olofsson_2018}.
Understanding these relations opened the path to including LPV pulsation into models of stellar populations \citep[e.g.][]{Trabucchi2021}.

Period-luminosity relations can only be revealed if the distances to the studied objects are known.
In the past, major advances were often based on data from the Magellanic Clouds and other extra-galactic systems.
The \Gaia all-sky survey is expected to add a major contribution to the study of LPVs during its five-year nominal mission plus extensions (mission extension has already been approved, to date, until the end of 2022), in particular by providing variability and distance information for Galactic field and halo stars and by covering the brightness range down to \gmag=21 mag for the complete sky. 
The high level of completeness of LPVs expected from the \Gaia survey will offer the opportunity of studying, among other subjects, the frequencies of various groups of LPVs in the extended solar neighbourhood and other parts of the Galaxy.
Thus, the \Gaia database of LPVs will be unique and will provide a major step forwards in understanding these variables. 

In the course of \Gaia Data Release 2 \citep[DR2][]{gaia_dr2_2018}, we published the first$^{}$ \Gaia catalogue of LPV candidates based on \Gaia data collected over a time span of 22 months \citep{mowlavi_etal_2018_dr2lpv}.
In this first$^{}$ catalogue, we place the priority on ensuring that the contamination is as low as possible, without targeting completeness.
This was a result of the limitations due to the relatively short total time span that was covered (compared to LPVs with periods that can exceed 1000 days) and the sparsity of \Gaia measurements due to the spacecraft scanning law.
Therefore, this catalogue
included only variables with amplitudes larger than 0.2~mag in the \Gaia \gmag band.
Small-amplitude red giant variables, detected as a large group in the OGLE database \citep[e.g.][]{Wray_etal_2004,2004AcA....54..129S}, were excluded at this stage.
The first$^{}$ \Gaia catalogue of LPVs with its more than 150\,000 entries still provided the largest collection of LPVs with amplitudes exceeding 0.2~mag until now.

The longer time series of measurements that form the base for \Gaia Data Release 3 (DR3) allows for a new, deeper search for LPV candidates.
In addition, access to spectroscopic information allows for a preliminary characterisation of catalogued stars in terms of their surface chemistry.
Therefore, we present with this paper the second$^{}$ \Gaia catalogue of LPVs.

A general description of \Gaia DR3 is provided in several separate papers produced by the \Gaia consortium \citep[see e.g.][and references therein]{DR3-DPACP-185}. An overview of the DR3 processing of variable stars is given in \citet{DR3-DPACP-162}.
In this paper, we focus on the DR3 catalogue of LPV candidates.

\section{Catalogue construction}
\label{sec:CatalogConstruction}

\begin{table}
\caption{Catalogue content in numbers.}
\label{tab:catalog_numbers}
\centering
\begin{tabular}{lr}
\hline\hline
Selector & number of objects\\
\hline
LPV candidates in \Gaia DR3 & 2\,326\,297 \\
in classification table & 2\,325\,775 \\
in SOS table (2$^{nd}$ \Gaia LPV catalogue) & 1\,720\,588 \\
\hspace{3mm} {\it including} & \\
\hspace{3mm} LPVs with $0<\sigma_\pi$/$\pi$ < 0.15 & 91\,912\\
\hspace{3mm} LPVs with periods & 392\,240\\
\hspace{3mm} LPVs with QR5 > 0.2 mag & 1\,219\,270\\
\hspace{3mm} LPVs with QR5 > 1.0 mag & 157\,029\\
\hspace{3mm} LPVs classified as C-stars & 546\,468\\
\hline
Sources in the 1$^{st}$ \Gaia LPV catalogue & 151\,761 \\
\hspace{3mm} with periods & 89\,617 \\
\hline
Sources in both \Gaia LPV catalogues & 145\,635\\
\hspace{3mm} with periods in both & 73\,362 \\

\hline
\end{tabular}
\end{table}

\begin{figure}
\centering
\includegraphics[width=\hsize]{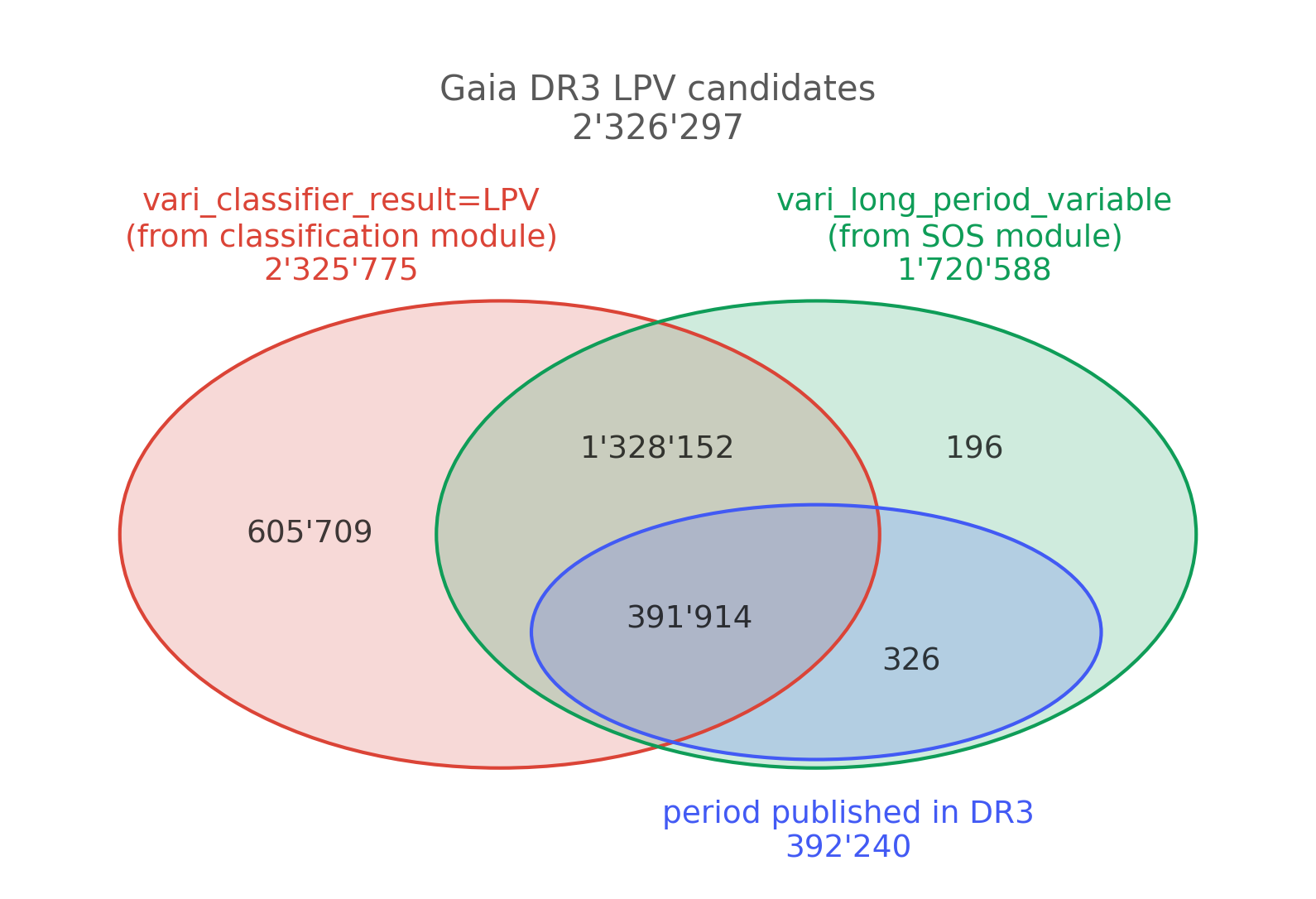}
\caption{Venn diagram of the repartition of the \Gaia DR3 LPV candidates in the \tableclas table with \texttt{best\_class\_name=LPV} (red, left in the figure) and \tablesos table (green, right in the figure) published in \Gaia DR3. The figure also illustrate the subset of sources whose period is published in \Gaia DR3, all of which are part of the \tablesos table.
All sources, except one, are present in \tableclas, but not necessarily with \texttt{best\_class\_name=LPV} (see text).
}
\label{Fig:VennDiagram}
\end{figure}

The identification of LPV candidates in \Gaia DR3
represents the convergence of two processing modules of the \Gaia variability pipeline. One module is tasked with the general classification of variable sources \citep[][]{DR3-DPACP-165}, while the other is a specific object study (SOS) module dedicated to the analysis of LPVs and the computation of LPV-specific attributes.
The existence of these two processing channels is reflected in the fact that LPV candidates are published in \Gaia DR3 in two separate, but largely overlapping, tables.
The results of the classification module are collected in the \tableclas table.
It contains 2\,325\,775 LPV candidates, identified in the table by the attribute \texttt{best\_class\_name=LPV} \citep[for a presentation of the classification pipeline and results see][]{DR3-DPACP-165}.
The results of the SOS module, on the other hand, are made available in the \tablesos table.
It contains 1\,720\,588 sources.
For simplicity, the \tableclas and \tablesos tables are referred to as ``classification table'' and ``SOS table'', respectively, in the rest of this paper.
Altogether, there are 2\,326\,297 LPV candidates in \Gaia DR3, of which almost 75\% are published in the second$^{}$ Gaia LPV catalogue (SOS table).
The numbers are summarised in Table~\ref{tab:catalog_numbers} and Fig.~\ref{Fig:VennDiagram}.
The main body of this paper focusses on the SOS table, while the classification table is briefly presented in Appendix\,\ref{app:ComparisonClassificationSOS}.

The vast majority of the sources in the SOS table were extracted from the set of LPV candidates among the classification table by applying the filters described in Sect.~\ref{sec:CatalogConstruction:FilteringCriteria}.
In addition, a small number of 522 \Gaia DR3 sources that were not classified as LPV candidates by the classification module but nonetheless fulfil all the SOS selection criteria for LPVs are present in the SOS table.
Of these, 502 were assigned \citep[see][Fig.~7]{DR3-DPACP-162} the \texttt{best\_class} type \texttt{SYST} (symbiotic stars) by the classification module, suggesting that they are indeed valid LPV candidates, and 19 were assigned the type \texttt{BE|GCAS|SDOR|WR} (B-type emission line star, $\gamma$~Cassiopeiae, S~Doradus, or Wolf-Rayet star), which, if correct (they are nevertheless very red), cannot be LPVs.
The remaining source, source \texttt{6715705353307835776,} does not have any  \texttt{best\_class} type.
It was initially classified as a Cepheid, as it was in DR2, but its classification has been questioned in the course of the \Gaia data analysis.
It was consequently removed from the classification results, but remained in the LPV table.
Its nature remains to be confirmed.

The attributes produced by the SOS module and published in the SOS table are described in Sect.~\ref{sec:CatalogConstruction:DataFields} and summarised in Table~\ref{tab:lpv_attributes}.
The period is published in the SOS table
in the form of the frequency and its associated error; see Sect.~\ref{sec:CatalogConstruction:DataFields}.
We note that the period of only a fraction of the LPV candidates is published in the SOS\ table of \Gaia DR3, as discussed in more detail in Sect.~\ref{sec:CatalogConstruction:PeriodAmplitudeDetermination}.
However, it is worth recalling that photometric time series are published for all \Gaia DR3 sources that are identified as variables, thereby enabling the users to independently study the variability of all \Gaia DR3 LPV candidates.

\subsection{Filtering criteria}
\label{sec:CatalogConstruction:FilteringCriteria}

We used the results of two sets of meta-classifiers from the classification pipeline to select the list of LPV candidates.
One set of meta-classifiers considered multi-class classifiers, and another set used one-versus-rest classifiers \citep[see][]{DR3-DPACP-165}.
Thresholds were used in these meta-classifiers to establish a preliminary set of LPV candidates, together with a mixed criterion involving the signal-to-noise ratio in the \gmag time series, the \gmag magnitude, and the variability amplitude. 
We then selected therein sources with \bpminrp colours larger than 0.5~mag and a 5-95\% quantile range larger than 0.1~mag in the $G$ band.
From the astrometric point of view, we restricted the sample to sources for which the re-normalised unit weight error (RUWE) is available, and for which the number of visibility periods used in the astrometric solution is greater than ten.

We finally imposed a lower threshold  on the number $N_G$ of data points in the cleaned \gmag time series, and a minimum value for the ratio of the number $N_{RP}$ of data points in the cleaned \grp time series to that in \gmag.
These limits are different for the classification and for the SOS tables.
We took $N_G \ge 9$ and $N_{RP}/N_G>0.5$ for the classification table, and $N_G \ge 12$ and $N_{RP}/N_G>0.8$ for the SOS table.
The limits were chosen to not exclude too many sources and at the same time have sufficient data points to determine period and colour. 
If more than 20\% of the \grp values were missed, the mean colour might be biased significantly and affect the C-star classification.
The limit on the number of good data points led to the exclusion of several bright nearby LPVs.
 
The catalogue of LPVs and the catalogue of short-timescale variables published in \texttt{vari\_short\_timescale} \citep{DR3-DPACP-162} have 3159 sources in common at the SOS\ level.
We verified that the positions of those with good parallaxes in the absolute colour-magnitude diagram (CMD) are compatible with the location of LPVs.

\begin{table*}
\caption{Data fields available in the \Gaia DR3 tables related to LPVs.}
\label{tab:lpv_attributes}
\centering
\begin{tabular}{ll}
\hline\hline
Data field name & Description \\
\hline
\multicolumn{2}{l}{*** Table \tableclas} \\
\texttt{source\_id}            & Unique source identifier of the LPV candidate \\
\texttt{best\_class}           & Equals \texttt{"LPV"} for all LPV candidates$^{(a)}$ \\
\texttt{best\_class\_score}    & Confidence (between 0 and 1) of the classifier in the identification of an LPV \\
\hline
\multicolumn{2}{l}{*** Table \tablesos} \\
\texttt{source\_id}            & Unique source identifier of the LPV candidate \\
\texttt{frequency}             & Frequency of variability of the LPV [day$^{-1}$] \\
\texttt{frequency\_error}      & Uncertainty on the frequency of the LPV [day$^{-1}$] \\
\texttt{amplitude}             & Half of the peak-to-peak amplitude of the best-fit model for the $G$-band light curve of the LPV [mag] \\
\texttt{median\_delta\_wl\_rp} & Median of the pseudo-wavelength separations between the two highest peaks in RP spectra \\
\texttt{is\_cstar}             & Flag to mark C-star candidates \\
\hline
\end{tabular}
\tablefoot{
  \tablefoottext{a}{With the exception of the 522 sources with a different type coming from the SOS processing channel.}
}
\end{table*}

\subsection{Data fields}
\label{sec:CatalogConstruction:DataFields}

The following data fields are provided in our second \Gaia catalogue of LPV candidates published in the \Gaia DR3 table \tablesos\footnote{For a description of the data fields published in the \tableclas table, see \citet{DR3-DPACP-165}.}.

\vspace{2mm}
\textbf{\scshape \large solution\_id \hypertarget{vari_long_period_variable-solution_id}}: Solution Identifier (long)

All Gaia data processed by the Data Processing and Analysis Consortium are tagged with a solution identifier. 
This is a numeric field attached to each table row that can be used to unequivocally identify the version of all the subsystems that where used in the generation of the data as well as the input data used.\footnote{To decode a given solution ID,  visit \url{https://gaia.esac.esa.int/decoder/solnDecoder.jsp}.}

\vspace{2mm}
\textbf{\scshape \large source\_id \hypertarget{vari_long_period_variable-source_id}}: Unique source identifier (long)

A unique single numerical identifier of the source obtained from gaia\_source (for a detailed description, see {\tt gaia\_source.source\_id}).

\vspace{2mm}
\textbf{\scshape \large frequency \hypertarget{vari_long_period_variable-frequency}}: Frequency of the LPV (double, Frequency[$day^{-1}$])

This field is the frequency found for the LPV candidate.
It is provided for a subset of the SOS table (see Fig.~\ref{Fig:VennDiagram} and Sect.~\ref{sec:CatalogConstruction:PeriodAmplitudeDetermination}).

\vspace{2mm}
\textbf{\scshape \large frequency\_error \hypertarget{vari_long_period_variable-frequency_error}}: Error on the frequency (float, Frequency[$day^{-1}$])

This field gives the error on the frequency for the LPV candidate. See Sect.\,\ref{sec:CatalogConstruction:PeriodAmplitudeDetermination}. 

\vspace{2mm}
\textbf{\scshape \large amplitude \hypertarget{vari_long_period_variable-amplitude}}: Amplitude of the LPV variability (float, Magnitude[mag])

This field gives the peak-to-peak semi-amplitude in magnitude, based on a best-fit model (see Sect.~\ref{sec:CatalogConstruction:PeriodAmplitudeDetermination}). It might differ from the observed magnitude range, which is stored as \texttt{trimmed\_range\_mag\_g\_fov} in the \Gaia archive table \texttt{vari\_time\_series\_statistics} and which can be retrieved as described in Appendix~\ref{app:catalogRetrieval}. 

\vspace{2mm}
\textbf{\scshape \large median\_delta\_wl\_rp \hypertarget{vari_long_period_variable-median_delta_wl_rp}}: Median of the pseudo-wavelength separations between the two highest peaks in the low-resolution RP spectra  (float, dimensionless). 
The wavelength difference between the two peaks is directly related to the presence or absence of specific molecular bands. 

This value is therefore used in the definition of the {\it isCstar} parameter; see Sect.~\ref{sec:CatalogConstruction:IdentificationCstars} for further explanation on its computation and usage. 
It is set to NaN when the spectrum does not allow automatically identifying two maxima in a reliable way.

\vspace{2mm}
\textbf{\scshape \large isCstar \hypertarget{vari_long_period_variable-isCstar}}: Flag to mark C stars  (Boolean, Dimensionless[see description])

The parameter {\it isCstar} is set to TRUE if a star has been identified as a C-star candidate based on the value of the {\it  median\_delta\_wl\_rp} parameter derived from the RP spectrum shape. It is set to FALSE if it is an O-rich candidate.
It is set to NULL when the shape of the spectrum does not allow an automatic classification between these two types of LPVs (i.e. when median\_delta\_wl\_rp is NaN).
See Sect.\,\ref{sec:CatalogConstruction:IdentificationCstars} for more details.

\vspace{2mm}
Some parameters that had been part of the first$^{}$ LPV catalogue are not included in this second$^{}$ catalogue, namely the bolometric correction, the absolute magnitude computed with the bolometric correction and the Gaia parallax, and the red supergiant flag.
All these parameters were problematic due to the lack of reliable values for the interstellar extinction and reddening at DR2.
This problem could not be solved for the DR3 analysis.
With the large number of LPVs in the Galactic disc and the extension of the catalogue towards lower brightness and thus larger distances, the lack of a reliable interstellar extinction would make values for the three mentioned parameters highly uncertain for a very large fraction of the catalogue stars.
However, these parameters are still kept in the internal data model and will likely be published with DR4.
For the meantime, we refer to the formula for computing bolometric corrections for cool giants from Gaia photometry, which our group has published in the appendix of \citet{lebzelter_etal_2018}.

\subsection{Period and amplitude determination}
\label{sec:CatalogConstruction:PeriodAmplitudeDetermination}

\begin{figure}
\centering
\includegraphics[width=\hsize]{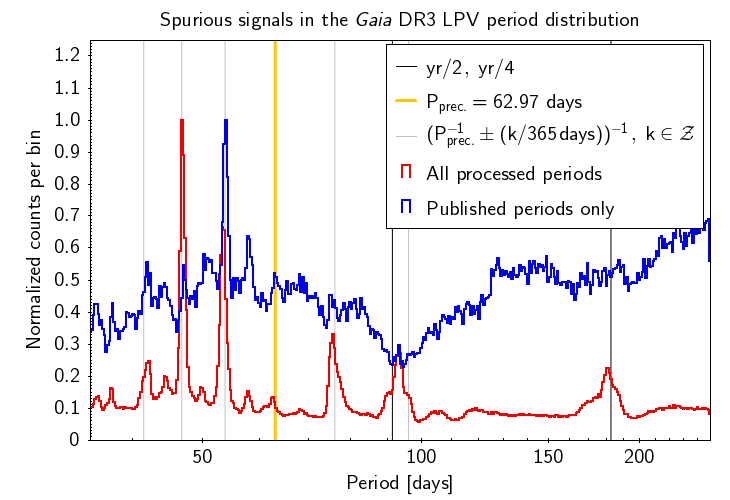}
\caption{Spurious signals in the period distribution. The histograms show the distribution of periods derived for all \Gaia DR3 LPV candidates (red) and limited to the published periods (blue), showing the substantial reduction of spurious signals in the latter. Vertical lines mark the periods equal to half a year and one quarter of a year (black), the alias caused by the \Gaia spacecraft precession period $P_{\rm prec.}=62.97$ days (yellow), and the spurious signals caused by the interaction of the latter with the yearly observing window of \Gaia.
For visualisation purposes, both histograms are limited to the range of 35-250 days, and normalised to their maximum value over that interval.}
\label{Fig:PeriodAliases}
\end{figure}

The periods of the \gmag light curves of the LPV candidates were searched using a standard least- squares method. 
For this, the \gmag time series were first cleaned from outliers as described in the Gaia DR3 release documentation\footnote{\url{https://gea.esac.esa.int/archive/documentation/GDR3/}}.
The frequency search range was set to 0.0007 to 0.1 d$^{-1}$ with a frequency step of 3.3$\cdot$10$^{-5}$\,d$^{-1}$.
The highest peak in the Fourier spectrum of each source was considered for the exported period. 
No period was exported when this result was below 35\,d or above the total duration of the obtained time series of that source.
This was done to reduce the risk of publishing spurious periods due to aliasing effects or highly uncertain values on the long-period end. 
Fig.~\ref{Fig:PeriodAliases} shows a close-up of the period distribution between 35 and 250 days, showing the distribution of published periods and the distribution of first periods determined for all LPV candidates in DR3 without filtering. In the latter case, a number of aliases are clearly visible, to begin with, the signals at 1/2 yr and 1/4 yr. 
At 62.97 days, we find the signal corresponding to the \Gaia precession period \citep[corresponding to 5.8 revolutions per year, i.e. a period of 365.25/5.8~days; see][]{gaia_mission_2016}. 
Most of the other peaks can be attributed to the interaction of the one-year period with the precession period, which produce aliases at periods
\begin{equation}\label{eq:precession_alias}
    P_{\rm alias}(k) = \left(\frac{1}{62.97\,{\rm days}}\pm\frac{k}{365\,{\rm days}}\right)^{-1} \,,
\end{equation}
for several integer values of $k$ \citep[see e.g.][]{vanderplas_2018}.
In total, 872\,693 stars do not have a period because of the short-period limit.
The computed period exceeded the total duration of the time series for 25 797 stars, and hence they also do not have a period entry in the catalogue. 

Furthermore, periods are not published if a Spearman correlation
higher than 0.75 is observed between the image parameter determination (IPD) goodness-of-fit time series and the \gmag time series. 
This would indicate a potential spurious variability in the G band \citep[see also][]{DR3-DPACP-164}.
Finally, the S/N in the G band has to exceed 15 to include a period in the catalogue.
In total, periods are provided for 392\,240 sources in the catalogue.
The filtering applied to select the periods to be published in DR3 is effective in rejecting a large number of spurious periods (see Fig.~\ref{Fig:PeriodAliases}), but a small fraction of them remain in the catalogue. 

For the amplitude estimation, a Fourier series with up to three harmonics is fit to the light curve using the previously derived period.
If a significant trend is present in the light curve, it is accounted for by adding a linear term to the model used to determine the amplitude.
The published amplitude corresponds to the amplitude of the fundamental harmonic of the Fourier series (amplitude of the sine function, i.e. half peak-to-peak).

\subsection{Identification of C stars}
\label{sec:CatalogConstruction:IdentificationCstars}

\begin{figure}
\centering
\includegraphics[width=\hsize]{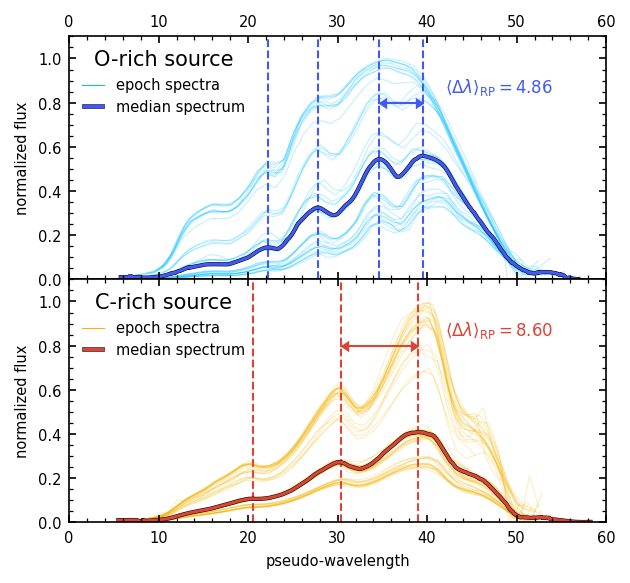}
\caption{Low-resolution RP spectra of the O-rich star T Aqr (top panel) and of the C-rich star RU Vir (bottom panel). Thin lines represent epoch spectra, and the thick, darker lines are median spectra. Vertical dashed lines indicate the value of the pseudo-wavelength at the most prominent peaks of the median spectra. The arrows mark the distance in pseudo-wavelength between the two highest peaks, $\mediandeltawl$, whose value is indicated. This is the adapted version of the figure published as Gaia image of the week IoW\_20181115.}
\label{Fig:example_spectra}
\end{figure}

\begin{figure*}
\centering
\includegraphics[width=.49\hsize]{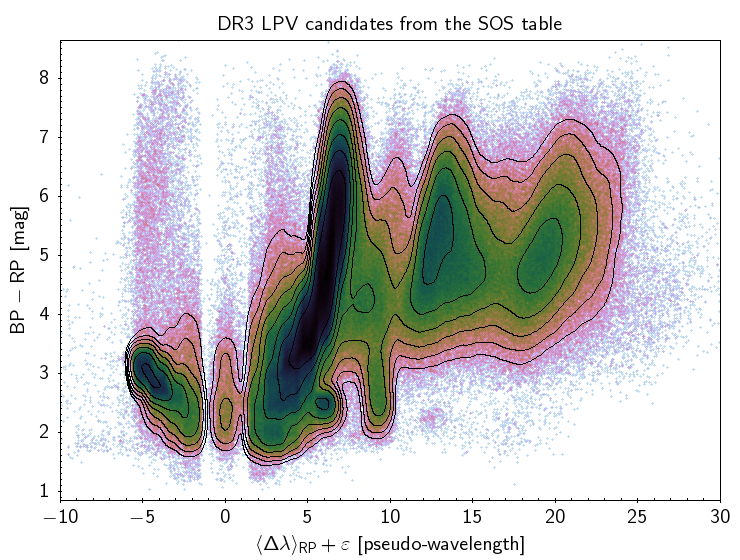}
\includegraphics[width=.49\hsize]{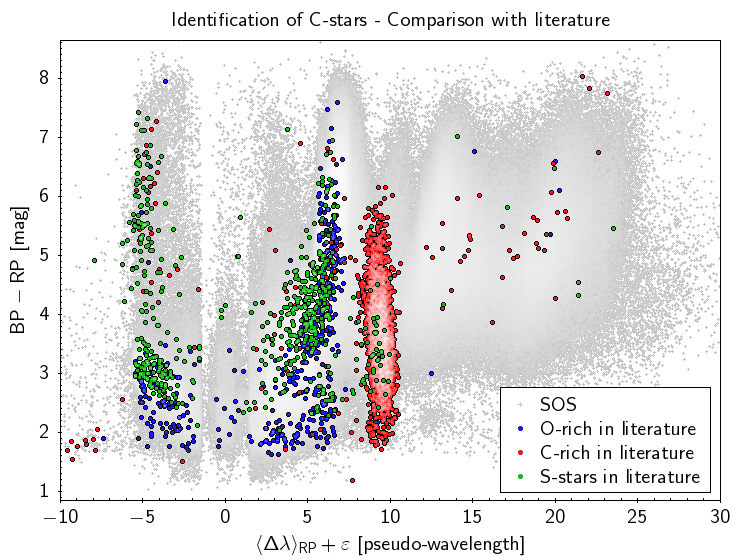}
\caption{Density-mapped distribution of the LPV candidates from the SOS table in the $\mediandeltawl$ vs $\gbp-\grp$ plane. Density contour lines are displayed in the left panel to highlight patterns in the diagram. In the right panel, sources spectroscopically identified in the literature as O-rich (blue), C-rich (red), or S-stars (green) are displayed on top of the data from the SOS table. Random artificial errors $-0.5\leq\varepsilon\leq0.5$ have been applied to $\mediandeltawl$ for visualisation purposes, in order to smooth out the artificial clustering of that parameter around a discrete set of value.}
\label{Fig:selectionCstars}
\end{figure*}

\begin{figure}
\centering
\includegraphics[width=\hsize]{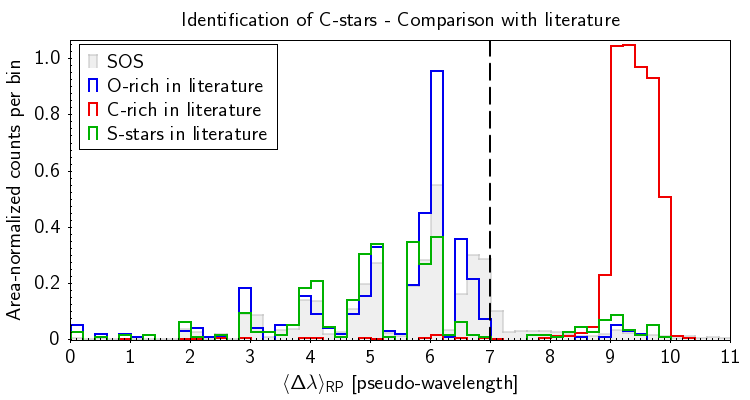}
\caption{Distribution of $\mediandeltawl$ for the SOS table (grey) and the sources identified in the literature as O-rich (blue), C-rich (red), or S stars (green).}
\label{Fig:hist_mediandeltawlrp}
\end{figure}

A significant new feature of this second$^{}$ \Gaia catalogue of LPVs with respect to the first catalogue published in \cite{mowlavi_etal_2018_dr2lpv}, in addition to the much higher number of candidates, is the identification of carbon stars (C stars). 
The C/O ratio in the atmospheres of these objects is higher than one as a result of nucleosynthesis and mixing events on the asymptotic giant branch (AGB). 
This results in signatures of carbon-bearing molecules such as CN or C$_{2}$ dominating the spectrum, while M-type stars are characterized by the molecular bands of TiO and VO. 

For the identification of C stars among the LPVs, we used an approach guided by a method based on narrow-band photometry \citep{1982AJ.....87.1739P}.
The wavelengths of the band heads, and thus the location of flux maxima and minima in the spectrum, are very different for C- and M-type stars.
The classical narrow-band method places one filter at 778 nm and another at 812 nm. In C stars, this corresponds to a pseudo-continuum point and a point within the depression due to a band head, respectively, while for M stars, the two filters measure the opposite. 

For the Gaia RP spectra, a direct application of this approach was not possible due to limits in resolution and wavelength calibration.
Figure~\ref{Fig:example_spectra} shows representative spectra of an M-type and a C-type LPV, where the broad features that can be seen are the result of molecular bands.
A clear difference in the shape of the spectra is obvious.
In particular, we note a difference in the distance between the two highest peaks.
For M-type stars, comparison with the spectral catalogue from \citet{2000A&AS..146..217L} suggests that the depression between the two highest peaks is caused by the TiO$\delta$ ($\Delta\nu$=0) band near 900 nm.
In the case of C stars, the two highest peaks within the spectral range covered by the RP spectra limit a strong CN band near 800 nm \citep[compare][]{2017A&A...601A.141G}. 

The SOS module examines this feature in an automated way by computing the pseudo-wavelength difference between the two highest peaks in each RP spectrum of a given source, taking the median value of the results, and storing it in the parameter \texttt{median\_delta\_wl\_rp}.
In this approach we assume all stars with \texttt{median\_delta\_wl\_rp} $>$7 to be C rich (isCstar = TRUE).

The distribution of the \texttt{median\_delta\_wl\_rp} parameter for all stars in the DR3 catalogue of LPV candidates is shown in the left panel of Fig.\,\ref{Fig:selectionCstars}. 
A random value of $\varepsilon$ between -0.5 and 0.5 was added to the parameter $\mediandeltawl$ in order to smear out the discretisation brought by the 
pixelisation and adopted number of significant digits in the pseudo-wavelengths.
The majority of the sources are O-rich candidates according to our classification, with \texttt{median\_delta\_wl\_rp} values lower than 7.
The selection of the limit of 7 to divide the two regimes in chemistry was guided by the theoretical considerations mentioned above and the location of the vast majority of spectroscopically identified C-rich stars in the literature (Figs.\,\ref{Fig:selectionCstars}, right panel, and \ref{Fig:hist_mediandeltawlrp}).
These latter sources were taken from \citet{christlieb_etal_2001}, \citet{MacConnell_2003}, \citet{Si_etal_2015}, and \citet{Li_etal_2018} for the C stars and from \citet{Speck_etal_2000} and \citet{yung_etal_2014} for the O-rich stars.

The candidates with $\texttt{median\_delta\_wl\_rp} > 12$ deserve some attention.
Very few objects at these large pseudo-wavelengths have a known chemistry from the literature, and they contain both C-rich and O-rich stars (see the right panel of Fig.~\ref{Fig:selectionCstars}).
A preliminary investigation of a subsample of sources in this region reveals that in these cases, the automatic algorithm detected as the second maximum a peak that corresponds to a molecular band farther to the blue. 
If the algorithm picks the incorrect peak, no conclusion about the C-rich or O-rich nature can be drawn.
This likely results from lower S/N or observations in crowded or highly reddened sky regions that might degrade the shape of the observed RP spectrum, thereby compromising the correct measurement of $\mediandeltawl$.
A large fraction of these stars have $\gbp>19$~mag and are found preferentially at low galactic latitudes.

We conclude that our selection method works very well for objects with spectra of sufficient S/N, but becomes increasingly uncertain for low-brightness stars.
Problematic cases are in particular those with $\texttt{median\_delta\_wl\_rp}$ > 12, for which we assume the classification to be highly uncertain.
For this reason, this parameter has been published together with the C-star flag.

We note a slight colour dependence for the lower limit of the $\texttt{median\_delta\_wl\_rp}$ range
for C stars, suggesting that for the reddest stars (that are affected by high interstellar reddening), the value of 7 may be slightly too low for the separation.
To select a sample of C stars from the SOS table, we recommend using 7 < $\texttt{median\_delta\_wl\_rp}$ < 11 and $\gbp<19$~mag. 
Tests show that limiting this sample further to objects with a period entry in the SOS table and adapting the limit for the coolest stars to 7.5 or 8 provides the best results.
In Appendix \ref{app:AdditionalDefinitions:RegionsMediandeltawlColor} we suggest boundaries in the $\gbp-\grp$ versus $\mediandeltawl$ plane to select C-rich and O-rich stars, with low contamination from the other types of sources.

We also found a number of sources with negative values of $\texttt{median\_delta\_wl\_rp}$.
In these cases, the peak on the blue side of the molecular absorption band is slightly higher than the peak on the red side. 
As a consequence, the $\texttt{median\_delta\_wl\_rp}$ value is mirrored.
This occurs more often for bluer objects where the two peaks are of similar height.
Comparison with our calibration sample from the literature shows that most of these objects are O rich, as expected.
Stars with a parameter value close to 0 hardly show any molecular bands and are likely of early spectral type, in agreement with their $\gbp-\grp$ colour.

Finally, we identify a small group of objects with $\gbp-\grp$ around 2.3 mag and $\texttt{median\_delta\_wl\_rp}$ = 13.
A preliminary study of objects in this area shows a significant fraction of stars with a strong emission peak in their spectra that most likely corresponds to the H$\alpha$ line. 
Comparison with literature studies on individual members of this group supports the identification as emission line stars, among them symbiotic stars and Bp stars. 

The case of S-type stars deserves a special note.
S stars are significantly enhanced in $^{12}$C and s-process elements due to the third dredge-up. 
While the optical spectrum is still dominated by TiO, they are identified by the appearance of ZrO bands in that wavelength range \citep{2017A&A...601A..10V}.
Depending on the amount of material that is mixed in the surface, S stars form a kind of continuum from M- to C-type with subtypes MS, S, SC, and CS.
Taking a randomly selected sample of 644 S stars from SIMBAD (those among the first 1000 S-type stars in SIMBAD that have a cross-match with our catalogue), we find the vast majority of them at $\texttt{median\_delta\_wl\_rp}$ < 7, that is, they appear as O rich in the catalogue (Figs.\,\ref{Fig:selectionCstars}, right panel, and \ref{Fig:hist_mediandeltawlrp}).
A few S stars are found in the parameter range attributed to C stars, and these may represent the fraction of S stars that is close to spectral type C. 
While the classification of S stars with this parameter will require further detailed analysis, we conclude that most S stars will be classified as oxygen-rich in the second$^{}$ Gaia LPV catalogue. 

\subsection{Comparison of DR3 and DR2}
\label{sec:CatalogOverview:ComparisonDR3-DR2}

\begin{figure}
\centering
\includegraphics[width=\hsize]{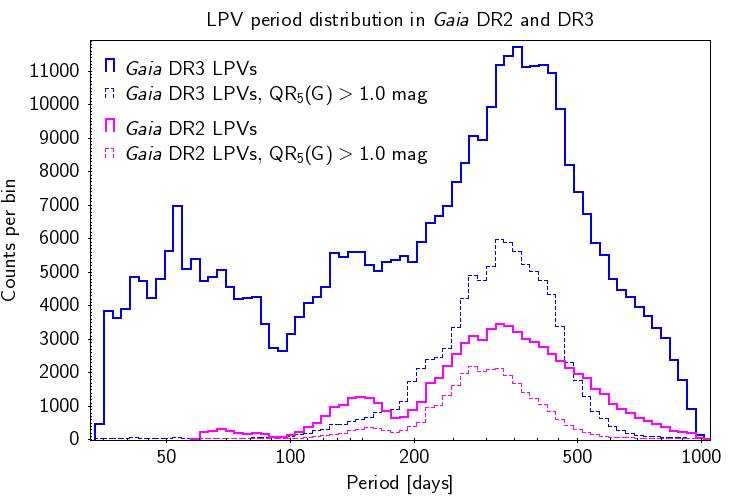}
\caption{Period distribution of LPV candidates from \Gaia DR2 (red) and \Gaia DR3 (blue). Thin dashed lines are limited to sources with a $G$-band amplitude larger than 1 mag (traced by the 5-95\% interquantile range, $\qrg$), which are mira candidates.}
\label{Fig:compare_P_DR2DR3_distrib}
\end{figure}

\begin{figure}
\centering
\includegraphics[width=\hsize]{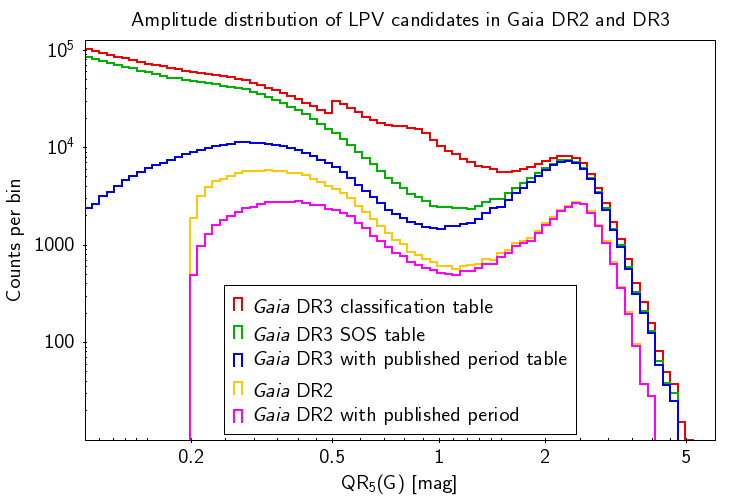}
\caption{Distribution of the $G$-band variability amplitude (traced by the 5-95\% interquantile range, $\qrg$) for the \Gaia DR3 LPV candidates of the classification table (red), the SOS table (green), and those with a published period (blue). The content of the \Gaia DR2 catalogue of LPV candidates (yellow), and its subset with periods published in DR2 (magenta), is also shown for comparison. A logarithmic scale is used along the vertical axis.}
\label{Fig:classif_SOS_qr5}
\end{figure}

\begin{figure}
\centering
\includegraphics[width=\hsize]{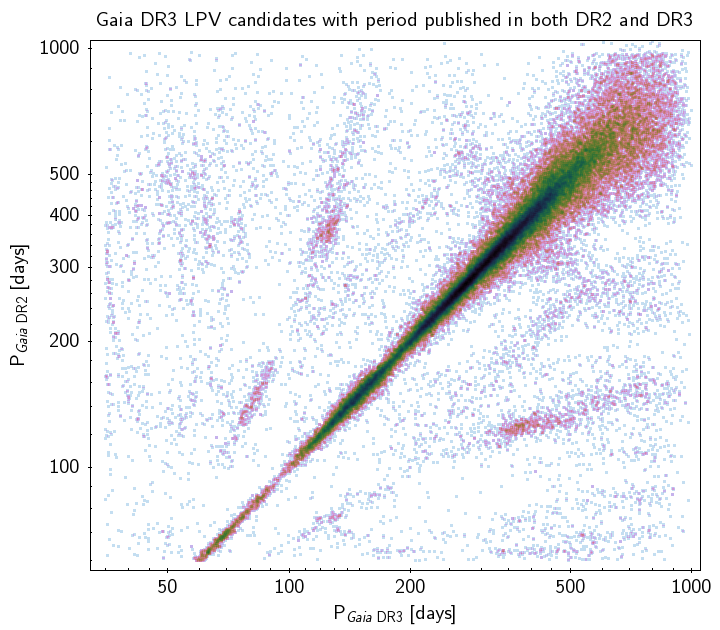}
\caption{Density-mapped comparison of the periods published in DR2 and DR3 for the LPV candidates common to the two data releases. The colour tone extends from blue to red and moves towards progressively more densely populated areas of the plot.}
\label{Fig:compare_P_DR2DR3}
\end{figure}

To compare the first$^{}$ and second$^{}$ Gaia catalogue of LPV candidates, we have to distinguish between the classification table and the more restrictive SOS table.
For a proper comparison of the two catalogues, we have to explain the two-table structure that is found in the first$^{}$ and second catalogue.
In the first$^{}$ catalogue, we provided a table of 151\,761 LPV candidates (\texttt{vari\_classifier\_result}) based on number of observations, variability amplitude, colour, and Abbe value. 
From this sample, a sub-sample was drawn for which periods and a few further parameters were exported (\texttt{vari\_long\_period\_variable}).
A small number of objects were added to that second table from a different classification path \citep[see][for details]{mowlavi_etal_2018_dr2lpv}, providing a sample of 89\,617 objects.
This latter sample can be considered the cleanest selection of LPV candidates based on DR2 data.

For the second$^{}$ catalogue, selection criteria were slightly different, and the output consists, if we wish to draw a comparison to the first$^{}$ catalogue, of three sub-samples.
This is schematically shown in Fig.\,\ref{Fig:VennDiagram}.
Depending on which (sub-)samples we compare, the increase in number from the first$^{}$ to the second$^{}$ catalogue lies between 4 and 15.
A better way of comparing the two catalogues is shown in Figs.\,\ref{Fig:compare_P_DR2DR3_distrib} and \ref{Fig:classif_SOS_qr5}.
In Fig.\,\ref{Fig:compare_P_DR2DR3_distrib}, the period distributions of the DR2 and DR3 sub-samples with an exported period are shown.
In addition, we identified the fraction of objects with an amplitude exceeding 1 mag in G.
The general shape of the distribution of the stars in the second$^{}$ catalogue is very similar to that from the first$^{}$ catalogue, except for the additional group of objects with short periods down to 35\,d and a significant extension on the long-period end.
The latter results mainly from the extended time coverage of the light curves available in DR3.
The various components visible in this distribution are discussed in Sect.\,\ref{sec:CatalogOverview:Variability}.
The distribution in amplitude (Fig.\,\ref{Fig:classif_SOS_qr5}) shows the most obvious difference again in the extended parameter range.
The number of objects that were added with small amplitudes is very significant and completely dominates the increase in object number for stars in the SOS table without exported periods (green line in Fig.\,\ref{Fig:classif_SOS_qr5}).
However, we note an increase among the large-amplitude variables as well, suggesting that in DR2, many of them had remained undetected.
The difference between the SOS table and the classification table in this diagram is discussed in Appendix \ref{app:ComparisonClassificationSOS}.

A comparison of the periods from DR3 with those from DR2 is presented in Fig.~\ref{Fig:compare_P_DR2DR3}.
The comparison was possible for 73\,362 sources for which periods in both data releases exist.
For 74\% of these objects, the periods agree within 10\%. 
For the remaining stars, various reasons for the observed difference can be detected in Fig.~\ref{Fig:compare_P_DR2DR3}.
A small fraction of stars, visible in the upper left and lower right corner, shows periods that differ by a factor of ten. 
In these cases, we suspect that the star varies with a short and a long secondary period, where one or the other had been detected in our two studies.
In addition, several structures are visible in Fig.~\ref{Fig:compare_P_DR2DR3} that are a result of the sampling of the light curves \citep[][see also Sect.\,\ref{sec:CatalogQuality:PeriodRecovery:OGLE3}]{vanderplas_2018}.
The quality of the periods is evaluated in Sect.\,\ref{sec:CatalogQuality:PeriodRecovery} by comparison with data from OGLE-III, ASAS-SN, and a sample of well-observed Galactic field stars.

\section{Catalogue overview}
\label{sec:CatalogOverview}

In this section, we give an overview on the content of the DR3 \Gaia LPV catalogue. 
We present the sky distribution of the LPV candidates in the catalogue in Sect.~\ref{sec:CatalogOverview:SkyDistribution} and the colour-magnitude diagram in Sect.~\ref{sec:CatalogOverview:CMD}.
We then summarise the variability properties in Sect.~\ref{sec:CatalogOverview:Variability}. 
The entire analysis is based on the SOS table.
The content of the classification table is briefly described in Appendix \ref{app:ComparisonClassificationSOS}.

\subsection{Sky distribution}
\label{sec:CatalogOverview:SkyDistribution}

\begin{figure}
\centering
\includegraphics[width=\hsize]{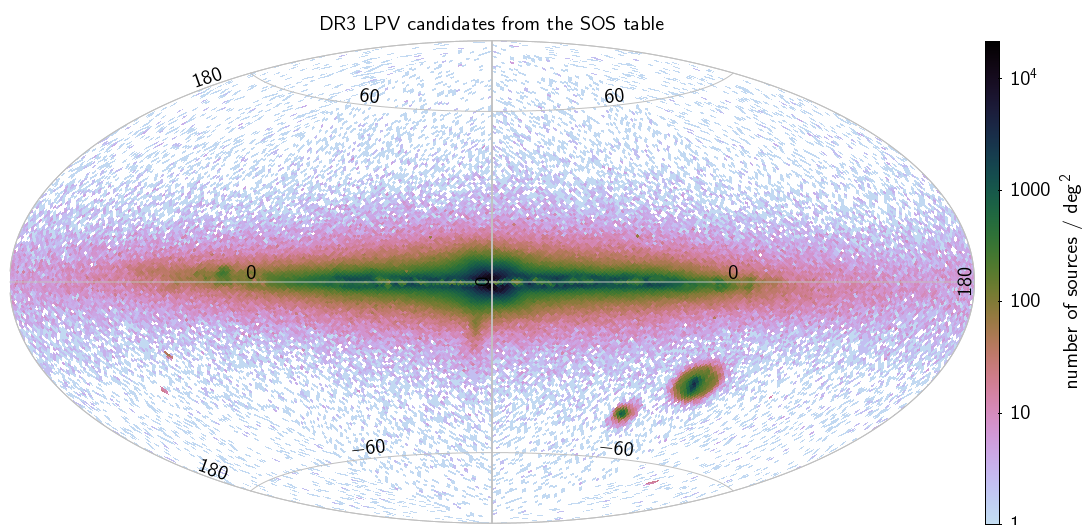}
\caption{Sky distribution of the sources from the SOS table. The basic parts of the Milky Way and several local group galaxies are clearly visible.}
\label{Fig:sky_distribution}
\end{figure}

\begin{figure}
\centering
\includegraphics[width=\hsize]{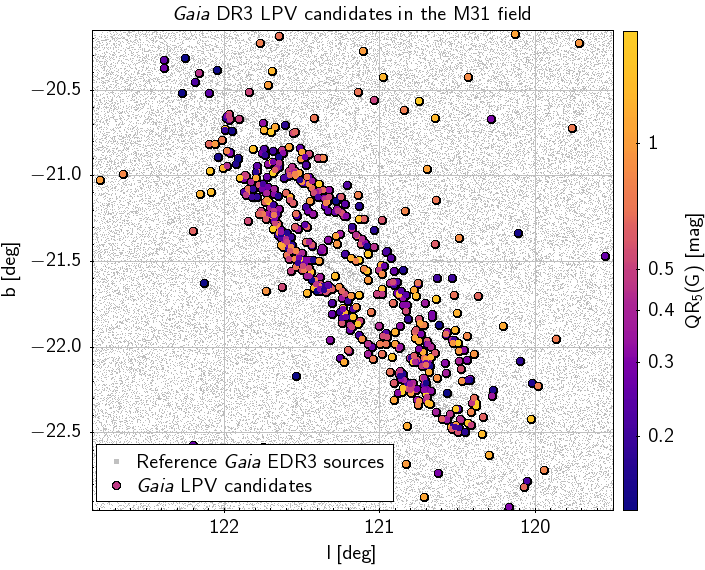}
\caption{\Gaia DR3 LPV candidates in the field of M31 (in Galactic coordinates), colour-coded according to their $G$-band amplitude (traced by the 5-95\% interquantile range). Grey symbols in the background are reference sources from \Gaia EDR3.}
\label{Fig:M31_zoom}
\end{figure}

\begin{figure}
\centering
\includegraphics[width=\hsize]{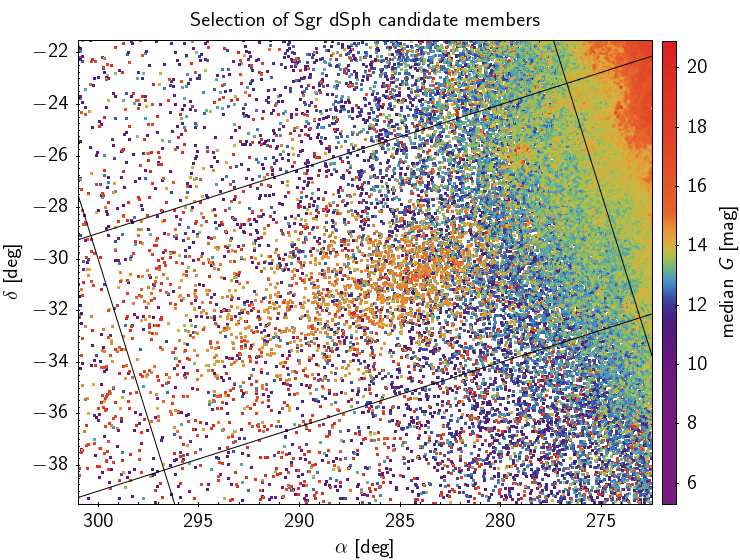}
\caption{Sky map towards Sgr dSph with \Gaia DR3 LPV candidates colour-coded according to their median $G$-band magnitude. Solid lines represent the cuts applied to select sources in that galaxy (see Table~\ref{tab:LocalGroupGalaxiesSelection}).}
\label{Fig:sky_Sgr}
\end{figure}

The sky distribution of all DR3 LPV candidates is shown in Fig.~\ref{Fig:sky_distribution}, where colour-coding indicates the object density. 
Several structures can be identified easily because they harbour a large number of LPVs:
the Galactic disc, bulge, and halo are clearly visible, as well as the extragalactic populations of the Magellanic Clouds (the two concentration areas below the Galactic plane at Galactic longitudes of about 285° for the Large Magellanic Cloud (LMC) and 303° for the Small Magellanic Cloud (SMC)) and of the Sagittarius dwarf galaxy (elongated tail below the Galactic bulge). 
On the lower left, M31 and M33 can be spotted. 
Figure~\ref{Fig:M31_zoom} shows a zoom into the region around M31.
The overall shape of the galaxy is well reproduced by the detected LPV candidates.
Colour-coding shows that this sub-sample does not contain a large fraction of miras, but many stars with a moderate light amplitude.  
As further illustrated in Sect.\,\ref{sec:CatalogQuality:SourcesWithKnownDistance:LocalGroup}, the sample is dominated by supergiants that are located primarily in the galaxy spiral arms, as shown in Fig.\,\ref{Fig:M31_zoom}. 

Another example of an interesting sky area with a high stellar density is shown in Fig.\,\ref{Fig:sky_Sgr}.
LPVs allow easily distinguishing the Sgr dSph from the Galactic bulge in the foreground by G-band brightness, as indicated by the colour-coding.
Again we refer to Sect.\,\ref{sec:CatalogQuality:SourcesWithKnownDistance:LocalGroup} for the further characterisation of the LPVs in that galaxy.
Further interesting sky maps in which the LPV candidates are marked are shown in Appendix \ref{app:AdditionalDefinitions:LocalGroupGalaxies}, Figs.\,\ref{Fig:sky_M31_M33_Fornax} and \ref{Fig:sky_Fornax_NGC6822_IC10_LeoI}.

\subsection{Colour-magnitude diagram}
\label{sec:CatalogOverview:CMD}

\begin{figure*}
\centering
\includegraphics[width=.49\hsize]{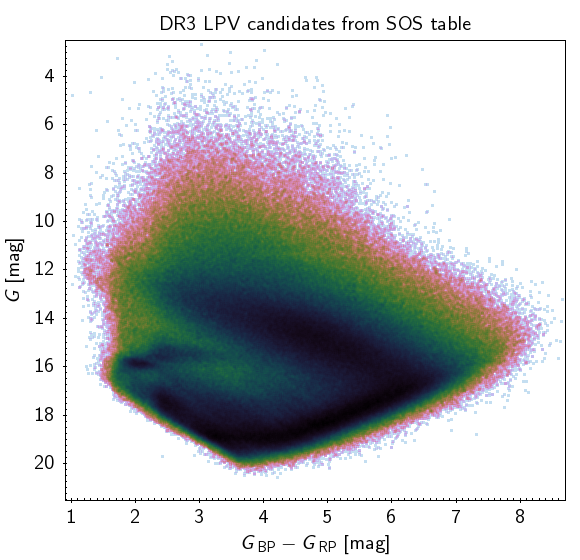}
\includegraphics[width=.49\hsize]{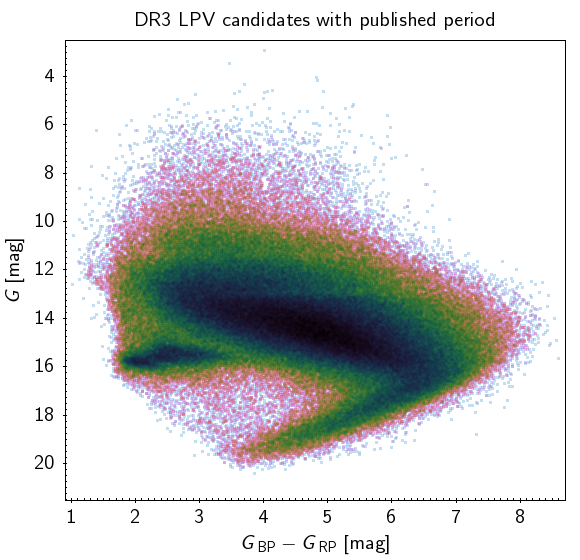}
\caption{Density-mapped colour-magnitude diagrams in the \Gaia passbands of the \Gaia DR3 LPV candidates. The panel on the left 
includes all sources in the SOS table, and in the right panel, the sample is limited to LPVs with periods published in \Gaia DR3. Sources belonging to the Magellanic Clouds are visible in both panels at $15\lesssim G\,/\,{\rm mag}\lesssim16$ and $1.5\lesssim(\gbp-\grp)\,/\,{\rm mag}\lesssim3.5$.}
\label{Fig:DR3_CM}
\end{figure*}

\begin{figure*}
\centering
\includegraphics[width=.49\hsize]{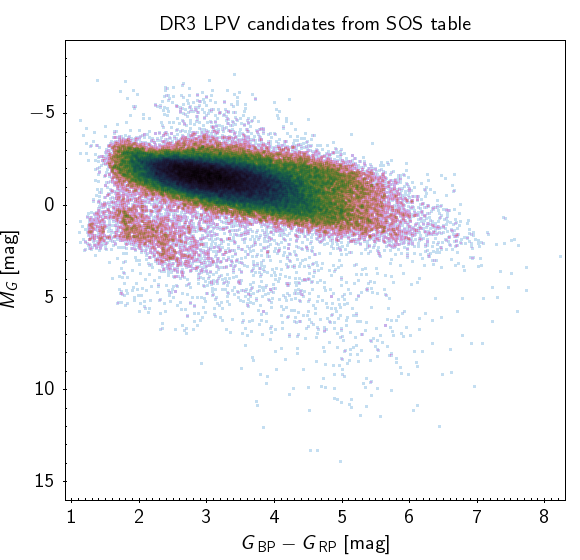}
\includegraphics[width=.49\hsize]{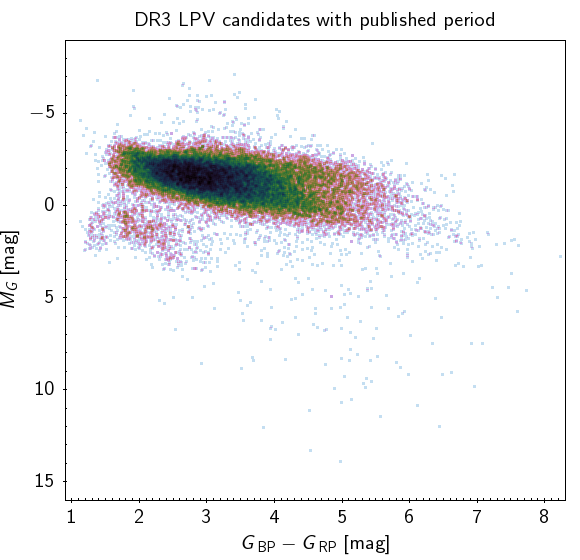}
\caption{Similar to Fig.~\ref{Fig:DR3_CM}, but showing the colour vs. absolute magnitude diagrams. In both panels, the sample is limited to sources with positive parallaxes and relative parallax uncertainties smaller than 15\%.}
\label{Fig:DR3_HRD}
\end{figure*}

In Fig.~\ref{Fig:DR3_CM} we present G versus $\gbp-\grp$ CMDs for the LPV candidates in the SOS table (left panel) and for only those LPVs for which a period could be determined (right panel). 
The spread in G magnitudes is largely due to distance effects. 
In both diagrams, the Magellanic Clouds can be clearly seen as a band at G$\approx$15-16 mag. 
Colours on the horizontal axis result from the temperature differences and interstellar and circumstellar reddening.
Filtering criteria limited the blue end of the LPV range to $\gbp-\grp$=0.5 mag.
However, no objects are bluer than $\gbp-\grp$=1.0 in our LPV database.
We applied a cut for objects in the lower left corner of Fig.\,\ref{Fig:DR3_CM}. 
The characteristics of objects that are blue, that is, not reddened, and that are faint suggests that these objects are not LPVs.
The removal was applied when filtering from the classification to the SOS table, so that these sources can be easily extracted from the classification table.
On the red end, $\gbp$ brightness drops below the Gaia sensitivity limit, leading to a lack of objects in the bottom right corner of the diagrams.

In the right panel of Fig.~\ref{Fig:DR3_CM}, only stars with an exported period (see \ref{sec:CatalogConstruction:PeriodAmplitudeDetermination}) are plotted.
Of the criteria that were applied to select periods for publication, the S/N limit of 15 has the largest selection effect. 
This affects, interestingly, both the brightest and the weakest objects, as can be seen in the CMD. This hinders the determination of a reliable period.
In addition, LPV candidates in the lower left part of the diagram are intrinsically blue and comparably weak.
The algorithm detected short periods in most of these cases, which are below the limit of 35 d that we set to export this parameter.
An interesting feature is the group of objects that is located along the sensitivity limit in the BP filter described above.
These are mainly C stars, as we discuss below. 

Almost 92\,000 of our LPV candidates have parallax uncertainties smaller than 15\%.
These are plotted in an observational Hertzsprung-Russell-diagram (HRD) in Fig.~\ref{Fig:DR3_HRD}.
No correction of interstellar extinction was applied.

The majority of the LPV candidates are found along the upper giant branch, which confirms their nature as highly evolved red giants. 
Several objects brighter than this sequence are variable red supergiants.
At G$\approx$1.0 lies a group of likely red clump stars.

The HRD gives a first impression of the content of the second$^{}$ Gaia LPV catalogue. 
The bulk of the sources lies in the expected region of AGB stars.
Blue stars fainter than $G_{abs}$=-0.5 amount to about 4\% of the LPV candidates.
We suspect them to be a mixture of red clump stars and red giants. 
The blue end consists of stars in the solar neighbourhood within a radius of 2 kpc, and their magnitude corresponds to the brightness of the red clump.
A few red supergiants are also visible on the bright side of the distribution.
Red supergiants and highly reddened stars will be more frequent in the catalogue than in the sample limited by parallax uncertainty. 

\subsection{Variability properties}
\label{sec:CatalogOverview:Variability}

\begin{figure}
\centering
\includegraphics[width=\hsize]{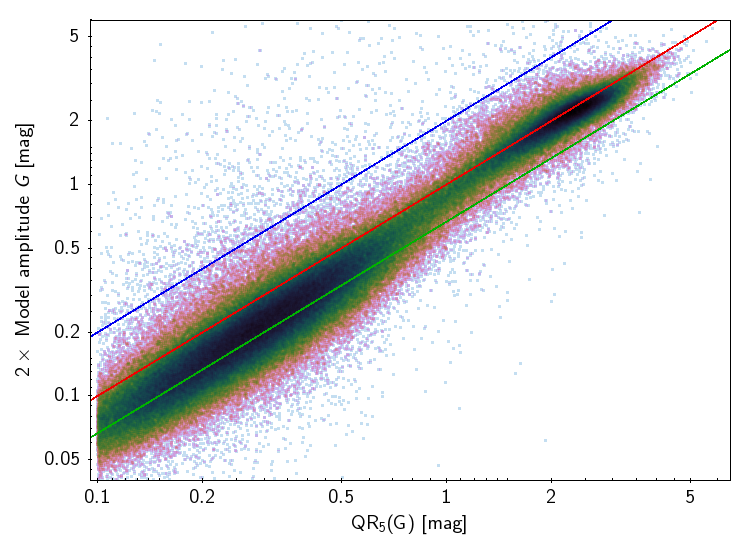}
\caption{Twice the best-fit model amplitude in the $G$ band compared to the 5-95\% interquantile range $\qrg$ for the \Gaia DR3 LPV sources whose period has been retained for publication. The blue, red, and green lines (from top to bottom) indicate where the model amplitude equals $2\,\qrg$, $\qrg,$ or $2/3\,\qrg$.}
\label{Fig:AmplitudeVSQR5}
\end{figure}

For all LPV candidates in the SOS table, the 5-95\% interquantile range $QR_{5}(G)$ can be extracted from the Gaia database.
For the subsample of 392\,252 objects with periods, the second$^{}$ Gaia LPV catalogue also includes the results of our model fits to the observed data, namely frequency, frequency error, and G-band model amplitude (half peak-to-peak; see \ref{sec:CatalogConstruction:PeriodAmplitudeDetermination}).
A comparison of $QR_{5}(G)$ and the model amplitude is presented in Fig.\,\ref{Fig:AmplitudeVSQR5}.
Because the model amplitude is half peak-to-peak, the general trend between the two quantities corresponds to an inclination of 0.5. 
Red lines in the plot indicate where the model amplitude equals $QR_{5}(G)$, $QR_{5}(G)/2$, and $QR_{5}(G)/3$. 
This plot includes only objects with a determined period.
For large-amplitude objects, the majority of the sample shows model amplitudes in agreement with $QR_{5}(G)/2$.
These are predominantly miras with single periods, for which the model amplitude represents the total light variation very well.
The small-amplitude variables, mostly multi-periodic, are mainly found below the $QR_{5}(G)/2$ line, because here the model amplitude represents only a part of the total light change. 
The majority is found between the $QR_{5}(G)/2$ and the $QR_{5}(G)/3$ line.
The mono-periodic model fit thus reproduces between 66 and 100\,\% of the light amplitude.
We conclude that there is a dominant period in most of our stars with exported periods, which is recovered by our model fit.
The fit of the observed light change with a monoperiodic model deteriorates when the smallest-amplitude variables in our catalogue are considered, as is shown in lower left corner of Fig.\,\ref{Fig:AmplitudeVSQR5}. 

\begin{figure}
\centering
\includegraphics[width=\hsize]{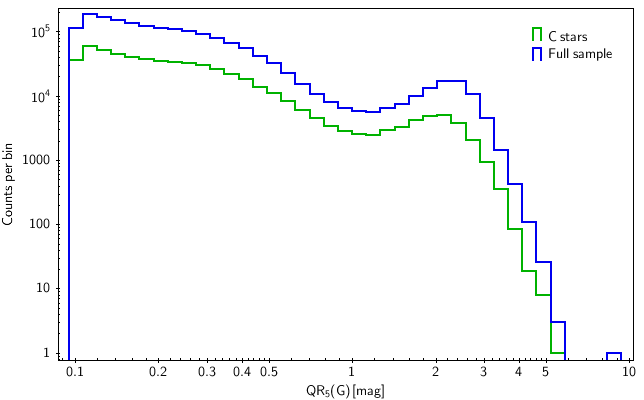}
\caption{Amplitude (5-95\% interquantile range) distributions of the LPV candidates. Blue shows all the sources in this second catalogue, and green shows those that are flagged as C-star candidates therein.
}
\label{Fig:range_distribution}
\end{figure}

Keeping the differences between model amplitude and the observed $QR_{5}(G)$ amplitude in mind, we used the latter in the following analysis because it is available for the complete SOS table. 
As illustrated in Sect.\,\ref{sec:CatalogOverview:ComparisonDR3-DR2}, Fig.\,\ref{Fig:classif_SOS_qr5}, the number of LPVs in the catalogue increases with decreasing amplitude for the complete SOS table, while the subsample of objects with amplitudes shows two peaks in its distribution around $QR_{5}(G)$=0.3 mag and $QR_{5}(G)$=3 mag.
  Fig.\,\ref{Fig:range_distribution} shows that the amplitude distribution of C stars is very similar to that of the complete sample.

As mentioned above, the longer time base of the second$^{}$ Gaia LPV catalogue allowed extending the range of periods for LPV candidates down to 35\,d (Fig.\,\ref{Fig:compare_P_DR2DR3_distrib}), which is different to the first$^{}$ catalogue, which stopped at 60\,d.
The distribution peaks around a period of 350\,d. 
However, the peak is a mixture of large-amplitude mira-type variables and small-amplitude stars ($\qrg$ < 1 mag).
Both groups contribute similarly to this peak. 
A second, less clearly expressed peak around 150\,d mainly consists of small-amplitude LPVs with a minor contribution from short-period miras.
A shoulder in the distribution at the long-period end indicates the increasing importance of long secondary periods that are now accessible with the longer time coverage of the DR3 light curves.

Except for the likely alias at 62.97\,d (see Fig.\,\ref{Fig:PeriodAliases}), the period distribution below 80\,d is mostly flat (Fig.\,\ref{Fig:compare_P_DR2DR3_distrib}).
An increase of objects towards shorter periods is expected, as seen in large surveys such as OGLE. 
However, the shorter-period objects will tentatively also be intrinsically weaker, which means that a larger fraction of these objects will be found at low apparent brightnesses, leading to S/N values that lie below our selection limit.
In addition, the irregularity of the light change is expected to increase for shorter-period objects, so that the amplitude of the strongest peak in the Fourier analysis may fall below the selection limit of 0.1 mag.
While these objects have no exported periods, publication of all light curves allows exploring periodicities within this group in detail. 

\begin{figure}
\centering
\includegraphics[width=\hsize]{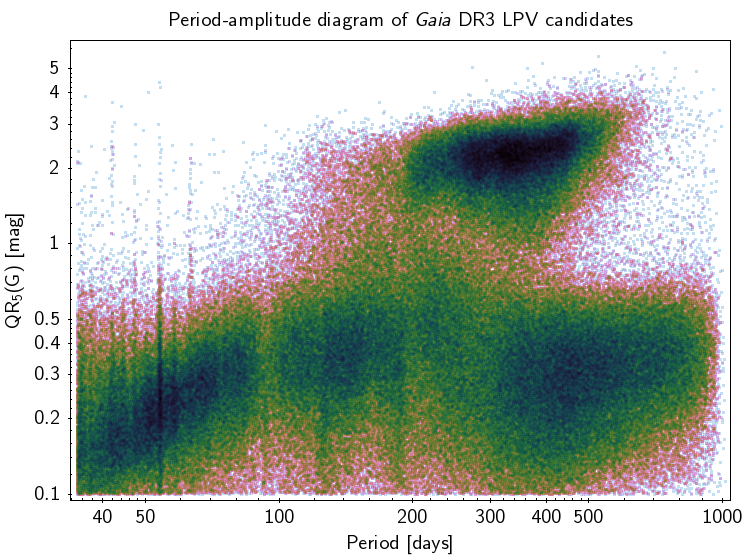}
\caption{Period-amplitude diagram of the \Gaia DR3 LPV sources whose period has been retained for publication. The amplitude is traced by the 5-95\% interquantile range $\qrg$. Several residual aliases are visible at periods shorter than 100 days (see Fig.~\ref{Fig:PeriodAliases}).}
\label{Fig:PeriodAmplitudeDiagram}
\end{figure}

The combination of the two parameters $\qrg$ and period gives a period-amplitude diagram of our subsample with periods (Fig.\,\ref{Fig:PeriodAmplitudeDiagram}). 
We can identify several features in this diagram.
First, for periods below 100\,d, there is a clear trend of amplitude with period.
The second obvious group of objects are the large-amplitude stars. 
These are mainly miras.
Starting at a period of about 140\,d, they also show a trend of increasing amplitude with increasing period, although this trend is less clear between 300 and 400\,d, where the trend is widened. 
This widening might result from objects at the upper end of the first-overtone pulsation sequence.
However, a complete parallel sequence extending towards shorter periods is expected.
An alternative explanation might be that these objects are similar to W Hya, which is likely an intermediate-mass fundamental-mode pulsator, which is thus showing a smaller variability amplitude \citep{2005A&A...431..623L}.
The low-amplitude regime above 100\,d shows no obvious correlation of period and amplitude.
The long-period end is of particular interest; it hosts the long secondary periods of LPVs.
Whether the long secondary period is indeed the dominant variability in these stars or if the long period is more likely to be discovered in the light-curve sampling will be investigated separately. 

\subsection{Light curves}

\begin{figure*}
\centering
\begin{tabular}{cc}
\subfloat{\includegraphics[width=.43\hsize]{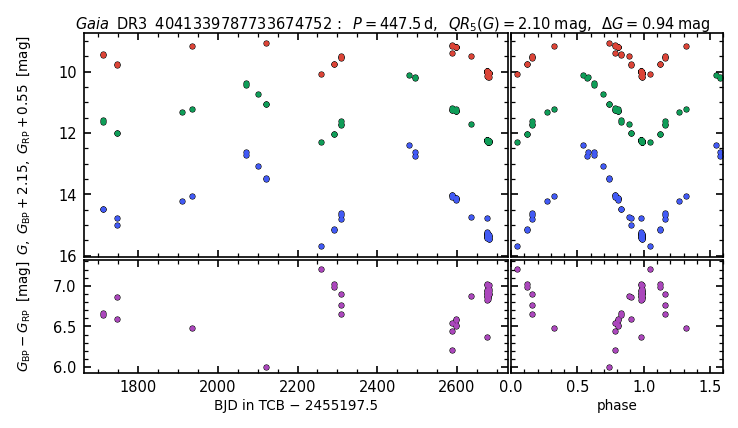}} &
\subfloat{\includegraphics[width=.43\hsize]{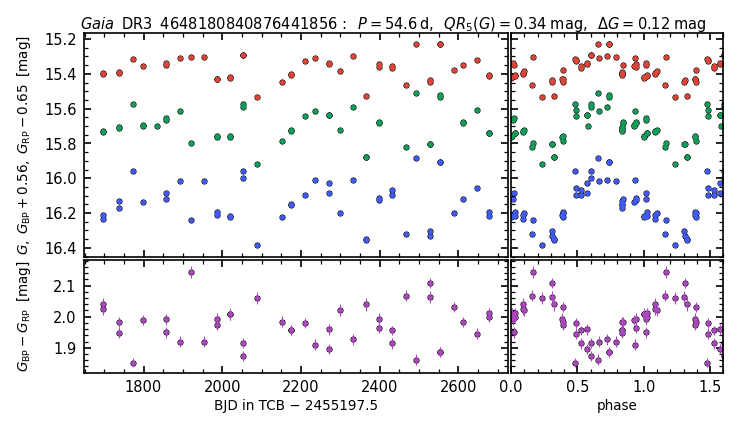}} \\
\subfloat{\includegraphics[width=.43\hsize]{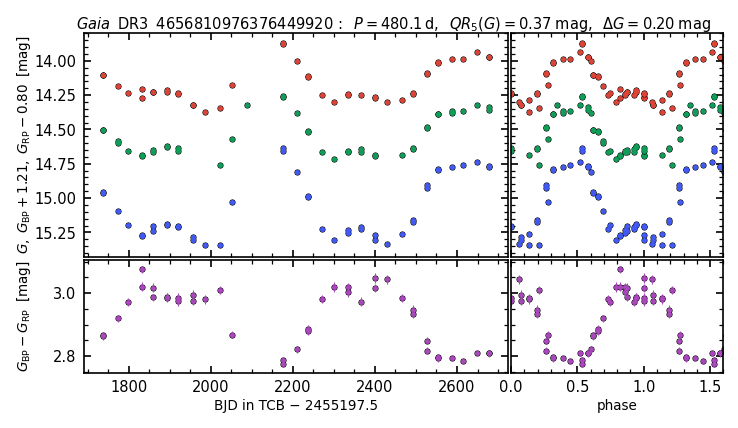}} &
\subfloat{\includegraphics[width=.43\hsize]{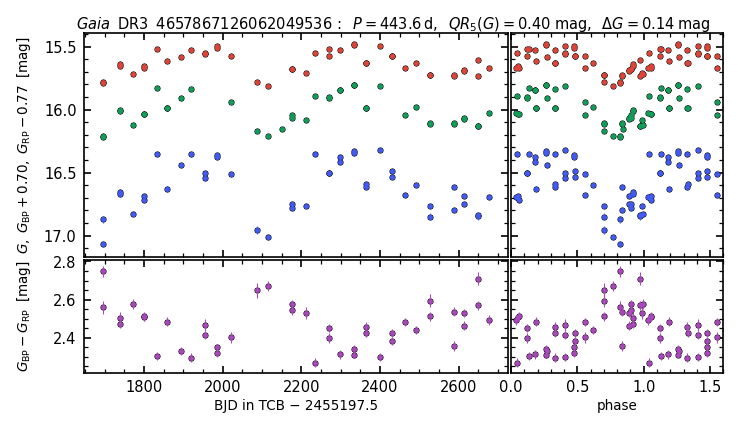}} \\
\subfloat{\includegraphics[width=.43\hsize]{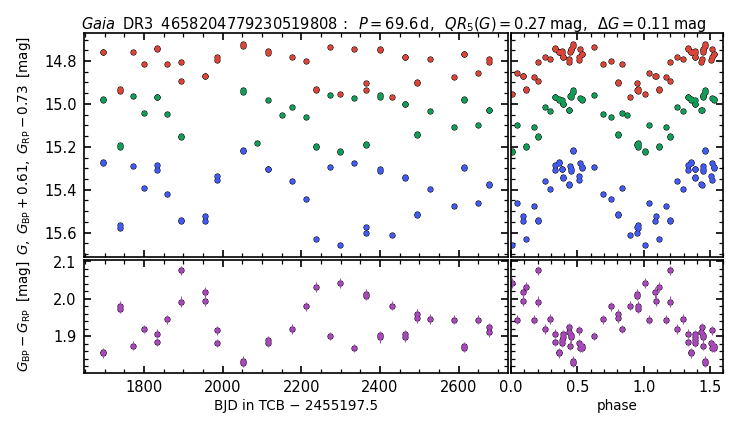}} &
\subfloat{\includegraphics[width=.43\hsize]{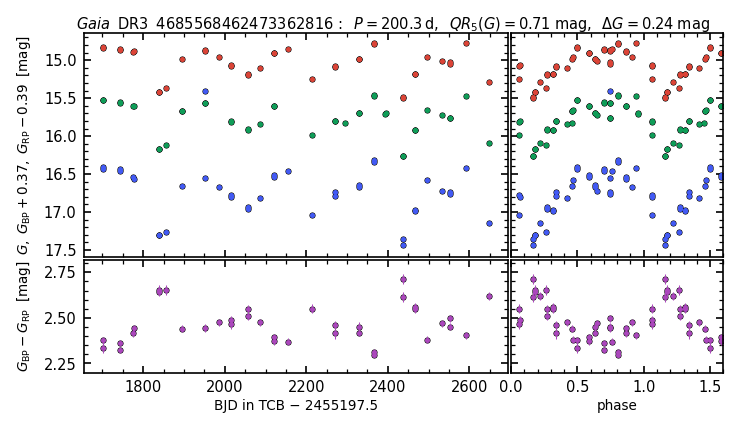}} \\
\subfloat{\includegraphics[width=.43\hsize]{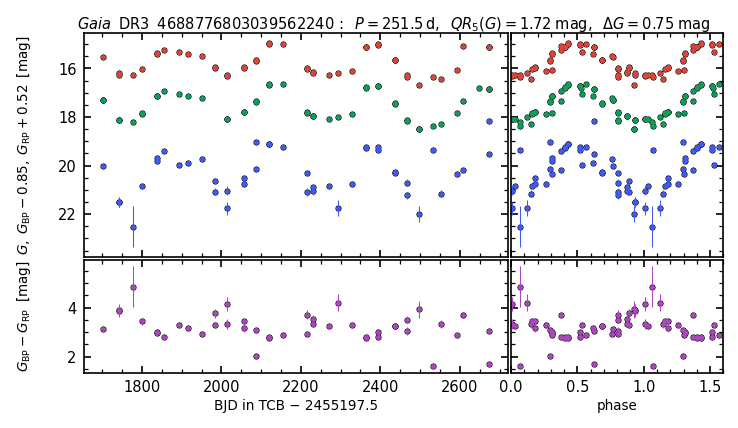}} &
\subfloat{\includegraphics[width=.43\hsize]{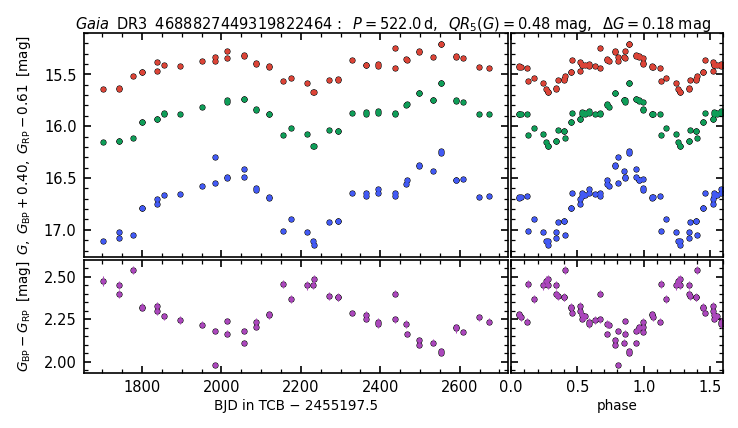}} \\
\end{tabular}
\caption{Example \Gaia DR3 time series of LPV candidates. The $\gbp$ and $\grp$ time series are offset by arbitrary amounts (indicated in the labels of the vertical axes) for visualisation purposes. The green, blue, and red symbols represent time series in the $\gmag$, $\gbp$, and $\grp$ filters. When the latter two are simultaneous (within 10 seconds of each other), the colour $\gbp-\grp$ is displayed in purple.}
\label{Fig:example_time_series1}
\end{figure*}

\begin{figure*}
\centering
\begin{tabular}{cc}
\subfloat{\includegraphics[width=.43\hsize]{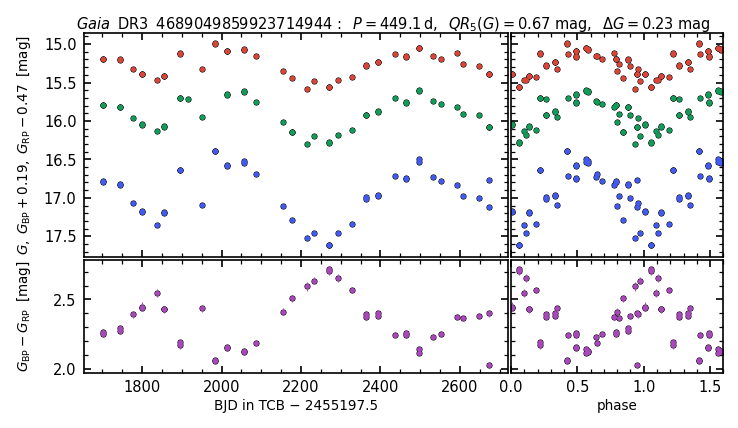}} &
\subfloat{\includegraphics[width=.43\hsize]{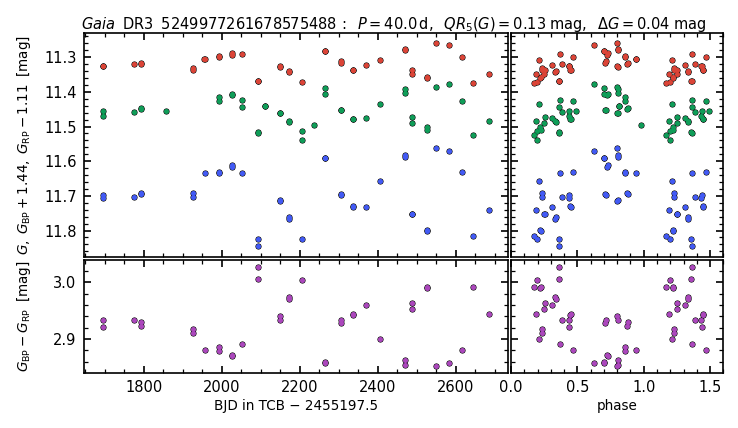}} \\
\subfloat{\includegraphics[width=.43\hsize]{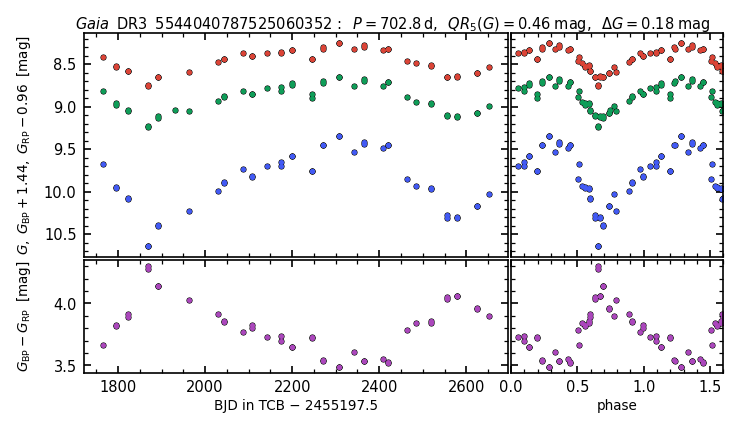}} &
\subfloat{\includegraphics[width=.43\hsize]{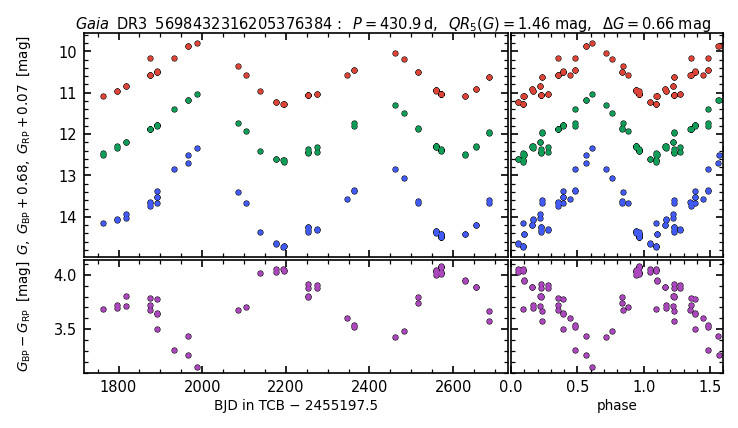}} \\
\subfloat{\includegraphics[width=.43\hsize]{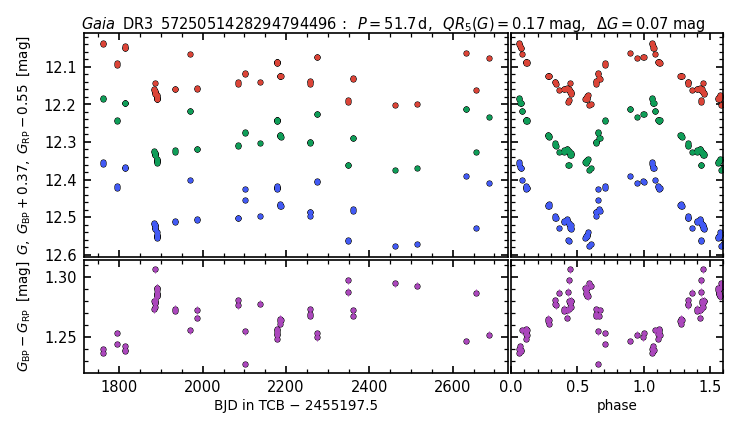}} &
\subfloat{\includegraphics[width=.43\hsize]{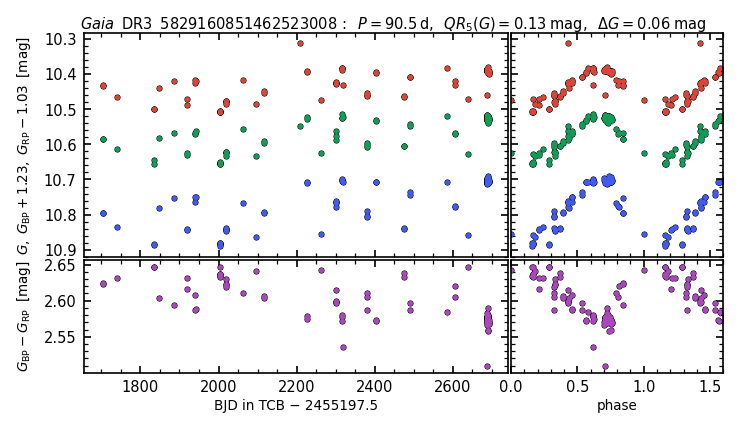}} \\
\subfloat{\includegraphics[width=.43\hsize]{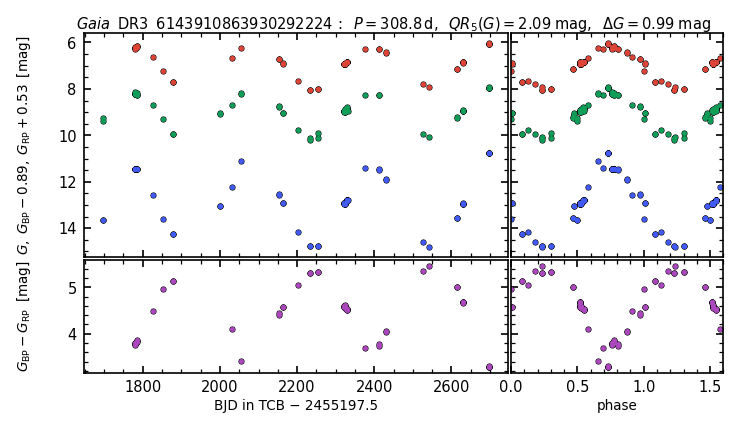}} &
\subfloat{\includegraphics[width=.43\hsize]{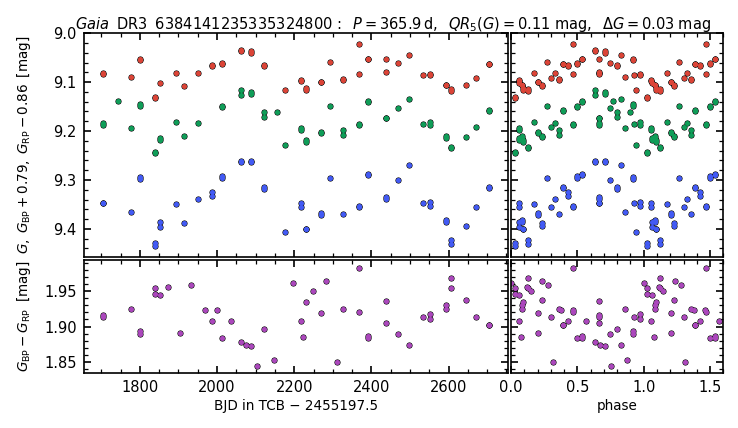}} \\
\end{tabular}
\caption{Similar as Fig.\,\ref{Fig:example_time_series1} for another set of LPV candidates.}
\label{Fig:example_time_series2}
\end{figure*}

\begin{figure}
\centering
\includegraphics[width=\hsize]{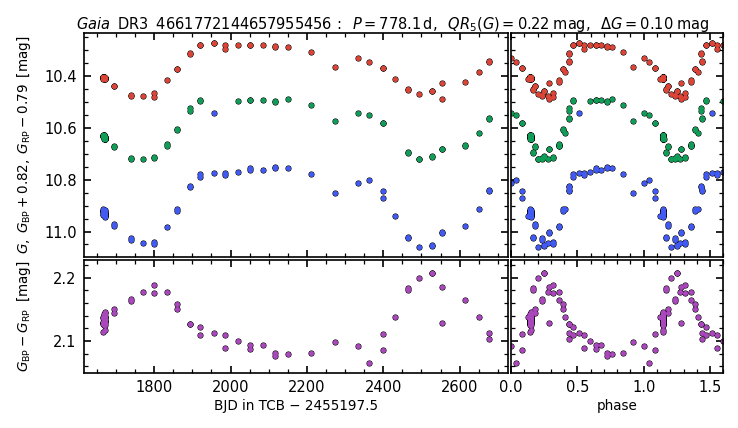}
\caption{Similar as Fig.\,\ref{Fig:example_time_series1} for another LPV source.}
\label{Fig:curious_light_curve}
\end{figure}

We close this section by presenting some representative light curves from the second$^{}$ Gaia LPV catalogue. 
For this presentation, we selected a randomly selected subset of light curves covering most of the parameter range of the catalogue (Figs.\,\ref{Fig:example_time_series1} to \ref{Fig:curious_light_curve}). 
Some remarkable light curves spotted in this subset are shown, but the complete dataset certainly contains many more interesting light changes awaiting their discovery by the user.

Gaia DR3 4041339787733674752: This is a nice example of a mira variable. The variation is well sampled in all three filters. This very red object can be found in the IRAS and WISE databases, but its long-period variability has only been reported in our first$^{}$ Gaia LPV catalogue. Although rather bright in G, it did not appear in any previous variability surveys referenced in SIMBAD.

Gaia DR3 4648180840876441856: Our second example is a small-amplitude variable, classically an SRV. Within the time span of Gaia measurements, the light change appears regular, as the phase diagram shows. The colour curve (bottom) agrees with expectations for a radially pulsating star.

Gaia DR3 4656810976376449920: The Gaia light curve shows a strange double minimum: the maximum is very broad and flat. Almost two complete cycles have been covered, and the phase plot confirms the regularity of this remarkable pattern. In agreement with the literature \citep[e.g.][]{1978A&AS...31...61W}, we classified this objects as a C star. Because it lies in the direction of the LMC, its variability has previously been reported by the MACHO, EROS-2, and OGLE surveys. The broad maximum is also visible in the OGLE data. \citet{soszynski_etal_2009_ogle3lmc} found three periods of 252, 487.6, and 3906\,d. Their second period agrees well with the period derived from the Gaia data. They also found a double-peaked minimum near JD 2454500, but not every minimum seems to show this feature. There is no signature of the very long period reported by OGLE in the Gaia data, but the reason may be that only one fourth of the expected LSP is covered. This object requires further investigation.

Gaia DR3 4657867126062049536: The long-period variability of this star has been reported by \citet{2008AJ....136.1242F}, who found from MACHO data a long primary period of 493\,d and a short secondary period of 55.86\,d. The period in the seocnd$^{}$ Gaia LPV catalogue is 443.6\,d, which agrees well with the primary period from the MACHO data and the value given in the first$^{}$ catalogue (415.11\,d). The phase plot suggests that a short period is present in addition to the primary period. 

Gaia DR3 4658204779230519808: This variable has previously detected both from OGLE and EROS data with a long period around 900\,d and a short period of 70\,d. The short period agrees very well with the Gaia period, and the long period corresponds to the total time span that is covered by the Gaia data. While we thus cannot confirm the long-period value, the presence of a trend cannot be excluded. This is a good example in which the periodicity is not obvious from the plot against time, but it is very clearly visible in the phase plot. The reason is that the period is  close to the precision period of Gaia.

Gaia DR3 4685568462473362816: In the literature, this SMC star has been classified as carbon rich \citep{1995A&AS..113..539M} and as a long-period variable \citep{soszynski_etal_2011_ogle3smc}, with two periods of 120 and 225\,d. The star has been identified as a C star in our analysis as well, and the period of 200.3\,d matches the longer period in the literature well, considering the semiregular variability of the star, which can be seen in the Gaia light curve.

Gaia DR3 4688776803039562240: This is another LPV in the SMC \citep{soszynski_etal_2011_ogle3smc}. Remarkably, the BP light change could be followed down to 22 mag, corresponding to the sensitivity limit of Gaia. The catalogue period agrees well with the period from the first$^{}$ Gaia LPV catalogue and that from the OGLE catalogue.

Gaia DR3 4688827449319822464: Variability occurs on a timescale of more than 500 days, and the Gaia light curve covers almost two cycles. The star was noted as an LPV in the first$^{}$ catalogue with a similar period, but nowhere else according to SIMBAD. Because the total amplitude is only about 0.5\,mag, it is not a mira, therefore the 522-day period might be a long secondary period of an SRV. However, indications for short time variations are rather weak.

Gaia DR3 4689049859923714944: The light change in this object is not completely regular. The period has been slightly revised from the first$^{}$ Gaia LPV catalogue value of 429.7 to 449.1\,d. The star is in the direction of the SMC and likely a member of it based on its location in the colour-magnitude diagram, as already noted by \citet{boyer_et_al_2011}. 

Gaia DR3 5249977261678575488: This variable has not been reported before in the literature according to SIMBAD. It was not part of the first$^{}$ Gaia LPV catalogue either, in agreement with its amplitude being lower than the selection limit. The phase diagram is compatible with a semiregular variability with multiple periods and amplitude variation. Using its period, parallax, and 2MASS $K$ magnitude,  we can place this star on sequence B in the period-luminosity diagram.

Gaia DR3 5544040787525060352: The derived period of 702.8\,d is based on only one completely covered light cycle. Even that was not possible in the first$^{}$ catalogue, which explains the period of more than 1000 days assumed there. The light curve is highly asymmetric, which can be seen both in the $G$ band and in the colour curve. This is a rather nearby object at a distance 1.46 kpc. In the period-luminosity diagram of solar neighbourhood LPVs (Sect. \ref{Sect:solarneighbor}), the star falls on sequence D, suggesting we see a long secondary period.

Gaia DR3 5698432316205376384: This star is V507 Pup, a well-known C star. Its spectral classification could be recovered with our classification method (median\_deltawl\_rp = 8.6). The period from the first$^{}$ catalogue has been revised by more than 10\%. Interestingly, this excellent example of a comparably bright mira-like variable has not been reported as an LPV before the Gaia mission, according to SIMBAD. The General Catalogue of Variable Stars reports Takamizawa as discoverer of the variability in 2000, but the publication is no longer accessible. 

Gaia DR3 5725051428294794496: Nothing could be found on this star in the literature. The variability amplitude is small, but the phased light curve shows a clear periodic variability with a period of 51.7\,d. 

Gaia DR3 5829160851462523008: The variability of this object has been unknown before, according to SIMBAD. The phased light curve reveals a regular light change of 90.5\,d. 

Gaia DR3 6143910863930292224: This is V372 Cen, a known mira. Several studies give consistent periods of about 315\,d, in agreement with the findings from Gaia. 

Gaia DR3 6384141235335324800: This is variable at the lower end of the amplitude range in our catalogue. Two cycles of its one-year period have been covered. A period in addition to the 365.9\,d is very likely. Variability has not been reported before, according to SIMBAD.

Gaia DR3 4661772144657955456: This star shows a very unusual light change with a long plateau, showing constant brightness for about 200 days. \citet{1981A&AS...43..267W} identified the star as an M-type supergiant. The spectral classification agrees with our result. From three radial velocity measurements at different epochs with a total velocity range of 11\,km/s, \citet{2021MNRAS.502.4890D} identified the star as a binary. Except for its entry in the first$^{}$ Gaia LPV catalogue, it has not been reported as variable in SIMBAD. The shape of our light curve recalls a binary, but the primary minimum is very broad. Further confirmations of radial velocity measurements are needed.

\section{Catalogue quality}
\label{sec:CatalogQuality}
\subsection{Completeness}
\label{sec:CatalogQuality:Completeness}

We estimated the completeness of the second$^{}$ \Gaia catalogue of LPV candidates by using a number of public LPV datasets as references, namely the catalogues from OGLE and from the All-Sky Automated Survey for SuperNovae (ASAS-SN). 
We recall that our analysis is limited to the \Gaia DR3 sources that were published as a result of the processing from the SOS module. 
As shown in the following, this dataset includes the majority of the LPVs that were previously known from the reference datasets. However, many of the missing sources can still be found in the \Gaia DR3 table of LPVs that is produced by the classification module \citep[as discussed in Sect.~\ref{sec:CatalogConstruction}; see][]{DR3-DPACP-165}, and their photometric time series are published in \Gaia DR3. 
A brief completeness analysis including the latter sources is presented in Appendix~\ref{app:ComparisonClassificationSOS:Completeness}.

\subsubsection{Comparison with OGLE}
\label{sec:CatalogQuality:Completeness:OGLE}

The catalogues of LPVs in the Magellanic Clouds and Galactic Bulge \citep{soszynski_etal_2009_ogle3lmc,soszynski_etal_2011_ogle3smc,soszynski_etal_2013_ogle3bulge} produced within the third phase of the Optical Gravitational Lensing Experiment (OGLE-III) represent one of the highest-quality datasets for studying these stars, and an appropriate reference for validating the second$^{}$ \Gaia catalogue of LPV candidates. They consist of a total 343 785 sources observed in the $V$ and $I$ filters, and provide mean magnitudes as well as three variability periods (and the three corresponding amplitudes in the $I$ band) for each source. Each LPV is assigned one of the three variability sub-types mira, SRV, and OGLE Small-Amplitude Red Giant \citep[OSARG;][]{Wray_etal_2004}. In addition, LPVs in the Magellanic Clouds are classified as O or C rich based on their optical and near-infrared photometric properties \citep{soszynski_etal_2009_ogle3lmc}.

\begin{figure}
\centering
\includegraphics[width=\hsize]{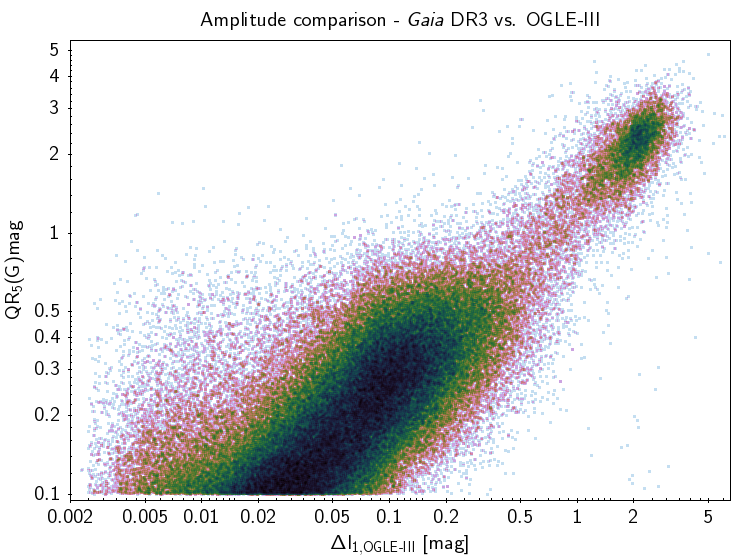}
\caption{Comparison between the \Gaia $G$-band amplitude (traced by the 5-95\% interquantile range, $\qrg$) of the \Gaia DR3 LPV candidates and the primary $I$-band amplitude, $\Delta I_1$, of their best-match OGLE-III sources.}
\label{Fig:A1_vs_trimmedrangeG_mathed}
\end{figure}

A preliminary cross-match of the \Gaia DR3 and OGLE-III LPV catalogues showed that the mean angular separation of the matched sources is about $0.1\arcsec$, and that a maximum cross-match radius of $2\arcsec$ is a reasonable choice to minimise mismatches. In order to consistently compare the two datasets, the OGLE-III sample needs to be restricted to sources with relatively large amplitude, thereby reproducing the filter applied in the construction our catalogue. To do this, we examined the matched sources in terms of the relation of the \Gaia $G$-band 5-95\% interquantile range, $\qrg$, and the OGLE-III primary amplitude in the $I$ band, $\Delta I_1$ (Fig.~\ref{Fig:A1_vs_trimmedrangeG_mathed}). For $0.1\leq\qrg/{\rm mag}<0.11$ (i.e. near the level of the amplitude cut applied to \Gaia LPV candidates), the distribution of $\Delta I_1$ peaks at about 0.03 mag. We therefore restricted the OGLE-III dataset to sources with a primary amplitude larger than this value.

Furthermore, the OGLE-III catalogue sky coverage is limited to a mosaic of fields of view towards the Magellanic Clouds and in the direction of the bulge. We restricted the OGLE-III and \Gaia DR3 samples to a set of common sky regions for the purpose of comparing the two catalogues (see Appendix~\ref{app:AdditionalDefinitions:CommonSkyOGLE3} for more details).

\begin{table}
\caption{Recovery rates of different sub-types of LPVs candidates with respect to the OGLE-III.}
\label{tab:ogle3_recovery_rates}
\centering
\begin{tabular}{c c c c}
\hline
Selection & OGLE-III & Matched $\leq2\arcsec$ & Recovery rate \\
\hline
\hline
\multicolumn{4}{c}{LPV candidates} \\ 
\hline
 All       &    84\,897 &    70\,395 &     82.9\% \\
 BLG       &    55\,644 &    45\,659 &     82.1\% \\
 LMC       &    25\,015 &    21\,370 &     85.4\% \\
 SMC       &     4\,238 &     3\,366 &     79.4\% \\
 Mira      &     5\,843 &     4\,679 &     80.1\% \\
 O-Mira    &        494 &        470 &     95.1\% \\
 C-Mira    &     1\,479 &     1\,006 &     68.0\% \\
 SRV       &    32\,630 &    30\,063 &     92.1\% \\
 O-SRV     &     6\,413 &     5\,874 &     91.6\% \\
 C-SRV     &     6\,461 &     6\,144 &     95.1\% \\
 OSARG     &    46\,424 &    35\,653 &     76.8\% \\
\hline
\multicolumn{4}{c}{with period published in \Gaia DR3} \\ 
\hline
 All       &    84\,897 &    29\,865 &     35.2\% \\
 BLG       &    55\,644 &    17\,039 &     30.6\% \\
 LMC       &    25\,015 &    10\,667 &     42.6\% \\
 SMC       &     4\,238 &     2\,159 &     50.9\% \\
 Mira      &     5\,843 &     4\,436 &     75.9\% \\
 O-Mira    &       494 &       452 &     91.5\% \\
 C-Mira    &     1\,479 &       851 &     57.5\% \\
 SRV       &    32\,630 &    15\,018 &     46.0\% \\
 O-SRV     &     6\,413 &     3\,384 &     52.8\% \\
 C-SRV     &     6\,461 &     3\,563 &     55.1\% \\
 OSARG     &    46\,424 &    10\,411 &     22.4\% \\
\hline
\end{tabular}
\end{table}
The OGLE-III sample limited in amplitude and sky coverage is hereafter referred to as the restricted OGLE-III catalogue. It consists of 84\,897 sources, 70\,395 of which have a \Gaia DR3 LPV match within $2\arcsec$, corresponding to a global recovery rate of 82.9\%. An overview of the numbers of successfully matched sources is given in Table~\ref{tab:ogle3_recovery_rates}. 
For each selection, the table gives the number of sources in the restricted OGLE-III catalogue (filtered by amplitude and sky position), the number of such sources in the second$^{}$ \Gaia catalogue of LPV candidates, and the corresponding recovery rate (the ratio of the latter to the former). The selections involving chemical types are further limited to the Magellanic Clouds and rely on the OGLE-III spectral type classification.

The recovery rates corresponding to OSARGs, SRVs, and miras are 76.8\%, 92.1\%, and 80.1\%, respectively. We recall that OSARG is an OGLE-specific variability type consisting of small-amplitude multi-periodic red giants that would be classified as SR or irregular variables according to the traditional scheme. Therefore, it is not surprising that more than 20\% of these sources are not recovered even after limiting the OGLE-III sample to stars with $\Delta I_1>0.03$ mag, especially since the latter condition cannot possibly reproduce exactly the $\qrg$ filter applied to the \Gaia dataset.

Because SRVs and miras have a larger amplitude than OSARGs, the corresponding recovery rates are also expected to be higher. This is the case for SRVs, while the recovery rate of miras is only moderately high, a fact that can be explained by analysing the miras of O- and C-rich composition separately. To do this, we adopted the chemical types assigned to LPVs in the MCs in the OGLE-III sample.
As O-rich miras have a 95.1\% recovery rate, it is clear that the issue lies with the C-rich miras, which are often faint at visual wavelengths. 
The mean $G$-band distribution of unmatched sources peaks at $\sim19.5$ mag, which is about 4 magnitudes fainter than the peak value for matched sources. This supports the idea that the optically faint sources have been rejected. We note that at 1.1$\mu$m, the OGLE $I$ band is still at $\sim20\%$ of its nominal peak transmissivity, whereas at that wavelength, the transmissivity of the \Gaia $G$ and $\grp$ bands is almost zero, which explains why a number of C-rich miras that are relatively faint in the optical are within reach of OGLE, but are less easily observed by \Gaia. We note that a non-negligible fraction of these lost C-rich miras can still be found among the \Gaia DR3 LPV candidates that are published as output of the classification module (see Appendix~\ref{app:ComparisonClassificationSOS:Completeness}).

The OGLE collection of LPVs has recently been updated with the publication of time series and mean properties for about 66\,000 miras in the Galactic bulge and disc \citep{ogle4miras} that have been observed as part of the fourth phase of OGLE (OGLE-IV). We examined this dataset separately from the OGLE-III catalogue as it is limited to miras, and therefore requires no amplitude limitation. However, we applied a selection to identify a common sky area, as was done for OGLE-III, and as detailed in Appendix~\ref{app:AdditionalDefinitions:CommonSkyOGLE3}.

This restricted OGLE-IV catalogue contains 50\,311 miras, 43\,232 of which have a counterpart within 2$\arcsec$ in the second$^{}$ \Gaia catalogue of LPV candidates. This corresponds to a recovery rate of 85.9\%, about 5\% better than the value relative to OGLE-III miras. The recovery rate is slightly better (about 90\%) in the Galactic disc than in the more crowded bulge (about 84\%).

\begin{table}
\caption{Similar to Table~\ref{tab:ogle3_recovery_rates}, but illustrating the recovery rate relative to the OGLE-IV catalogue of miras.}
\label{tab:ogle4_recovery_rates}
\centering
\begin{tabular}{c c c c}
\hline
Selection & OGLE-IV & Matched $\leq2\arcsec$ & Recovery rate \\
\hline
\hline
\multicolumn{4}{c}{LPV candidates} \\ 
\hline
Mira     &    50\,311 &    43\,232 &     85.9\% \\
BLG-Mira &    33\,806 &    28\,292 &     83.7\% \\
GD-Mira  &    16\,505 &    14\,940 &     90.5\% \\
\hline
\multicolumn{4}{c}{with period published in \Gaia DR3} \\ 
\hline
Mira     &    50\,311 &    41\,696 &     82.9\% \\
BLG-Mira &    33\,806 &    27\,138 &     80.3\% \\
GD-Mira  &    16\,505 &    14\,558 &     88.2\% \\
\hline
\end{tabular}
\end{table}

\subsubsection{Comparison with ASAS-SN}
\label{sec:CatalogQuality:Completeness:ASASSN}

Using time series collected by the All-Sky Automated Survey for SuperNovae \citep[ASAS-SN;][]{asassn_survey}, \citet[][see also references therein]{asassn_variables} have being progressively building a catalogue of variable stars that is suitable for validating the second$^{}$ catalogue of \Gaia LPV candidates over the whole sky. It consists of 687\,695 sources that were primarily observed in the $V$ band. Variability types were assigned by means of a random forest classifier. The classification includes mira (M), SR, and irregular (L) variables (which can collectively be considered as LPVs), as well as a number of other variability types. The variability amplitude in the $V$ band is provided for each source, but the period is only given for certain variability types. In particular, a period is provided for both miras and SRs, but not for L variables.

\begin{figure*}
\centering
\includegraphics[width=.355\hsize]{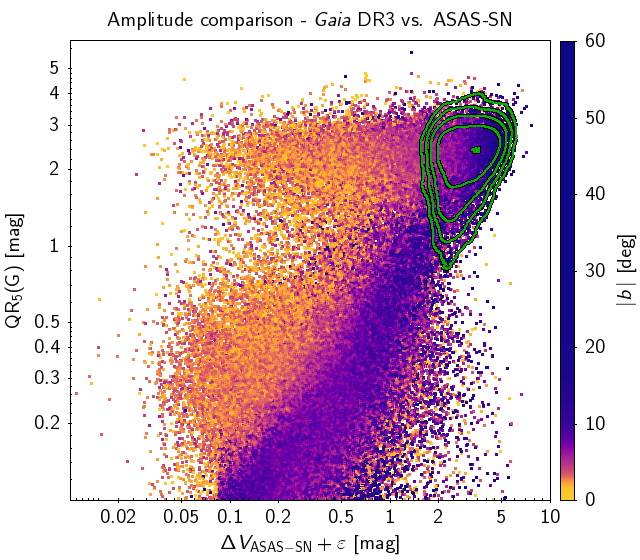}
\includegraphics[width=.640\hsize]{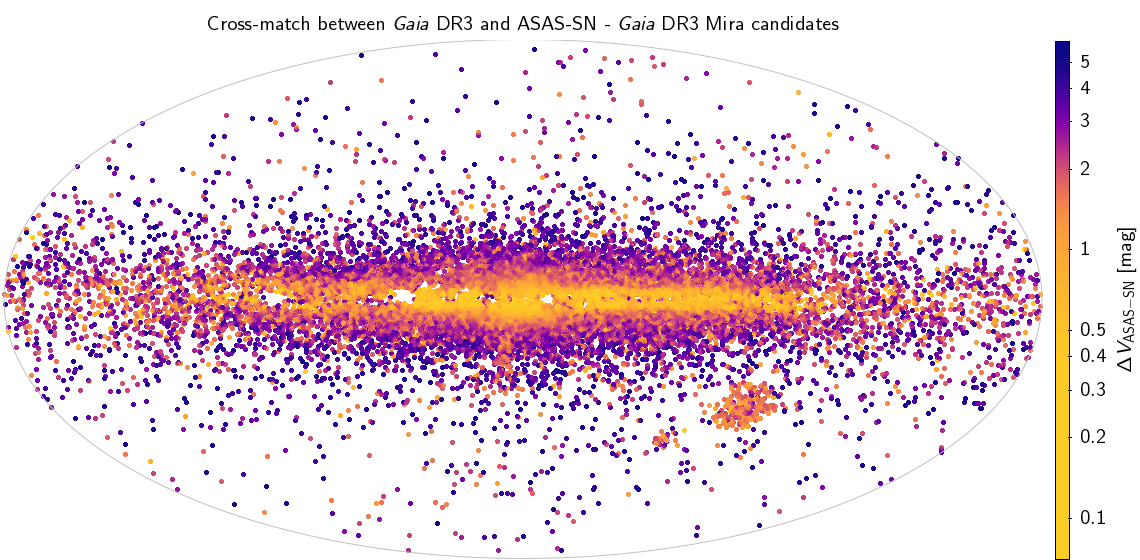}
\caption{Comparison of the $G$-band amplitude ($\qrg$) of \Gaia DR3 LPV candidates and the $V$-band amplitude $\Delta V$ of their best-match ASAS-SN sources. The left panel shows a direct comparison, limited to sources that are classified as LPVs in ASAS-SN. Data points are colour-coded according to the absolute value of their Galactic latitude, and the green density contour lines show the location of ASAS-SN sources that are classified as miras. The right panel shows the distribution in Galactic coordinates of the same sample, but limited to $\qrg>1$ mag (i.e. sources with a large enough \Gaia amplitude to be considered mira candidates), colour-coded by the value of the ASAS-SN amplitude. We note that random artificial errors $-0.01\leq\varepsilon\leq0.01$ have been added to the ASAS-SN amplitudes in the left panel to smooth out artefacts (evident in logarithmic scale) caused by the limited precision with which these values are provided.}
\label{Fig:amp_crowd_asassn}
\end{figure*}

\begin{table}
\caption{Number of sources in different selections in the ASAS-SN sample (limited to amplitudes larger than 0.15 mag), and number of these sources that were successfully matched with the \Gaia DR3 catalogue of LPVs. The ratio of the latter to the former gives the recovery rate.}
\label{tab:asassn_recovery_rates}
\centering
\begin{tabular}{c c c c}
\hline
Selection & ASAS-SN & Matched $\leq2\arcsec$ & Recovery rate \\
\hline
\hline
\multicolumn{4}{c}{LPV candidates} \\ 
\hline
LPV        &   225\,726 &   177\,346 &     78.6\% \\
Mira       &    11\,249 &    10\,018 &     89.1\% \\
SR         &   139\,980 &   111\,375 &     79.6\% \\
L          &    74\,497 &    55\,953 &     75.1\% \\
\hline
\multicolumn{4}{c}{with period published in \Gaia DR3} \\ 
\hline
LPV        &   225\,726 &    98\,219 &     43.5\% \\
Mira       &    11\,249 &     9\,871 &     87.8\% \\
SR         &   139\,980 &    60\,746 &     43.4\% \\
L          &    74\,497 &    27\,602 &     37.1\% \\
\hline
\multicolumn{4}{c}{with period published in ASAS-SN} \\ 
\hline
LPV        &   225\,726 &   123\,513 &     54.7\% \\
Mira       &    11\,249 &    10\,432 &     92.7\% \\
SR         &   139\,980 &   113\,081 &     80.8\% \\
\hline
\multicolumn{4}{c}{with period published in \Gaia DR3 and in ASAS-SN} \\ 
\hline
LPV        &   225\,726 &    70\,409 &     31.2\% \\
Mira       &    11\,249 &     9\,663 &     85.9\% \\
SR         &   139\,980 &    60\,746 &     43.4\% \\
\hline
\end{tabular}
\end{table}

In Table~\ref{tab:asassn_recovery_rates} we provide an overview of the recovery rates associated with distinct selections of LPVs. We find a fairly good recovery rate for miras and moderately high recovery rates for SR and L variables. We recall that the ASAS-SN SR variables are not the same type of object as the OGLE-III SRVs, even though the two overlap significantly. Similarly, OGLE-III OSARGs overlap the SR and L groups of ASAS-SN. All four groups are characterised by amplitudes that are systematically smaller than those of miras, and from this point of view, the recovery rates with respect to ASAS-SN sources are globally consistent with those resulting from the comparison with the OGLE-III data.

However, we note that the comparison between \Gaia DR3 and ASAS-SN in terms of variability amplitude is more complex than in the case of OGLE-III sources. This is shown in Fig.~\ref{Fig:amp_crowd_asassn}, which shows the relation of the \Gaia $G$-band 5-95\% interquantile range, $\qrg$, and the ASAS-SN $V$-band amplitude, $\Delta V$, for the matched sources. In contrast with the case of the OGLE-III cross match (see Fig.~\ref{Fig:A1_vs_trimmedrangeG_mathed}), a large fraction of sources exists whose ASAS-SN amplitude is unexpectedly small given the amplitude derived from \Gaia DR3 light curves. These sources are frequently located in the neighbourhood of the Galactic plane, where ASAS-SN LPVs indeed have a strong tendency to display a small variability amplitude. This behaviour can be attributed to the fact that the observed light curve of a variable star can appear to be compressed if the star lies in a crowded region of the sky, so that the measured amplitude is smaller than the true value \citep{Riess_etal_2020}. Owing to its higher spatial resolution, \Gaia provides light curves that are substantially less affected by crowding and better represent the true variability of the observed LPVs.

Because of this effect, a fraction of miras are probably incorrectly classified as SR or L variables in the ASAS-SN catalogue. The 89.1\% recovery rate of ASAS-SN miras therefore should probably rather be considered as an upper limit, and the corresponding recovery rates of SR and L variables represent lower limits.


\subsubsection{Comparison with LPVs from the Bright Star Catalogue}
\label{sec:CatalogQuality:Completeness:WellObservedLPVs}

\begin{figure}
\centering
\includegraphics[width=\hsize]{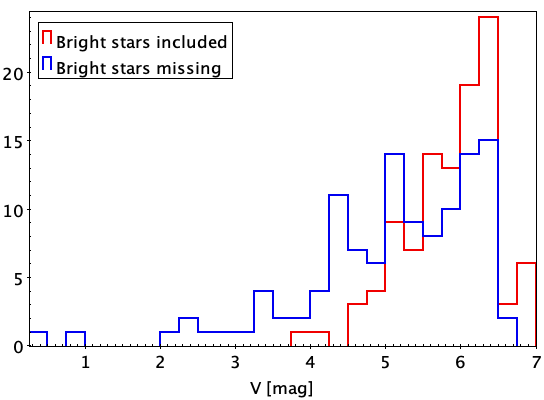}
\caption{Brightness distribution of Bright Star catalogue stars included in (red) or missing from (blue) the Gaia DR3 LPV catalogue.}
\label{fig:BrightStars_V}
\end{figure}

The Bright Star catalogue \citep{vizier:v/50} includes 251 objects that are classified either as mira or as LPV in SIMBAD.
We cross-correlated them with median \gmag <9\,mag\footnote{The Bright Star catalogue also includes stars when they are brighter than $V$=7 at some phase of their light curve. This is true for some miras. We thus set the limit to \gmag =9.} objects from the Gaia DR3 LPV catalogue with a search radius of 5" (however, except for one prominent star, o Cet, the coordinates matched to better than 2"). 
In this way, we identified 40\% of the bright stars in the SOS table and an additional 11\% in the classification table. 
The bright stars that we did not find in our catalogue include prominent supergiants such as $\alpha$\,Ori or $\mu$\,Gem, but a total of 23 (74\%) of the bright miras are present in the SOS table, 22 of them with periods.
One obvious reason for the exclusion of a star was that the stellar flux was close to or exceeded the Gaia limits on the very bright side.
This is illustrated in Fig.\,\ref{fig:BrightStars_V}, which shows the distribution of V magnitudes of the LPVs from the Bright Star catalogue.
In the transition region V$=$5 to 6.5 mag lies an increasing fraction of bright stars that are included in our LPV catalogue.
Mira itself (o Cet) is not included in the SOS table because of the missing \texttt{RUWE} parameter and because there are too few visibility periods.

The parallax errors of the included bright stars are typically smaller than 10\% of the parallax.
The Gaia DR3 distances of these stars fall between 100 and 1000 pc, which seems reasonable. 

\subsection{New LPV candidates}
\label{sec:CatalogQuality:NewLPVCandidates}

Our catalogue includes a large number of newly discovered LPV candidates. In order to characterise these new discoveries, we again relied on a comparison with the OGLE-III and ASAS-SN catalogues, limited to sources that are classified as LPVs there, and filtered by amplitude and sky position. Table~\ref{tab:newdiscoveries} gives an overview of this comparison and provides estimated discovery rates.
As references, we took the OGLE-III catalogue of LPVs in the bulge, LMC, and SMC, and the LPVs in the ASAS-SN catalogue of pulsating stars. 
We selected only the sources with a large enough amplitude to be compared with \Gaia DR3 (see Sect.~\ref{sec:CatalogQuality:Completeness:OGLE} and~\ref{sec:CatalogQuality:Completeness:ASASSN}), and performed the comparison over sky areas in which \Gaia and the reference catalogues overlap. 
For each catalogue, we present the number of known such sources and of the new sources from the \Gaia DR3 LPV catalogue. 
The discovery rates are expressed as the ratio of new discoveries to the selected sources (regardless of whether they have a match in the \Gaia DR3 LPV catalogue)

\begin{figure*}
\centering
\includegraphics[width=.49\hsize]{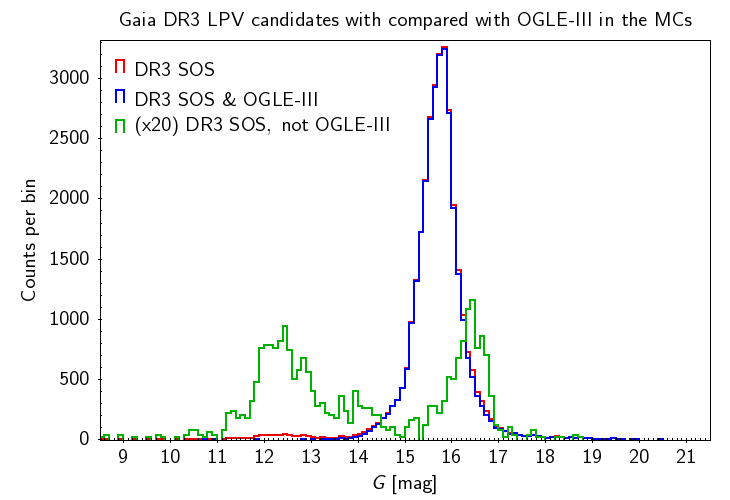}
\includegraphics[width=.49\hsize]{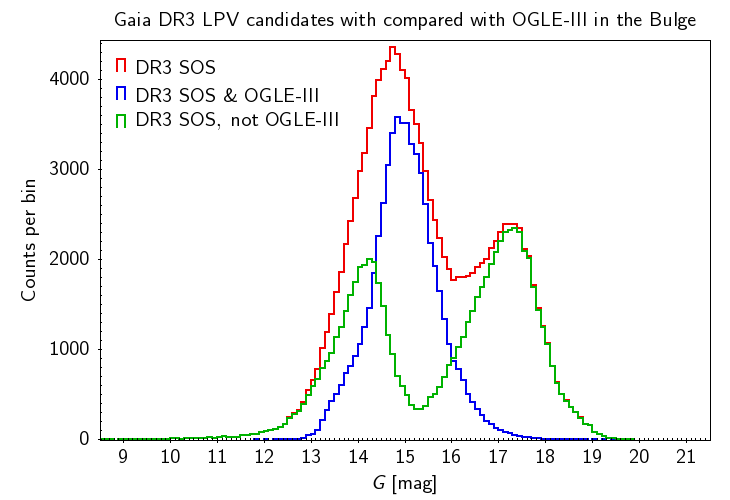}
\caption{Brightness distribution in the \Gaia $G$ band of the LPV candidates in the selected sky area towards the MCs (left panel) and the Galactic bulge (right panel) in common with the OGLE-III catalogue. The blue and green curve correspond to the recovered sources (matched with OGLE-III within 2$\arcsec$) and the new candidates (not matched with OGLE-III), respectively, and the red curve includes them both. In the left panel, the counts associated with the green curve have been enhanced by a factor 20 for visual clarity.}
\label{fig:hist_mediang_XMandNew_dr3sos_ogle3}
\end{figure*}

The number of newly discovered LPV candidates towards the fields explored by OGLE-III is comparable with the number of sources that are known from this catalogue. The rate of new discoveries is about 88\%.
A closer inspection reveals that the vast majority of them are located towards the Galactic bulge, whereas the newly discovered LPV candidates towards the Magellanic Clouds represent just about 4-5\% of the previously known sources.
As illustrated in the right panel of Fig.~\ref{fig:hist_mediang_XMandNew_dr3sos_ogle3}, most  of these new sources are either relatively faint ($G\gtrsim16$~mag) or bright  ($G\lesssim 14.5$~mag).
In contrast, most of the LPVs in the brightness range covered by the OGLE-III sample are already known from OGLE-III.

\begin{figure}
\centering
\includegraphics[width=\hsize]{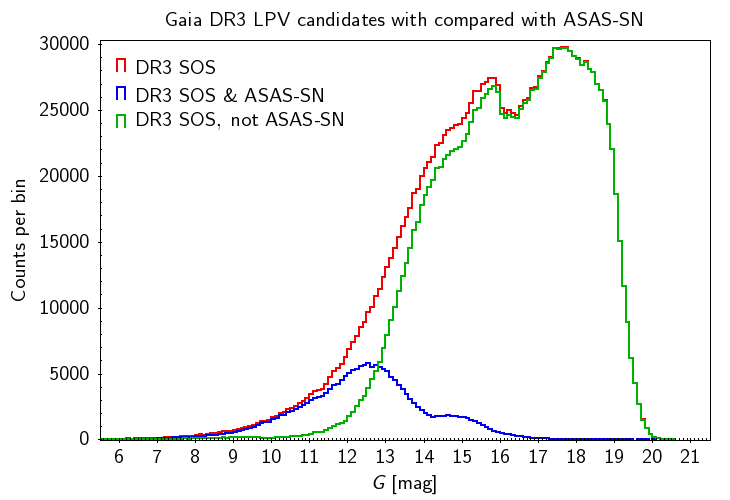}
\caption{Similar to Fig.~\ref{fig:hist_mediang_XMandNew_dr3sos_ogle3}, but for the cross match with the ASAS-SN catalogue of pulsating stars.}
\label{fig:hist_mediang_XMandNew_dr3sos_asassn}
\end{figure}

In our comparison to the ASAS-SN catalogue, we find 6.7 times more new candidates than known LPVs in that survey.
They are also relatively faint sources (see Fig.~\ref{fig:hist_mediang_XMandNew_dr3sos_asassn}) and tend to be located in the direction of the Galactic bulge and plane. Their observation was possible through the deeper reach and higher spatial resolution of \Gaia compared with the ASAS-SN telescopes.

\begin{table*}
\caption{Overview of newly discovered LPV candidates with respect to literature data.
}
\label{tab:newdiscoveries}
\centering
\begin{tabular}{lrrrrr}
\hline\hline
\multicolumn{2}{c}{Literature data} & \multicolumn{4}{c}{\Gaia DR3} \\
\cmidrule(lr){1-2}\cmidrule(lr){4-6}
\multirow{2}{*}{Catalogue} & \multirow{2}{*}{Selection} & \multicolumn{2}{c}{LPV candidates} &\multicolumn{2}{c}{with published period} \\
\cmidrule(lr){3-4}\cmidrule(lr){5-6}
                &            &         New & Discovery rate &         New & Discovery rate \\
\hline
OGLE-III        &    84\,897 &     74\,630 &         87.9\% &     13\,746 &         16.2\% \\
\hspace{6mm}LMC &    25\,015 &      1\,008 &          4.0\% &         284 &          1.1\% \\
\hspace{6mm}SMC &     4\,238 &         215 &          5.1\% &         138 &          3.3\% \\
\hspace{6mm}BLG &    55\,644 &     73\,407 &        131.9\% &     13\,324 &         23.9\% \\
ASAS-SN$^{a}$   &   225\,726 & 1\,522\,105 &        674.3\% &    285\,202 &        126.3\% \\
\hline

\end{tabular}
\tablefoot{
\tablefoottext{a}{Limited to sources hat are tclassified as LPVs (mira, SRV, irregular) in ASAS-SN.}}
\end{table*}

\subsection{Contamination}
\label{sec:CatalogQuality:Contamination}

The ASAS-SN catalogue of pulsating stars is not limited to LPVs, but rather includes a number of sources of other variability types. Under the assumption that the ASAS-SN automated classification is correct, this would allow us to assess the contamination rate from non-LPV sources in the \Gaia catalogue.

Of the LPV candidates that are matched with the restricted ASAS-SN catalogue, 3\,628 are not classified in ASAS-SN as LPVs, which corresponds to a global contamination rate of 2.0\% (estimated as the ratio of the number of such sources to the total number of matched LPV candidates). However, the most numerous group of non-LPV sources (2\,349) are classified in ASAS-SN as generic variable stars (VAR), which realistically include unclassified LPVs (Tab.\,\ref{tab:asassn_contamination_rates}). If all of them were in fact LPVs, the contamination rate would be as low as 0.7\%. Therefore, the true contamination rate lies between these 0.7\% and 2.0\%. When we exclude these generic variables, the most numerous class of contaminants consists of young stellar objects (YSOs), followed by rotational variables (ROT), with contamination rates of 0.2\% and 0.3\%, respectively. These sources of contamination are expected because they are variables and their colours are similar to those of LPVs. We caution that because these are comparatively faint sources, the fraction of such contaminants will be larger in volume-limited samples of relatively nearby stars.

Finally, we considered the case of other relevant types of variable stars, namely classical pulsators (classical Cepheids and RR Lyrae) and eclipsing binaries. These sources, whose behaviour might be mistaken as long-period variability, result in  a collective contamination rate of 0.04\%. All remaining variability times listed in the ASAS-SN catalogue produce a contamination at a 0.1\% level.

\begin{table}
\caption{Number of \Gaia DR3 LPV candidates that are matched with the restricted ASAS-SN, but are not classified as LPVs there. The right column gives the fraction of such stars with respect to the total number of matched sources and is an indicator of the contamination rates.}
\label{tab:asassn_contamination_rates}
\centering
\begin{tabular}{c c c}
\hline
ASAS-SN type & Matched $\leq2\arcsec$ & Percentage \\
\hline\hline
YSO                         &       544 &      0.3\% \\
ROT                         &       335 &      0.2\% \\
Ecl., Cep., RRL             &        63 &     0.04\% \\
Other                       &       337 &      0.1\% \\
\hline
Total                       &    1\,279 &      0.7\% \\
unclassified (VAR)          &    2\,349 &      1.3\% \\
\hline
\end{tabular}
\end{table}

\subsection{Period recovery}
\label{sec:CatalogQuality:PeriodRecovery}

When comparing the periods in the second$^{}$ \Gaia LPV catalogue with values from the literature, we have to keep in mind that LPVs are known to show cycle-to-cycle variations in period length. 
SRVs (and OSARGs) are by definition known to have changing periods and to alternate between different pulsation modes (which entails changes in period in the sense that the observed primary period becomes different). 
Most miras show changes in periods that \citet{ZijlstraBedding_2002} described as a normal period jitter of about 5\% in period length.
In addition, these authors identified continuous, sudden, or meandering \citep[see also][]{Uttenthaler_etal_2011} period variations in a few percent of the miras with an increasing fraction among long-period miras.
This can lead to real period changes of several percent over a decade, a time span that corresponds to the difference between OGLE-III and \Gaia.
Therefore, we chose a deviation of less than 10\% between \Gaia and literature values as an appropriate indicator for period compatibility.
This should also account for differences in the method of period determination and in the sampling of the analysed time series between \Gaia and OGLE or ASAS-SN.

\subsubsection{Comparison with OGLE-III}
\label{sec:CatalogQuality:PeriodRecovery:OGLE3}

\begin{figure}
\centering
\includegraphics[width=\hsize]{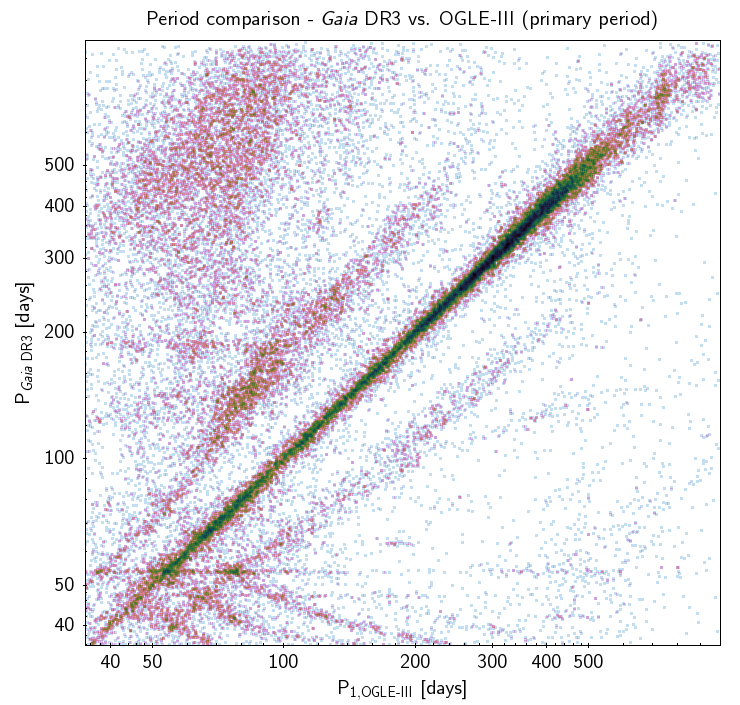}
\caption{Comparison of the period published for \Gaia DR3 LPV candidates and the primary period of their best-match counterparts in the restricted OGLE-III catalogue.}
\label{Fig:scatter_Pgaia_P1ogle}
\end{figure}

As indicated in Table~\ref{tab:ogle3_recovery_rates}, for 29\,865 of the LPV candidates that matched with OGLE-III, the period was retained for publication in \Gaia DR3 (Table~\ref{tab:ogle3_recovery_rates}). This number corresponds to 41.1\% of the matched sources and 35.2\% of the sources in the restricted OGLE-III catalogue. This apparently small fraction is entirely expected because the photometric time series collected by the OGLE program are substantially longer than \Gaia observations, and they are often more densely sampled. Nonetheless, we find a reasonably good compatibility of periods, as illustrated in Fig.~\ref{Fig:scatter_Pgaia_P1ogle}. In order to quantify this, we considered the relative difference between the \Gaia DR3 period and the a reference value from OGLE-III, which can be either the primary period $P_{\rm1,OGLE}$ or the period $P_{\rm c,OGLE}$ (of the three published by OGLE) whose value is closest to the \Gaia DR3 period,
\begin{equation}\label{eq:deltaPogle}
    \delta P_1=\frac{\left|P_{\Gaia}-P_{\rm1,OGLE}\right|}{P_{\rm1,OGLE}} \,,\;\;
    \delta P_{\rm c}=\frac{\left|P_{\Gaia}-P_{\rm c,OGLE}\right|}{P_{\rm c,OGLE}} \,.
\end{equation}
The adoption of the latter indicator allows us to account for the fact that because OGLE-III provides three periods for each source, there is no guarantee that the strongest periodic signal seen by \Gaia corresponds to the variability that is observed by OGLE, especially given the different sampling and coverage of the two surveys. Moreover, the time intervals during which Gaia and OGLE observed do not overlap, so there is a reasonable chance that the dominant pulsation period of some SRVs and OSARGs has changed. We also considered the differences between the corresponding frequencies,
\begin{equation}\label{eq:deltanuogle}
    \delta\nu_1 = \left|\nu_{\Gaia}-\nu_{\rm1,OGLE}\right| \,,\;\;
    \delta\nu_{\rm c} = \left|\nu_{\Gaia}-\nu_{\rm c,OGLE}\right| \,,
\end{equation}
which we can compare with the frequency uncertainty published for DR3 LPV candidates. We indicate them as $\varepsilon_{\nu,\Gaia}$ (corresponding to the quantity \texttt{frequency\_error} of Table~\ref{tab:lpv_attributes}).

An overview of the period compatibility between \Gaia DR3 and OGLE-III is provided in Table~\ref{tab:ogle3_period_comparison}. About half of the LPV candidates that  are matched with OGLE-III and have a published period in DR3 are within 20\% of the primary OGLE-III period, while this compatibility is achieved for two out of three LPV candidates when the comparison is made with the closest OGLE-III period. As expected, the vast majority of miras has a high degree of period compatibility, as they display well-developed mono-periodic variations. For SRVs, which are often multi-periodic, only about 37\% of the LPV candidates have a period within 20\% of the primary OGLE-III period. However, when the comparison is made with respect to the closest OGLE-III period, the fraction of compatible SRVs is almost doubled. We find a similar situation for the OSARG variables.

Overall, the fraction of sources whose period is compatible with the OGLE-III data decreases when sources with progressively less regular light curves are considered, that is, when instead of miras, SRVs and OSARGs are considered. However, the compatibility tends to be slightly higher for OSARGs than for SRVs, even though the former have more regular light curves on average. The reason for this result is likely that OSARGs oscillate in several radial and non-radial modes that have similar periods, in particular, when two modes with the same radial order and different angular degree are compared. Hence there is a relatively high probability that one of the OGLE-III periods and the \Gaia DR3 period are compatible according to our metrics, even though they do not result from the same oscillation mode.

\begin{figure}
\centering
\includegraphics[width=\hsize]{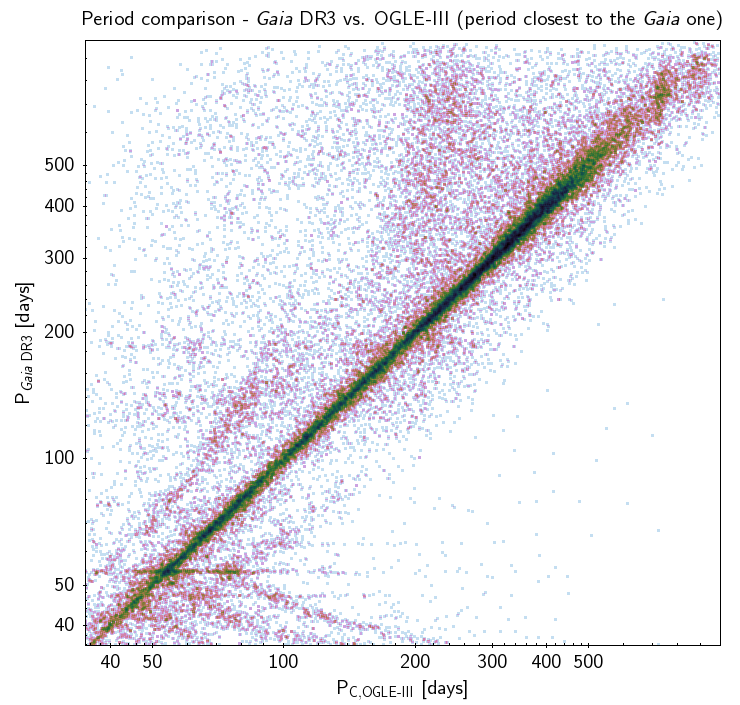}
\caption{Similar to Fig.~\ref{Fig:scatter_Pgaia_P1ogle}, but showing the OGLE-III period whose value is closest to the \Gaia DR3 period in each source, rather than the primary period from OGLE-III.}
\label{Fig:scatter_Pgaia_PclosestOgle}
\end{figure}

\begin{figure}
\centering
\includegraphics[width=\hsize]{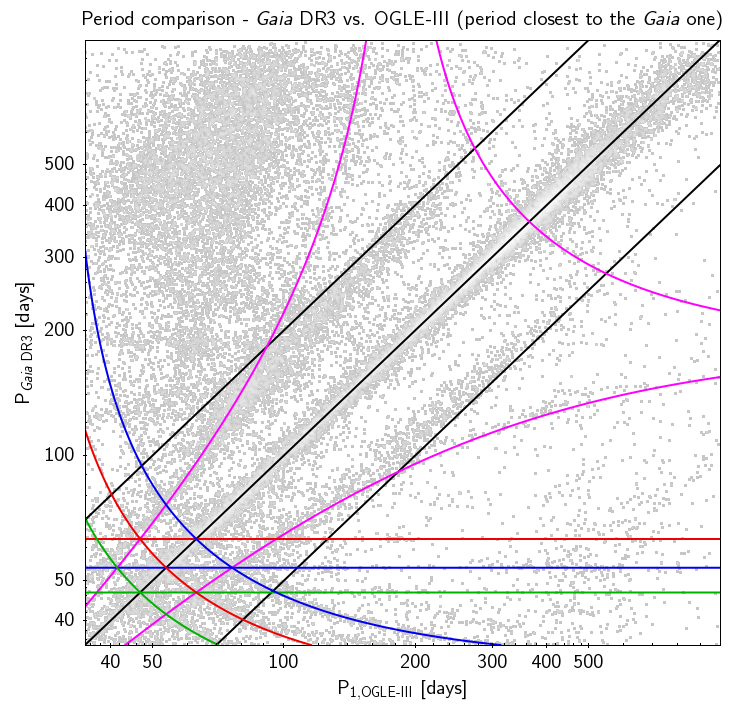}
\caption{Similar to Fig.~\ref{Fig:scatter_Pgaia_P1ogle}, but highlighting the patterns caused by aliases and failure modes (see Sect.~\ref{sec:CatalogQuality:PeriodRecovery:OGLE3}).}
\label{Fig:scatter_Pgaia_P1ogle_failureModes}
\end{figure}

Figures.~\ref{Fig:scatter_Pgaia_P1ogle} and~\ref{Fig:scatter_Pgaia_PclosestOgle} illustrate the compatibility between the \Gaia DR3 and OGLE-III periods. It is worth clarifying the patterns appearing in these diagrams, which are highlighted in Fig.~\ref{Fig:scatter_Pgaia_P1ogle_failureModes}. The diagonal trends (straight black lines) indicate that the \Gaia period is the same as the OGLE-III period, or that it is half or twice that value (meaning that the light curve deviates significantly from a purely sinusoidal shape and has strong higher harmonic components). The coloured curves in Fig.~\ref{Fig:scatter_Pgaia_P1ogle_failureModes} correspond to failure modes associated with aliases, and have the expressions \citep[e.g.][]{vanderplas_2018}
\begin{equation}
    P_{Gaia} = \left|\frac{m}{P_{\rm1,OGLE}}\pm n\delta\nu\right|^{-1} \;\;.
\end{equation}
In particular, the magenta curves indicate the pattern resulting from $\delta\nu=1/$yr (with $m=1$, $n=2$), the alias due to the seasonal gap in the OGLE ground-based observations. The red, green, and blue curves, on the other hand, are caused by aliases affecting the \Gaia time series, which are indicated by the horizontal lines of the same colours. They corresponds to $\delta\nu=1/62.97$ days, resulting from the \Gaia precession period and its associated features at $\sim47$ days and $\sim54$ days (corresponding to $k=2,3$ in Eq.~\ref{eq:precession_alias}).

\begin{table*}
\caption{Number $N_{\rm xm}$ of LPV candidates with a published period in DR3 that have a match within 2$\arcsec$ with the restricted OGLE-III sample, and the number and fraction of them whose period is compatible with the OGLE-III value. The period compatibility is evaluated in terms of the difference in period ($\delta P_1$, $\delta P_{\rm c}$, see Eq.~\ref{eq:deltaPogle}) or frequency ($\delta\nu_1$, $\delta\nu_{\rm c}$, see Eq.~\ref{eq:deltanuogle}).}
\label{tab:ogle3_period_comparison}
\centering
\begin{tabular}{cccccccc}
\hline\hline
Selection & $N_{\rm xm}$ & $\delta P_1<0.1$ & $\delta P_{\rm c}<0.1$ & $\delta P_1<0.2$ & $\delta P_{\rm c}<0.2$ & $\delta \nu_1<\varepsilon_{\nu}^{Gaia}$ & $\delta \nu_{\rm c}<\varepsilon_{\nu}^{Gaia}$ \\
\hline
All       &    29\,865 &    12\,797 &    16\,754 &    14\,457 &    19\,710 &    13\,560 &    19\,691 \\
          &           & ( 42.8\%) & ( 56.1\%) & ( 48.4\%) & ( 66.0\%) & ( 45.4\%) & ( 65.9\%) \\
Mira      &     4\,436 &     4\,290 &     4\,310 &     4\,340 &     4\,360 &     3\,948 &     3\,970 \\
          &           & ( 96.7\%) & ( 97.2\%) & ( 97.8\%) & ( 98.3\%) & ( 89.0\%) & ( 89.5\%) \\
SRV       &    15\,018 &     4\,856 &     7\,441 &     5\,523 &     8\,953 &     5\,031 &     9\,018 \\
          &           & ( 32.3\%) & ( 49.5\%) & ( 36.8\%) & ( 59.6\%) & ( 33.5\%) & ( 60.0\%) \\
OSARG     &    10\,411 &     3\,651 &     5\,003 &     4\,594 &     6\,397 &     4\,581 &     6\,703 \\
          &           & ( 35.1\%) & ( 48.1\%) & ( 44.1\%) & ( 61.4\%) & ( 44.0\%) & ( 64.4\%) \\
\hline
\end{tabular}
\end{table*}

\begin{figure}
\centering
\includegraphics[width=\hsize]{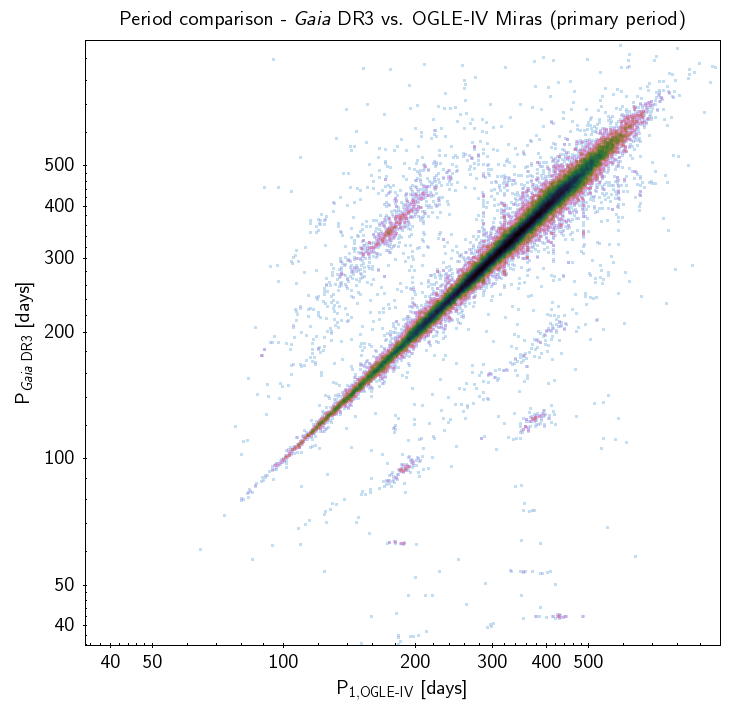}
\caption{Similar to Fig.~\ref{Fig:scatter_Pgaia_P1ogle}, but showing the comparison relative to the restricted OGLE-IV catalogue of miras.}
\label{Fig:scatter_Pgaia_Pogle4}
\end{figure}

Figure~\ref{Fig:scatter_Pgaia_Pogle4} illustrates the period comparison of the \Gaia DR3 LPV candidates and the miras from the restricted OGLE-IV catalogue. There is very little scatter around the diagonal, indicating a high degree of period compatibility. Out of 41\,696 miras in the restricted OGLE-IV catalogue that have a counterpart DR3 LPV candidate with a published period in DR3, more than 96\% have periods that are compatible within 20\%, and for more than 94\%, the period differs from the OGLE-IV value by less than 10\% (Tab.\,\ref{tab:ogle4_period_comparison}).

\begin{table*}
\caption{Similar to Table~\ref{tab:ogle3_period_comparison}, but compared with the periods of miras from the restricted OGLE-IV catalogue.}
\label{tab:ogle4_period_comparison}
\centering
\begin{tabular}{ccccc}
\hline\hline
Selection & $N_{\rm xm}$ & $\delta P_1<0.1$ & $\delta P_1<0.2$ & $\delta \nu_1<\varepsilon_{\nu}^{Gaia}$ \\
\hline
Mira     &    41\,696 &    39\,360 &    40\,169 &    35\,929 \\
         &           & ( 94.4\%) & ( 96.3\%) & ( 86.2\%) \\
BLG-Mira &    27\,138 &    24\,995 &    25\,725 &    22\,767 \\
         &           & ( 92.1\%) & ( 94.8\%) & ( 83.9\%) \\
GD-Mira  &    14\,558 &    14\,365 &    14\,444 &    13\,162 \\
         &           & ( 98.7\%) & ( 99.2\%) & ( 90.4\%) \\
\hline
\end{tabular}
\end{table*}

\subsubsection{Comparison with ASAS-SN}
\label{sec:CatalogQuality:PeriodRecovery:ASASSN}

The ASAS-SN catalogue of variable stars \citep{asassn_variables} provides the period only a subset of its sources. In particular, among the DR3 LPV candidates matched within 2$\arcsec$ with the restricted ASAS-SN catalogue, 98\,219 have a period published in \Gaia DR3, 123\,513 have aperiod published in ASAS-SN, and 70\,409 have both (considering only sources that are classified as LPVs in ASAS-SN, see Table~\ref{tab:asassn_recovery_rates}). They consist of 9\,663 miras and 60\,746 SRVs (irregular variables do not have a period published in ASAS-SN). We take this sample as a reference to estimate the rate of period recovery in comparison with ASAS-SN, and adopt same approach as in the OGLE-III case by analyzing the period and frequency differences between the \Gaia DR3 LPV candidates and the ASAS-SN sources,
\begin{equation}\label{eq:deltaPasassn}
    \delta P=\frac{\left|P_{\Gaia}-P_{\textrm{ASAS-SN}}\right|}{P_{\textrm{ASAS-SN}}} \,,
\end{equation}
\begin{equation}\label{eq:deltanuasassn}
    \delta\nu_1 = \left|\nu_{\Gaia}-\nu_{\textrm{ASAS-SN}}\right| \,.
\end{equation}

We considered the fraction of sources whose \Gaia and ASAS-SN period matched within 10\% or 20\%, as summarised in Table~\ref{tab:asassn_period_comparison}. Globally, about 60\% of the selected LPVs are compatible in period within 20\%. This fraction is as high as 90\% for miras, and it is roughly 56\% for SR variables. When we recall the differences between the OGLE-III and ASAS-SN variability type classifications and that a significant fraction of ASAS-SN SR stars may actually be miras (see Sect.~\ref{sec:CatalogQuality:Completeness:ASASSN}), these recovery rates are consistent with the values obtained from the comparison with OGLE-III. We note that the chance of detecting a different period still exists, but cannot be assessed directly as both the \Gaia and ASAS-SN catalogues provide only one period per source. However, it is interesting to examine the \Gaia sources whose period is twice (or half) as long as the ASAS-SN period.
Globally, about 10\% of the selected LPVs are within 20\% of a 2:1 ratio of the periods from the two catalogues. While strictly speaking these sources do not add up to the recovery rate, we can interpret this result as an indication that the light curves of \Gaia DR3 LPV candidates are suitable for characterising the variability period of more sources than the nominal 60\% period recovery rate emerging from the comparison with ASAS-SN, even though the automatic processing tends to pick up a harmonic component rather than the true dominant signal \citep[see also the discussion in Sect.~6.3 of][]{mowlavi_etal_2018_dr2lpv}.

\begin{table*}
\caption{Number $N_{\rm xm}$ of LPV candidates with period published in DR3 and in ASAS-SN that match the restricted ASAS-SN sample within 2$\arcsec$   and the number and fraction of them whose period is compatible with the ASAS-SN value. The period compatibility is evaluated in terms of the difference in period ($\delta P$, see Eq.~\ref{eq:deltaPasassn}) or frequency ($\delta\nu$, see Eq.~\ref{eq:deltanuasassn}).}
\label{tab:asassn_period_comparison}
\centering
\begin{tabular}{ccccc}
\hline\hline
Selection & $N_{\rm xm}$ & $\delta P<0.1$ & $\delta P<0.2$ & $\delta \nu<\varepsilon_{\nu}^{Gaia}$ \\
\hline
LPV        &    70\,409 &    38\,150 &    42\,515 &    36\,979 \\
           &            &  ( 54.2\%) &  ( 60.4\%) &  ( 52.5\%) \\
Mira       &     9\,663 &     8\,329 &     8\,698 &     6\,441 \\
           &            &  ( 86.2\%) &  ( 90.0\%) &  ( 66.7\%) \\
SR         &    60\,746 &    29\,821 &    33\,817 &    30\,538 \\
           &            &  ( 49.1\%) &  ( 55.7\%) &  ( 50.3\%) \\
\hline
\end{tabular}
\end{table*}

\begin{figure}
\centering
\includegraphics[width=\hsize]{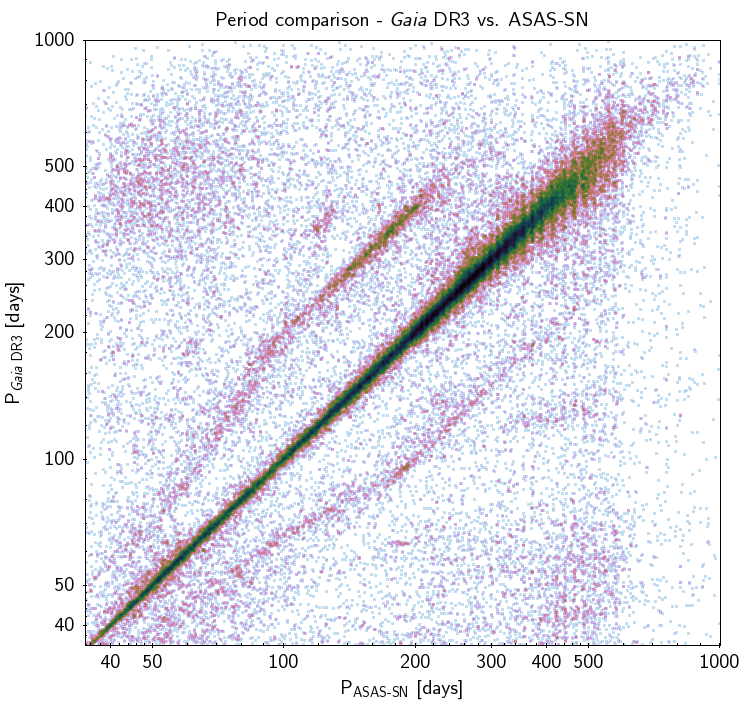}
\caption{Comparison of the period published for \Gaia DR3 LPV candidates and the period of their best-match counterparts in the ASAS-SN catalogue (limited to sources classified as LPVs and with $\Delta V_{\rm ASAS-SN}>0.15$ mag).}
\label{Fig:scatter_Pgaia_Pasassn}
\end{figure}

\subsubsection{Comparison with well-observed LPVs}
\label{sec:CatalogQuality:PeriodRecovery:WellObservedLPVs}

We selected a list of 98 well-observed LPVs covering a wide range in spectral type, period, and variability type.
This sample was used as test cases for our analysis pipeline, similar to a small sample used for the first$^{}$ catalogue,
where the selection was mainly driven by choosing the LPVs with the largest number of references in SIMBAD.
This indicated their interest for the research community.
Here we selected a sample of 98 stars that are present in the first but not necessarily in the second catalogue, with the aim of having a sample that covers various types of stellar parameters.
Thus, the analysis of this sample, which we list in Table \ref{tab:wellknownLPVs} in the appendix, allows insight into prominent LPVs in the catalogue and into the period recovery. 

From this sample, eight stars are neither present in the SOS table nor in the larger classification table. In all these cases, the condition on $N_{RP}$/$N_{G}$>0.5 filter was decisive, that is, they all have a comparably low number of \grp measurements. 
We suspect that because these stars are bright for Gaia, \grp may have been above or very close to the saturation level, resulting in a rejection of \grp measurements.
Seven objects from our sample can only be found in the classification table due to $N_{RP}$/$N_{G}$<0.8. 

Comparing the cases where a literature period exists and a period has been exported into the 2$^{nd}$ Gaia LPV catalogue, we find an agreement within 10\% for 79\% of the stars. 
In two cases, the Gaia period is much shorter than the literature value, and in one case, the Gaia period is several times longer.
We conclude that the period recovery rate for this sample is similar to the results found for the comparison with OGLE-III (Tab.\,\ref{tab:ogle3_period_comparison}), suggesting no significant dependency of the recovery rate on the stellar brightness.

In Tab.\,\ref{tab:wellknownLPVs} we also list periods that were determined by our algorithm, but did not meet our final selection criteria for the catalogue. These are marked with an asterisk (we list the periods only for sources whose light curves are available in DR3). It is, however, interesting to take a look at these values as well. 
In several cases, these Gaia periods agree very well with the literature value. 
For the SRV g Her, a long secondary period of about 900\,d has been reported in the literature \citep{1963AJ.....68..253H}, which agrees well with the 834\,d derived from the Gaia data.
This illustrates that even for stars without an exported period in the catalogue, the Gaia light curves might harbour usable results.

We also indicate the C-star classification in Table \ref{tab:wellknownLPVs}. 
Except for one case, TU Gem, all M and C stars were correctly classified.
As expected, part of the S stars were classified as C rich as well.

\section{Illustrative examples}
\label{sec:Illustrative_examples}

\subsection{Solar neighbourhood}\label{Sect:solarneighbor}

\begin{figure}
\centering
\includegraphics[width=\hsize]{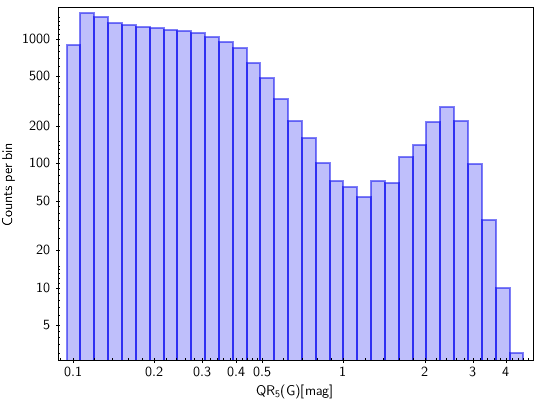}
\caption{Amplitude distribution for solar neighbourhood LPVs (2 kpc radius around the sun).}
\label{Fig:solarneighborhood_qr5}
\end{figure}

\begin{figure}
\centering
\includegraphics[width=\hsize]{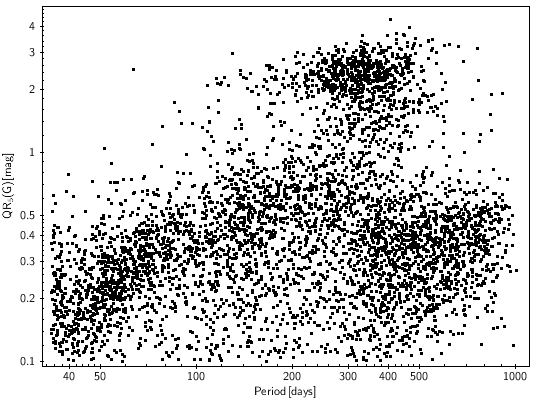}
\caption{Period-amplitude diagram for solar neighbourhood LPVs.}
\label{Fig:solarneighborhood_p_qr5}
\end{figure}

\begin{figure}
\centering
\includegraphics[width=\hsize]{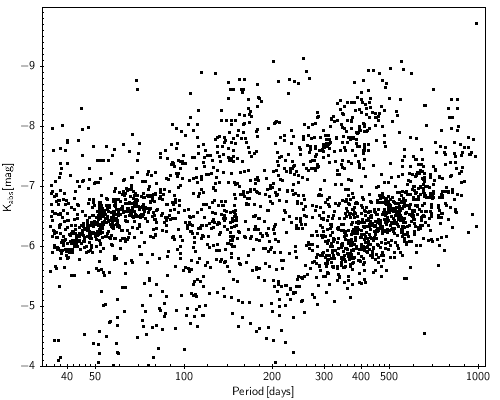}
\caption{logP-K diagram for solar neighbourhood LPVs. For this diagram, only stars with 0<$\sigma_{\pi}/\pi$<0.05 have been chosen.}
\label{Fig:solarneighborhood_pld}
\end{figure}

We define the solar neighbourhood with a distance of 2~kpc around the Sun because 3D reddening maps exist for this radius \citep{2017A&A...606A..65C}.
For this illustrative presentation, however, we did not take the reddening into account.

Much improvement has been made in the EDR3/DR3 astrometric solution compared to that in DR2, not only due to the longer time baseline of the observations, but also due to improved data processing.
In particular, source colour-dependent calibrations were improved in DR3 for sources that had well-determined colours in DR2 \citep{Lindegren_etal21}.
However, and especially for objects such as LPVs, which are both very red and have a wide range of colour changes with time, some strong effects still need to be considered.
The large variability, for example, requires chromatic effects to be taken into account on a per-epoch basis in the image parameter determination and astrometric solution derivation, which is not yet done in DR3 \citep{Lindegren_etal21}.
Further astrometric improvements will occur in future \Gaia data releases for these objects.

 Stars with relative parallax uncertainties better than 15\% (0<$\sigma_\pi/\pi$<0.15) include 21\,218 LPV candidates in the solar neighbourhood from the catalogue.
We further restricted this sample to stars with an absolute 2MASS $K$-band luminosity brighter than -4 mag. This excludes sources that are clearly fainter than the tip of the RGB.
The nature of these intrinsically fainter stars is to be confirmed.
They include a number of YSOs that show some similarity to LPVs in their variability behaviour and colour, a distinct group of relatively blue stars with \gbp-\grp between 1.0 and 2.0, and a few red low brightness stars.
This leaves us with a total sample of 18\,739 stars.
Fig.~\ref{Fig:solarneighborhood_qr5} shows the amplitude distribution of this sample.
It is very similar to the distribution we found for the complete SOS table.
Of the 18\,739 LPVs, 5\,074 have periods in the SOS table. 
The period-amplitude diagram (Fig.\,\ref{Fig:solarneighborhood_p_qr5}) 
allows easily spotting the distinct group of local miras with approximately 600 members and the large number of short-period small-amplitude objects, again similar in their parameter range to the complete SOS sample.
These LPVs are within reach for a variety of high-resolution observing techniques.

Finally, we present in Fig.\,\ref{Fig:solarneighborhood_pld} the P-L diagram for solar neighbourhood LPVs. 
In this case, we restricted the parallax uncertainty to only 5\%, limiting the number of stars to 2\,216.
The P-L diagram is one of the key tools for studying LPV pulsation and shows several parallel pulsation sequences that represent various pulsation modes plus the long secondary period sequence D.
For our sample, we used 2MASS K magnitudes as a substitute for the luminosity.
Sequences B, C$^{\prime}$, C, and D are clearly visible.
This is the most accurate P-L-diagram of LPVs for the galactic field provided up to now.

\subsection{LPVs in globular clusters}
\label{sec:CatalogQuality:SourcesWithKnownDistance:GCs}

Long-period variables in globular clusters have the advantage that their distance can be determined independently from the stellar populations in the clusters. 
Several studies, mainly in the past two decades, led to the detection of a considerable number of LPVs in various stellar clusters.
Detecting LPVs in these clusters allows us to estimate the completeness of the catalogue in sky regions of high stellar density and to estimate the quality of the distance determination in the case of LPVs by comparing them with other cluster members.

Of the 59 cluster stars listed in the catalogue of variable stars in globular clusters \citep{2017yCat.5150....0C} as miras, 54 are included in the second$^{}$ Gaia LPV catalogue. 
For 85\% of these objects, the periods agree well  in the Clement catalogue and the Gaia results.
A more detailed analysis is possible for individual clusters.
For the prominent cluster $\omega$ Cen, \citet{lebzelter_wood_2016} presented a thorough search for LPVs.
Of the 34 LPVs they identified, 23 were recovered by Gaia, including all 15 objects with periods in \citet{lebzelter_wood_2016}.
Ten of the missing 11 objects have visual light amplitudes of about or below 0.1 mag in V, which makes it likely that they were excluded due to our amplitude limit of 0.1 mag in G. 
The reason for the lack of the cluster star LW 23 in our catalogue is not clear.
In addition, the Gaia LPV catalogue includes at least five further LPV candidates that were not listed in \citet{lebzelter_wood_2016}.

We chose the globular cluster 47 Tuc as our main test case because of its closeness, the large number of LPVs that are detected in past surveys, and existing studies on its parallax, including a study that is based on Gaia DR2 \citep{chen_2018}. 
According to the catalogue of \citet{lebzelter_wood_2005}, 45 LPVs are known in this cluster. 
Of this list, 43 objects  are included in the LPV DR3 catalogue, and only V22 and LW22 could not be identified. This provides an excellent recovery rate even in the comparably dense areas of a globular cluster.
Fig.\,\ref{Fig:47Tuc_CMD} shows a CMD of Gaia EDR3 stars selected on the basis of the position on the sky and proper motion data. 
Black circles indicate the LPVs in the cluster area.
A group of LPVs between G = 14 and 16 mag do not belong to the cluster, but are SMC stars.
All the other LPVs, including 5 previously undetected LPVs, nicely form the upper giant branch of the cluster.

\begin{figure}
\centering
\includegraphics[width=\hsize]{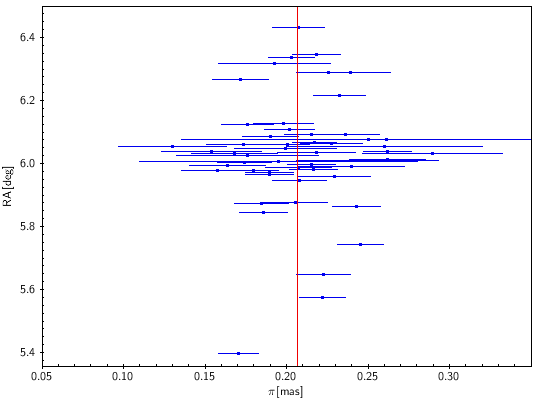}
\caption{Right ascension of 47 Tuc LPVs against parallax in mas. The plotted sources have all been considered members in the past. The vertical red line marks the $\pi_{47Tuc}-\pi_{SMC}$ value derived by \citet{chen_2018} from a large set of stars on the lower giant branch.}
\label{Fig:47Tuc_DM}
\end{figure}

\begin{figure}
\centering
\includegraphics[width=\hsize]{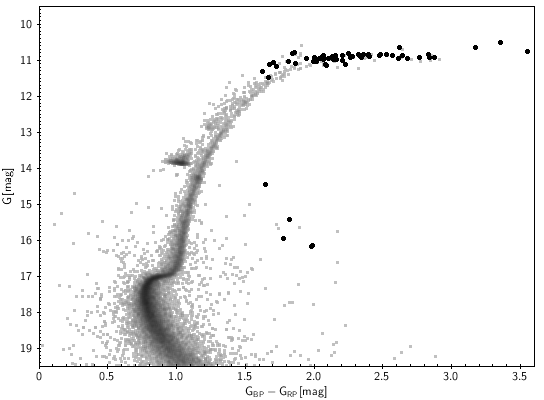}
\caption{CMD for a sample of EDR3 stars in a 0.4° field of view around the centre of 47 Tuc (small points). The large black points mark the LPVs from our catalogue that likely belong to the cluster. Their membership is confirmed by their location in the CMD. The lower brightness LPVs are SMC stars.}
\label{Fig:47Tuc_CMD}
\end{figure}

Excluding the SMC stars, we used the 47 Tuc sample for a test of the accuracy of Gaia distances for LPVs. 
In Fig.\,\ref{Fig:47Tuc_DM} we plot the stellar RA values against the parallax.
The crowding of objects in the plot centre corresponds to the central part of the cluster.
The vertical line marks the $\pi_{47Tuc}-\pi_{SMC}$ value derived by \citet{chen_2018} from a large set of stars on the lower giant branch.
For the majority of the LPVs, the parallax agrees with the mean parallax from all cluster stars within the parallax uncertainties. 
Significant offsets seen for stars at larger distances from the cluster centre may be due to the stars being foreground or background stars, which disagrees with the locations of these stars in the CMD, however.
Many of the stars close to the cluster centre with large deviations from the mean parallax are found in the innermost regions of the cluster, suggesting that the reliability of the Gaia parallax measurements decreases in regions of high stellar density. 
We conclude that parallaxes and parallax uncertainties for LPVs derived in Gaia DR3 have to be used with some caution, in particular in crowded areas.
However, the average parallax of the stars plotted in Fig.\,\ref{Fig:47Tuc_DM} agrees with $\pi$=0.209$\pm$0.033 almost exactly with the value from \citet{chen_2018}.

\begin{figure}
\centering
\includegraphics[width=\hsize]{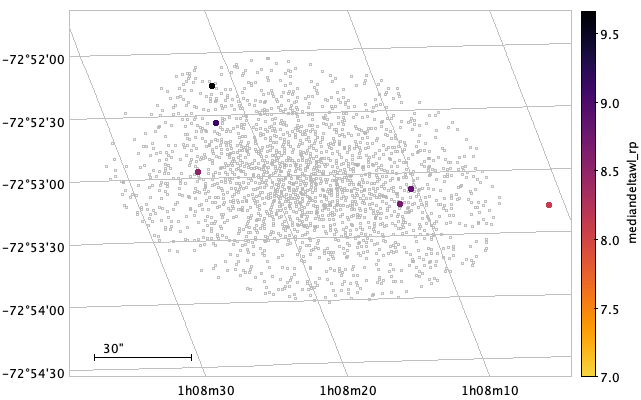}
\caption{LPVs in NGC 419. Colours indicate our selection criterion for C stars.}
\label{Fig:NGC419_field}
\end{figure}

The SMC cluster NGC 419 is known to be rich in C stars \citep[e.g.][]{kamath_wood_2010}.
Fig.\,\ref{Fig:NGC419_field} shows the sky area around the cluster. 
Gaia EDR3 sources within 1 arcmin around the cluster centre were added with small open symbols to guide the eye on the location of the cluster.
Large symbols mark objects from the EDR3 LPV catalogue. The colour indicates the $\mediandeltawl$ value.
As discussed above, values above 7 in this parameter are classified as C stars. 
Of the 16 C-type LPVs known in the cluster, 6 are in the catalogue, all of them showing values above 7, which confirms the ability of identifying C stars with our chosen method.
The 6 stars are not the brightest stars in the cluster, but we see a lack of LPV identifications in the central regions of the cluster, where the stellar density was obviously too high.

\subsection{LPVs in Local Group galaxies}
\label{sec:CatalogQuality:SourcesWithKnownDistance:LocalGroup}

In addition to the Magellanic Clouds, we identified seven Local Group galaxies with at least ten candidate members among the \Gaia DR3 LPV catalogue. They are listed in Table~\ref{tab:lpv_local_group}. They include M31 and M33\footnote{
    We note that additional LPV candidates in M31 are realistically present in the \Gaia Andromeda Photometric Survey \citep{DR3-DPACP-142}, for which the time series are available.
}, as well as several dwarf spheroidal galaxies. Compared with the previous version of the \Gaia LPV catalogue, we find an increase of $\sim$3 to $\sim$8 times in the number of candidate members for these galaxies, except for M31 and M33, in which cases the increase is more than two orders of magnitudes because of the very small numbers in the DR2 catalogue.

We note that when we considered sources with a published period, the increase factor from DR2 to DR3 is smaller. The fraction of DR2 LPV candidates with a published period in each galaxy is larger than the corresponding fraction in DR3. This means that the criterion for the inclusion of sources in the \Gaia DR3 catalogue has been relaxed with respect to DR2 to a higher degree than what was effectively done for the criterion concerning the publication of periods.

\begin{figure}
\centering
\includegraphics[width=\hsize]{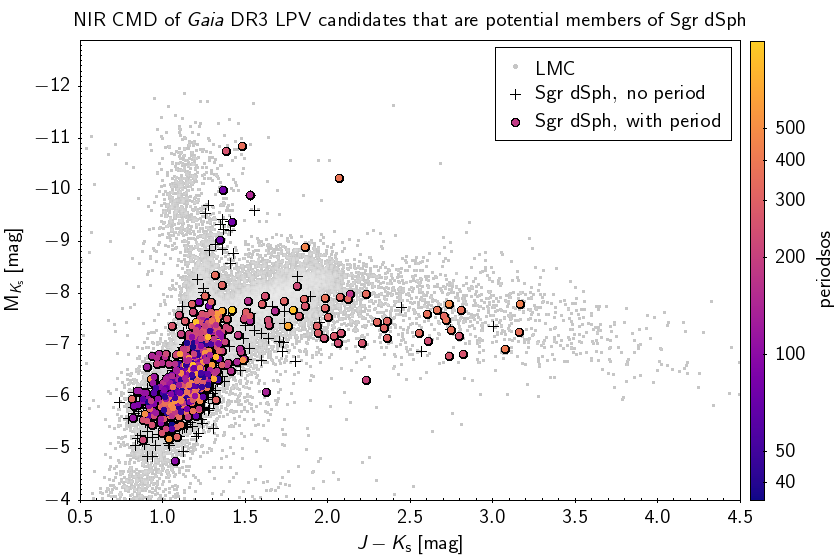}
\caption{Colour vs absolute magnitude diagram in the NIR $J$ and $\ks$ filters of 2MASS of \Gaia DR3 LPVs that are candidate members of Sgr dSph. Plus symbols show sources whose period was not retained for publication, and circles are colour-coded according to the period. Grey symbols in the background are \Gaia DR3 LPVs in the LMC for reference.}
\label{Fig:JK_Sgr}
\end{figure}

\begin{figure*}
\centering
\includegraphics[width=.325\hsize]{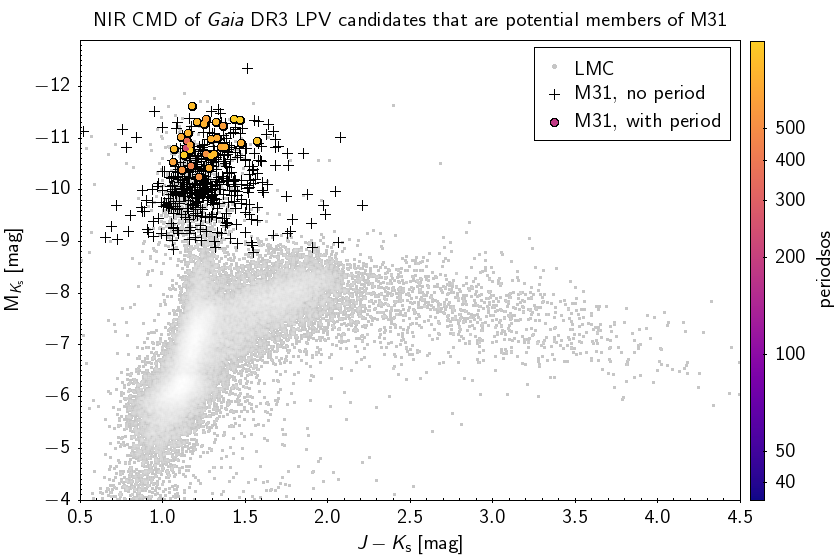}
\includegraphics[width=.325\hsize]{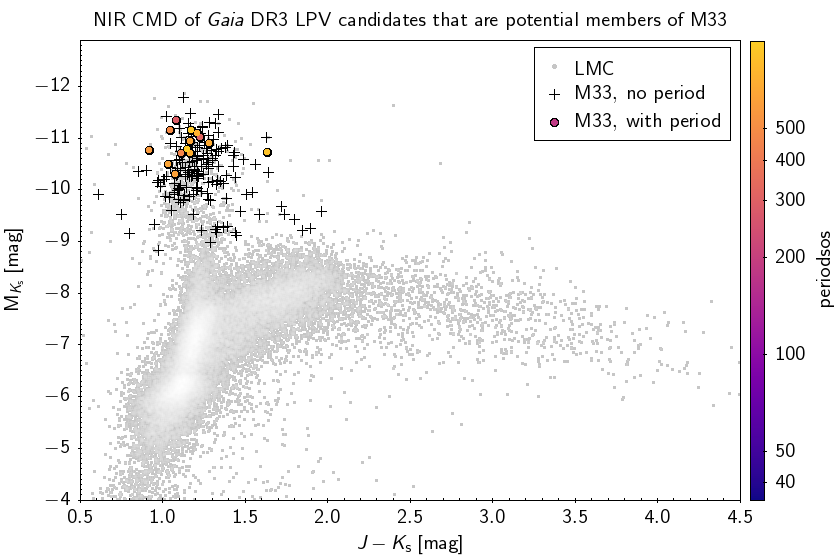}
\includegraphics[width=.325\hsize]{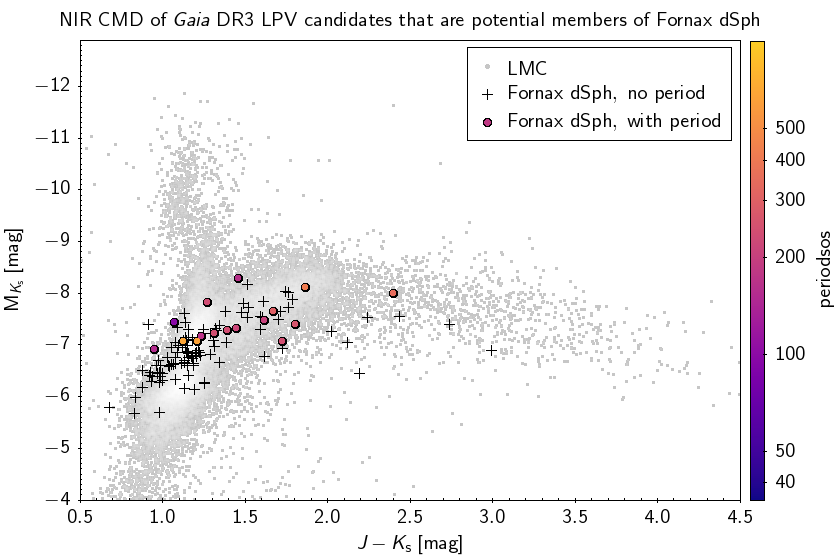}
\caption{Similar to Fig.~\ref{Fig:JK_Sgr}, but showing (from left to right) M31, M33, and Fornax dSph.
}
\label{Fig:JK_M31_M33_Fornax}
\end{figure*}

\begin{figure*}
\centering
\includegraphics[width=.325\hsize]{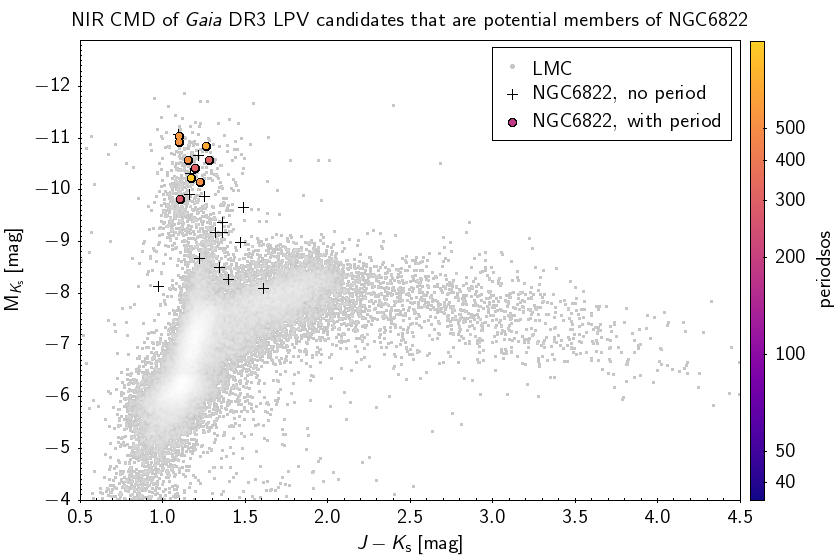}
\includegraphics[width=.325\hsize]{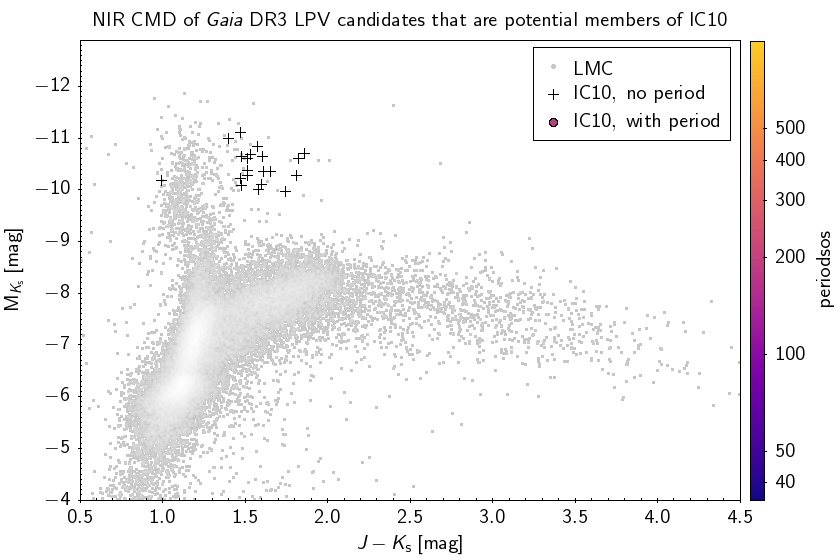}
\includegraphics[width=.325\hsize]{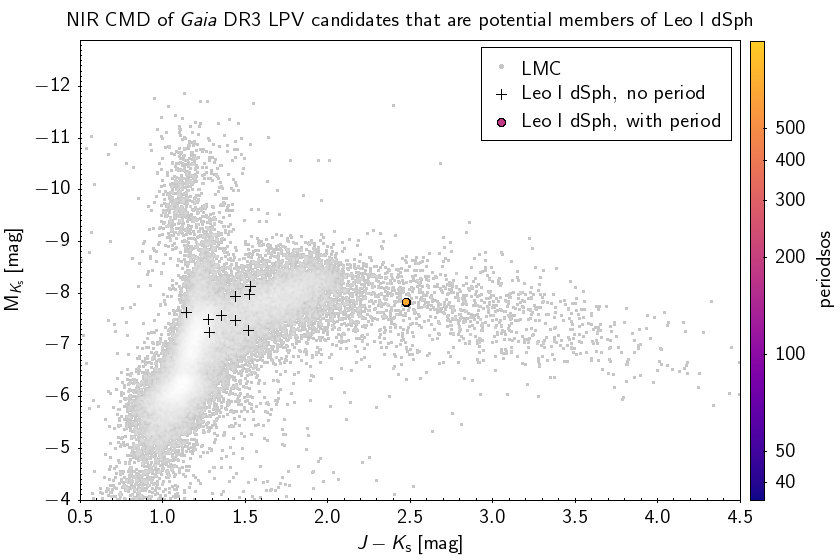}
\caption{Similar to Fig.~\ref{Fig:JK_Sgr}, but showing (from left to right) NGC 6822, IC 10, and Leo I dSph.}
\label{Fig:JK_NGC6822_IC10_LeoI}
\end{figure*}

Figs.\,\ref{Fig:JK_Sgr} to \ref{Fig:JK_NGC6822_IC10_LeoI} show the 2MASS colour-absolute magnitude diagram for the LMC, and the location of the LPVs is overplotted for each galaxy. 
In the Sgr dSph, the detected LPVs cover a similar range as the LMC stars. Periods are available for almost 40\% of these objects. At the upper giant branch, the bifurcation into C stars (large $J-K$) and M stars is clearly visible.
The Fornax dSph lacks massive AGB stars, as expected. 
Summarising various studies, \citet{2010A&A...520A..55Y} characterised the dwarf irregular galaxy IC 10 as a low-mass, metal poor ([Fe/H]=$-$1.1), and actively star-forming galaxy. The presence of luminous LPVs in this galaxy, possibly including supergiants, is thus not surprising. 
At this large distance, the fraction of LPVs with periods becomes very low. This is also true for M31 and M33. The colour-absolute magnitude diagrams (Fig.\,\ref{Fig:JK_M31_M33_Fornax}) reveal that our catalogue includes only the brightest objects there.

\begin{figure}
\centering
\includegraphics[width=\hsize]{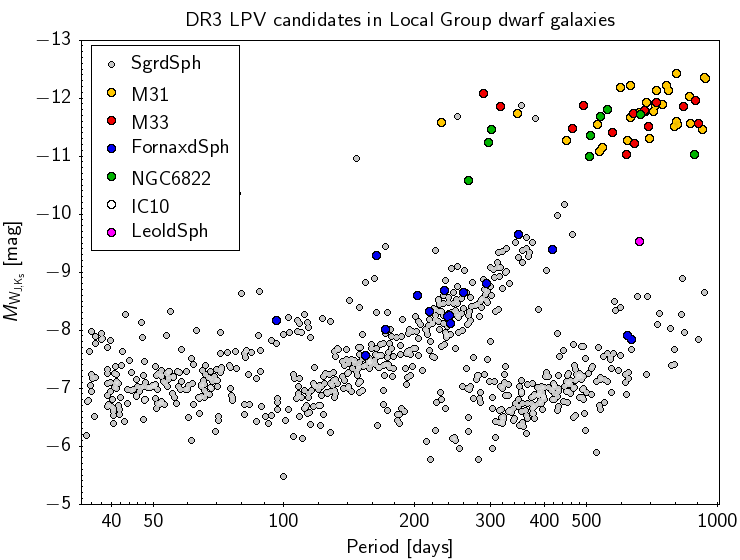}
\caption{Period-luminosity diagram, with the 2MASS NIR Wesenheit index $\wjk$, of \Gaia DR3 LPVs with published period that are candidate members of Local Group galaxies (excluding the Magellanic Clouds).}
\label{Fig:PLD_LGGs_WJK}
\end{figure}

\begin{figure*}
\centering
\includegraphics[width=.49\hsize]{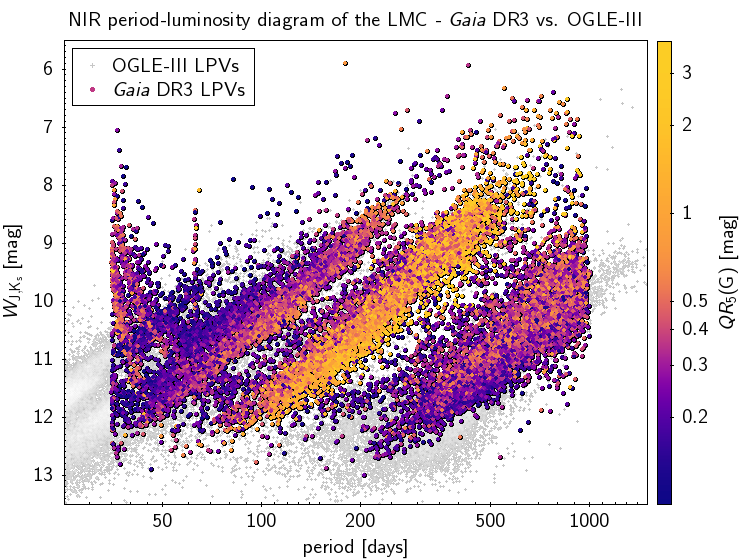}
\includegraphics[width=.49\hsize]{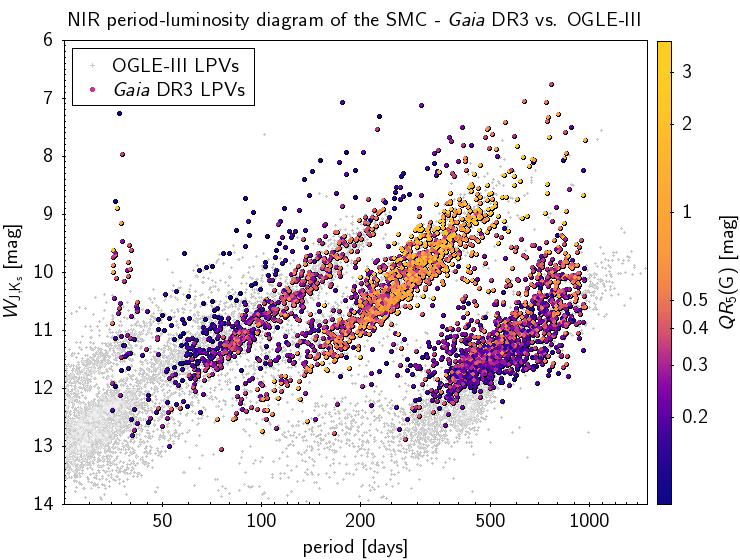}
\caption{Period-luminosity diagram, with the 2MASS NIR Wesenheit index $\wjk$, of \Gaia DR3 LPVs that are candidate members of the LMC (left panel) and SMC (right panel) \citep[using the selection criteria of][]{mowlavitrabucchilebzelter_2019}. Data points are colour-coded by their $G$-band amplitude (5-95\% interquantile range). Data from the OGLE-III catalogue are shown as grey symbols in the background for reference.}
\label{Fig:PLD_MCs}
\end{figure*}

Figure~\ref{Fig:PLD_LGGs_WJK} displays the period-luminosity diagram of candidate members of Sgr dSph, M31, M33, Fornax dSph, NGC 6822, and Leo I dSph using the 2MASS NIR Wesenheit index, $\wjk$, as a luminosity indicator. 
The Sgr dSph LPVs clearly form the usual period-luminosity relations C$^{\prime}$, C, and D, which are known from the Magellanic Clouds \citep{wood_etal_1999}. These relations are evident in the period-luminosity diagrams obtained with data from the second$^{}$ \Gaia catalogue of LPV candidates combined with 2MASS data (Fig.\,\ref{Fig:PLD_MCs}). In addition to these three sequences, the present catalogue includes LPVs populating the period-luminosity sequence B in the two Magellanic Clouds, as well as the bright end of sequence A in the LMC. Neither of these two sequences were visible with data from the first$^{}$ \Gaia catalogue of LPV candidates \cite[e.g.][]{lebzelter_etal_2019}.
The few LPVs with periods in the Fornax dSph fit these sequences very well.
The luminous sources detected in M31, M33, and NGC 6822 fit the extensions of the sequences in the supergiant regime.

\begin{table*}
\caption{Number of sources, selected among the \Gaia DR3 LPV catalogue, that are candidate members of Local Group galaxies, as well the number (and percentage) of them with a published period value in DR3. The same quantities are provided for sources that were part of the \Gaia DR2 LPV catalogue. We also show the number of sources found in the classification table.
}
\label{tab:lpv_local_group}
\centering
\begin{tabular}{c c r r l r r l c }
\hline\hline
\multirow{2}{*}{Galaxy} & \multirow{2}{*}{$(m-M)_0$} &                        \multicolumn{3}{c}{\Gaia DR3} &             \multicolumn{3}{c}{\Gaia DR2} &       including sources \\
\cmidrule(lr){3-5}\cmidrule(lr){6-8}
                        &                            & members &            \multicolumn{2}{c}{with period} & members & \multicolumn{2}{c}{with period} & in classification table \\
\hline
Sgr dSph                &             17.10 $^{(b)}$ &   1 772 &                             719 & (38.7\%) &     342 &                  149 & (43.6\%) &                   1 860 \\
LMC                     &             18.49 $^{(a)}$ &  40 059 &                          17 814 & (42.4\%) &  11 022 &                6 787 & (61.6\%) &                  42 047 \\
SMC                     &             18.96 $^{(a)}$ &   5 027 &                           3 073 & (59.6\%) &   1 789 &                1 352 & (75.6\%) &                   5 153 \\
Fornax dSph             &             20.72 $^{(c)}$ &      46 &                              16 & (12.9\%) &      15 &                   15 &  (100\%) &                     124 \\
Leo I dSph              &             22.15 $^{(f)}$ &       1 &                               1 & (10.0\%) &       0 &                    0 &      (-) &                      10 \\
NGC 6822                &             23.40 $^{(d)}$ &      13 &                               9 &    (9\%) &       4 &                    2 & (50.0\%) &                      29 \\
IC 10                   &             24.36 $^{(a)}$ &       4 &                               0 &      (-) &       0 &                    0 &      (-) &                      20 \\
M31                     &             24.45 $^{(a)}$ &     123 &                              32 &  (6.0\%) &       4 &                    4 &  (100\%) &                     531 \\
M33                     &             24.67 $^{(a)}$ &      51 &                              14 &  (7.6\%) &       1 &                    0 &      (-) &                     185 \\
\hline
\end{tabular}
\tablefoot{$^{(a)}$\citet{degrijs_etal_2017}, $^{(b)}$\citet{monaco_etal_2004}, $^{(c)}$\citet{rizzi_2007}, $^{(d)}$\citet{feast_et_al_1989}, $^{(e)}$\citet{kim_etal_2009}, $^{(f)}$\citet{stetson_etal_2014}.\\}
\end{table*}

\section{Summary and conclusions}

In this paper we discussed the production of the second$^{}$ \Gaia catalogue of LPV candidates, explored its completeness, validated the determined parameters, and illustrated its content.
This is currently the largest database for LPVs with 1\,720\,588 entries in the SOS table and more than 600\,000 additional candidates in the classification table. Our analysis shows it to be a valid resource for the study of LPVs and AGB stars.
In Appendix \ref{app:catalogRetrieval} we provide the necessary codes to retrieve the full catalogue or parts of the database from the \Gaia archive.

However, like all catalogues, it has some limitations that need to be taken into account when the data are used.
We summarise here the main caveats and options for the usage.
Several of these limitations are expected to be removed with Data Release 4.

\begin{itemize}
    \item Completeness has been discussed in Sect.\,\ref{sec:CatalogQuality:Completeness}. The typical recovery rate for LPVs derived both from LMC and field stars is about 80\%. Although this a high degree of completeness, some LPVs will be missing from the catalogue.
    \item Contamination from non-LPV sources is estimated to be about 2\%.

    \item Periods are provided for a sub-sample of 392\,240 LPVs with periods longer than 35~days and excluding signals that could be identified as spurious.
    Our results agree well with those available in the literature.
    The light curves of all the sources in the catalogue are available in the \Gaia archive, so that the users can perform an independent period search.
    In particular, LPV types that are known or suspected to be multi-periodic deserve further investigations. 
    
    \item Periods and amplitudes in the catalogue were derived from light curves with a maximum duration of 1000\,d. Therefore, sources with variability timescales exceeding 500\,d should be considered with care because only only one complete light cycle is covered by the DR3 data.
    
    \item As a major advance compared to the first catalogue, C-star candidates were identified among the LPVs from their (variable) RP spectra.
    We refer users to Sect.~\ref{sec:CatalogConstruction:IdentificationCstars} for a description of the C-star flag and recommendations on its usage.
\end{itemize}

\begin{acknowledgements}
This work presents results from the European Space Agency (ESA) space mission \Gaia.
\Gaia data are being processed by the \Gaia Data Processing and Analysis Consortium (DPAC).
Funding for the DPAC is provided by national institutions, in particular the institutions participating in the \Gaia MultiLateral Agreement (MLA).
The \Gaia mission website is \url{https://www.cosmos.esa.int/gaia}.
The \Gaia archive website is \url{https://archives.esac.esa.int/gaia}.
Acknowledgements are given in Appendix~\ref{app:acknowledgements}.
\end{acknowledgements}

\bibliographystyle{aa}
\bibliography{referencesDR3.bib}

\begin{thebibliography}{61}
\expandafter\ifx\csname natexlab\endcsname\relax\def\natexlab#1{#1}\fi

\bibitem[{{Boyer} {et~al.}(2011){Boyer}, {Srinivasan}, {van Loon}, {McDonald},
  {Meixner}, {Zaritsky}, {Gordon}, {Kemper}, {Babler}, {Block}, {Bracker},
  {Engelbracht}, {Hora}, {Indebetouw}, {Meade}, {Misselt}, {Robitaille},
  {Sewi{\l}o}, {Shiao}, \& {Whitney}}]{boyer_et_al_2011}
{Boyer}, M.~L., {Srinivasan}, S., {van Loon}, J.~T., {et~al.} 2011, \aj, 142,
  103

\bibitem[{{Capitanio} {et~al.}(2017){Capitanio}, {Lallement}, {Vergely},
  {Elyajouri}, \& {Monreal-Ibero}}]{2017A&A...606A..65C}
{Capitanio}, L., {Lallement}, R., {Vergely}, J.~L., {Elyajouri}, M., \&
  {Monreal-Ibero}, A. 2017, \aap, 606, A65

\bibitem[{{Chen} {et~al.}(2018){Chen}, {Richer}, {Caiazzo}, \&
  {Heyl}}]{chen_2018}
{Chen}, S., {Richer}, H., {Caiazzo}, I., \& {Heyl}, J. 2018, \apj, 867, 132

\bibitem[{{Christlieb} {et~al.}(2001){Christlieb}, {Green}, {Wisotzki}, \&
  {Reimers}}]{christlieb_etal_2001}
{Christlieb}, N., {Green}, P.~J., {Wisotzki}, L., \& {Reimers}, D. 2001, \aap,
  375, 366

\bibitem[{{Clement}(2017)}]{2017yCat.5150....0C}
{Clement}, C.~M. 2017, VizieR Online Data Catalog, V/150

\bibitem[{{de Grijs} {et~al.}(2017){de Grijs}, {Courbin},
  {Mart{\'\i}nez-V{\'a}zquez}, {Monelli}, {Oguri}, \&
  {Suyu}}]{degrijs_etal_2017}
{de Grijs}, R., {Courbin}, F., {Mart{\'\i}nez-V{\'a}zquez}, C.~E., {et~al.}
  2017, \ssr, 212, 1743

\bibitem[{{Dorda} \& {Patrick}(2021)}]{2021MNRAS.502.4890D}
{Dorda}, R. \& {Patrick}, L.~R. 2021, \mnras, 502, 4890

\bibitem[{{Evans et al.}(2022)}]{DR3-DPACP-142}
{Evans et al.} 2022, \aap\ in prep.

\bibitem[{{Eyer et al.}(2022)}]{DR3-DPACP-162}
{Eyer et al.} 2022, \aap, in prep.

\bibitem[{{Feast} {et~al.}(1989){Feast}, {Glass}, {Whitelock}, \&
  {Catchpole}}]{feast_et_al_1989}
{Feast}, M.~W., {Glass}, I.~S., {Whitelock}, P.~A., \& {Catchpole}, R.~M. 1989,
  \mnras, 241, 375

\bibitem[{{Fraser} {et~al.}(2008){Fraser}, {Hawley}, \&
  {Cook}}]{2008AJ....136.1242F}
{Fraser}, O.~J., {Hawley}, S.~L., \& {Cook}, K.~H. 2008, \aj, 136, 1242

\bibitem[{{Gaia Collaboration} {et~al.}(2018){Gaia Collaboration}, {Brown},
  {Vallenari}, {Prusti}, {de Bruijne}, {Babusiaux}, {Bailer-Jones}, {Biermann},
  {Evans}, {Eyer}, \& et~al.}]{gaia_dr2_2018}
{Gaia Collaboration}, {Brown}, A.~G.~A., {Vallenari}, A., {et~al.} 2018, \aap,
  616, A1

\bibitem[{{Gaia Collaboration} {et~al.}(2016){Gaia Collaboration}, {Prusti},
  {de Bruijne}, {Brown}, {Vallenari}, {Babusiaux}, {Bailer-Jones}, {Bastian},
  {Biermann}, {Evans}, \& et~al.}]{gaia_mission_2016}
{Gaia Collaboration}, {Prusti}, T., {de Bruijne}, J.~H.~J., {et~al.} 2016,
  \aap, 595, A1

\bibitem[{{Gaia Collaboration} {et~al.}(2022){Gaia Collaboration}, {Vallenari},
  \& {et al.}}]{DR3-DPACP-185}
{Gaia Collaboration}, {Vallenari}, A., \& {et al.} 2022, \aap, in prep.

\bibitem[{{Gonneau} {et~al.}(2017){Gonneau}, {Lan{\c{c}}on}, {Trager},
  {Aringer}, {Nowotny}, {Peletier}, {Prugniel}, {Chen}, \&
  {Lyubenova}}]{2017A&A...601A.141G}
{Gonneau}, A., {Lan{\c{c}}on}, A., {Trager}, S.~C., {et~al.} 2017, \aap, 601,
  A141

\bibitem[{Hoffleit~D.(1991)}]{vizier:v/50}
Hoffleit~D., W. J.~W. 1991, {VizieR Online Data Catalog: Bright Star Catalogue,
  5th Revised Ed.}

\bibitem[{{H{\"o}fner} \& {Olofsson}(2018)}]{hoefner_olofsson_2018}
{H{\"o}fner}, S. \& {Olofsson}, H. 2018, \aapr, 26, 1

\bibitem[{{Holl et al.}(2022)}]{DR3-DPACP-164}
{Holl et al.} 2022, \aap\ in prep.

\bibitem[{{Houk}(1963)}]{1963AJ.....68..253H}
{Houk}, N. 1963, \aj, 68, 253

\bibitem[{{Iwanek} {et~al.}(2022){Iwanek}, {Soszy{\'n}ski}, {Koz{\l}owski},
  {Poleski}, {Pietrukowicz}, {Skowron}, {Wrona}, {Mr{\'o}z}, {Udalski},
  {Szyma{\'n}ski}, {Skowron}, {Ulaczyk}, {Gromadzki}, {Rybicki}, \&
  {Ratajczak}}]{ogle4miras}
{Iwanek}, P., {Soszy{\'n}ski}, I., {Koz{\l}owski}, S., {et~al.} 2022, arXiv
  e-prints, arXiv:2203.16552

\bibitem[{{Jayasinghe} {et~al.}(2021){Jayasinghe}, {Kochanek}, {Stanek},
  {Shappee}, {Holoien}, {Thompson}, {Prieto}, {Dong}, {Pawlak}, {Pejcha},
  {Pojmanski}, {Otero}, {Hurst}, \& {Will}}]{asassn_variables}
{Jayasinghe}, T., {Kochanek}, C.~S., {Stanek}, K.~Z., {et~al.} 2021, \mnras,
  503, 200

\bibitem[{{Kamath} {et~al.}(2010){Kamath}, {Wood}, {Soszy{\'n}ski}, \&
  {Lebzelter}}]{kamath_wood_2010}
{Kamath}, D., {Wood}, P.~R., {Soszy{\'n}ski}, I., \& {Lebzelter}, T. 2010,
  \mnras, 408, 522

\bibitem[{{Kim} {et~al.}(2009){Kim}, {Kim}, {Hwang}, {Lee}, {Im}, {Karoji},
  {Noumaru}, \& {Tanaka}}]{kim_etal_2009}
{Kim}, M., {Kim}, E., {Hwang}, N., {et~al.} 2009, \apj, 703, 816

\bibitem[{{Kochanek} {et~al.}(2017){Kochanek}, {Shappee}, {Stanek}, {Holoien},
  {Thompson}, {Prieto}, {Dong}, {Shields}, {Will}, {Britt}, {Perzanowski}, \&
  {Pojma{\'n}ski}}]{asassn_survey}
{Kochanek}, C.~S., {Shappee}, B.~J., {Stanek}, K.~Z., {et~al.} 2017, \pasp,
  129, 104502

\bibitem[{{Lan{\c{c}}on} \& {Wood}(2000)}]{2000A&AS..146..217L}
{Lan{\c{c}}on}, A. \& {Wood}, P.~R. 2000, \aaps, 146, 217

\bibitem[{{Lebzelter} {et~al.}(2005){Lebzelter}, {Hinkle}, {Wood}, {Joyce}, \&
  {Fekel}}]{2005A&A...431..623L}
{Lebzelter}, T., {Hinkle}, K.~H., {Wood}, P.~R., {Joyce}, R.~R., \& {Fekel},
  F.~C. 2005, \aap, 431, 623

\bibitem[{{Lebzelter} {et~al.}(2018){Lebzelter}, {Mowlavi}, {Marigo},
  {Pastorelli}, {Trabucchi}, {Wood}, \&
  {Lecoeur-Ta{\"i}bi}}]{lebzelter_etal_2018}
{Lebzelter}, T., {Mowlavi}, N., {Marigo}, P., {et~al.} 2018, \aap, 616, L13

\bibitem[{{Lebzelter} {et~al.}(2019){Lebzelter}, {Trabucchi}, {Mowlavi},
  {Wood}, {Marigo}, {Pastorelli}, \&
  {Lecoeur-Ta{\"\i}bi}}]{lebzelter_etal_2019}
{Lebzelter}, T., {Trabucchi}, M., {Mowlavi}, N., {et~al.} 2019, \aap, 631, A24

\bibitem[{{Lebzelter} \& {Wood}(2005)}]{lebzelter_wood_2005}
{Lebzelter}, T. \& {Wood}, P.~R. 2005, \aap, 441, 1117

\bibitem[{{Lebzelter} \& {Wood}(2016)}]{lebzelter_wood_2016}
{Lebzelter}, T. \& {Wood}, P.~R. 2016, \aap, 585, A111

\bibitem[{{Li} {et~al.}(2018){Li}, {Luo}, {Du}, {Zuo}, {Wang}, {Zhao}, {Jiang},
  {Zhang}, {Liu}, {Qin}, {Wang}, {Du}, {Guo}, {Wang}, {Han}, {Xiang}, {Huang},
  {Chen}, {Chen}, {Kong}, {Hou}, {Song}, {Wang}, {Wu}, {Zhang}, {Zhang},
  {Wang}, {Cao}, {Hou}, \& {Zhao}}]{Li_etal_2018}
{Li}, Y.-B., {Luo}, A.~L., {Du}, C.-D., {et~al.} 2018, \apjs, 234, 31

\bibitem[{{Lindegren} {et~al.}(2021){Lindegren}, {Klioner}, {Hern{\'a}ndez},
  {Bombrun}, {Ramos-Lerate}, {Steidelm{\"u}ller}, {Bastian}, {Biermann}, {de
  Torres}, {Gerlach}, {Geyer}, {Hilger}, {Hobbs}, {Lammers}, {McMillan},
  {Stephenson}, {Casta{\~n}eda}, {Davidson}, {Fabricius}, {Gracia-Abril},
  {Portell}, {Rowell}, {Teyssier}, {Torra}, {Bartolom{\'e}}, {Clotet},
  {Garralda}, {Gonz{\'a}lez-Vidal}, {Torra}, {Abbas}, {Altmann}, {Anglada
  Varela}, {Balaguer-N{\'u}{\~n}ez}, {Balog}, {Barache}, {Becciani}, {Bernet},
  {Bertone}, {Bianchi}, {Bouquillon}, {Brown}, {Bucciarelli}, {Busonero},
  {Butkevich}, {Buzzi}, {Cancelliere}, {Carlucci}, {Charlot}, {Cioni},
  {Crosta}, {Crowley}, {del Peloso}, {del Pozo}, {Drimmel}, {Esquej}, {Fienga},
  {Fraile}, {Gai}, {Garcia-Reinaldos}, {Guerra}, {Hambly}, {Hauser},
  {Jan{\ss}en}, {Jordan}, {Kostrzewa-Rutkowska}, {Lattanzi}, {Liao}, {Licata},
  {Lister}, {L{\"o}ffler}, {Marchant}, {Masip}, {Mignard}, {Mints}, {Molina},
  {Mora}, {Morbidelli}, {Murphy}, {Pagani}, {Panuzzo}, {Pe{\~n}alosa Esteller},
  {Poggio}, {Re Fiorentin}, {Riva}, {Sagrist{\`a} Sell{\'e}s}, {Sanchez
  Gimenez}, {Sarasso}, {Sciacca}, {Siddiqui}, {Smart}, {Souami}, {Spagna},
  {Steele}, {Taris}, {Utrilla}, {van Reeven}, \&
  {Vecchiato}}]{Lindegren_etal21}
{Lindegren}, L., {Klioner}, S.~A., {Hern{\'a}ndez}, J., {et~al.} 2021, \aap,
  649, A2

\bibitem[{{MacConnell}(2003)}]{MacConnell_2003}
{MacConnell}, D.~J. 2003, \pasp, 115, 351

\bibitem[{{Monaco} {et~al.}(2004){Monaco}, {Bellazzini}, {Ferraro}, \&
  {Pancino}}]{monaco_etal_2004}
{Monaco}, L., {Bellazzini}, M., {Ferraro}, F.~R., \& {Pancino}, E. 2004,
  \mnras, 353, 874

\bibitem[{{Morgan} \& {Hatzidimitriou}(1995)}]{1995A&AS..113..539M}
{Morgan}, D.~H. \& {Hatzidimitriou}, D. 1995, \aaps, 113, 539

\bibitem[{{Mowlavi} {et~al.}(2018){Mowlavi}, {Lecoeur-Ta{\"i}bi}, {Lebzelter},
  {Rimoldini}, {Lorenz}, {Audard}, {De Ridder}, {Eyer}, {Guy}, {Holl},
  {Jevardat de Fombelle}, {Marchal}, {Nienartowicz}, {Regibo}, {Roelens}, \&
  {Sarro}}]{mowlavi_etal_2018_dr2lpv}
{Mowlavi}, N., {Lecoeur-Ta{\"i}bi}, I., {Lebzelter}, T., {et~al.} 2018, \aap,
  618, A58

\bibitem[{{Mowlavi} {et~al.}(2019){Mowlavi}, {Trabucchi}, \&
  {Lebzelter}}]{mowlavitrabucchilebzelter_2019}
{Mowlavi}, N., {Trabucchi}, M., \& {Lebzelter}, T. 2019, in The Gaia Universe,
  62

\bibitem[{{Palmer} \& {Wing}(1982)}]{1982AJ.....87.1739P}
{Palmer}, L.~G. \& {Wing}, R.~F. 1982, \aj, 87, 1739

\bibitem[{{Riess} {et~al.}(2020){Riess}, {Yuan}, {Casertano}, {Macri}, \&
  {Scolnic}}]{Riess_etal_2020}
{Riess}, A.~G., {Yuan}, W., {Casertano}, S., {Macri}, L.~M., \& {Scolnic}, D.
  2020, \apjl, 896, L43

\bibitem[{{Rimoldini et al.}(2022)}]{DR3-DPACP-165}
{Rimoldini et al.} 2022, \aap\ in prep.

\bibitem[{{Rizzi} {et~al.}(2007){Rizzi}, {Held}, {Saviane}, {Tully}, \&
  {Gullieuszik}}]{rizzi_2007}
{Rizzi}, L., {Held}, E.~V., {Saviane}, I., {Tully}, R.~B., \& {Gullieuszik}, M.
  2007, \mnras, 380, 1255

\bibitem[{{Samus} {et~al.}(2017){Samus}, {Kazarovets}, {Durlevich}, {Kireeva},
  \& {Pastukhova}}]{gcvs}
{Samus}, N.~N., {Kazarovets}, E.~V., {Durlevich}, O.~V., {Kireeva}, N.~N., \&
  {Pastukhova}, E.~N. 2017, Astronomy Reports, 61, 80

\bibitem[{{Si} {et~al.}(2015){Si}, {Li}, {Luo}, {Tu}, {Shi}, {Zhang}, {Wei},
  {Zhao}, {Wu}, {Wu}, \& {Zhao}}]{Si_etal_2015}
{Si}, J.-M., {Li}, Y.-B., {Luo}, A.~L., {et~al.} 2015, Research in Astronomy
  and Astrophysics, 15, 1671

\bibitem[{{Soszy{\'n}ski} {et~al.}(2004){Soszy{\'n}ski}, {Udalski}, {Kubiak},
  {Szyma{\'n}ski}, {Pietrzy{\'n}ski}, {{\.Z}ebru{\'n}}, {Szewczyk}, \&
  {Wyrzykowski}}]{2004AcA....54..129S}
{Soszy{\'n}ski}, I., {Udalski}, A., {Kubiak}, M., {et~al.} 2004, \actaa, 54,
  129

\bibitem[{{Soszy{\'n}ski} {et~al.}(2009){Soszy{\'n}ski}, {Udalski},
  {Szyma{\'n}ski}, {Kubiak}, {Pietrzy{\'n}ski}, {Wyrzykowski}, {Szewczyk},
  {Ulaczyk}, \& {Poleski}}]{soszynski_etal_2009_ogle3lmc}
{Soszy{\'n}ski}, I., {Udalski}, A., {Szyma{\'n}ski}, M.~K., {et~al.} 2009,
  \actaa, 59, 239

\bibitem[{{Soszy{\'n}ski} {et~al.}(2011){Soszy{\'n}ski}, {Udalski},
  {Szyma{\'n}ski}, {Kubiak}, {Pietrzy{\'n}ski}, {Wyrzykowski}, {Ulaczyk},
  {Poleski}, {Koz{\l}owski}, \& {Pietrukowicz}}]{soszynski_etal_2011_ogle3smc}
{Soszy{\'n}ski}, I., {Udalski}, A., {Szyma{\'n}ski}, M.~K., {et~al.} 2011,
  \actaa, 61, 217

\bibitem[{{Soszy{\'n}ski} {et~al.}(2013){Soszy{\'n}ski}, {Udalski},
  {Szyma{\'n}ski}, {Kubiak}, {Pietrzy{\'n}ski}, {Wyrzykowski}, {Ulaczyk},
  {Poleski}, {Koz{\l}owski}, {Pietrukowicz}, \&
  {Skowron}}]{soszynski_etal_2013_ogle3bulge}
{Soszy{\'n}ski}, I., {Udalski}, A., {Szyma{\'n}ski}, M.~K., {et~al.} 2013,
  \actaa, 63, 21

\bibitem[{{Speck} {et~al.}(2000){Speck}, {Barlow}, {Sylvester}, \&
  {Hofmeister}}]{Speck_etal_2000}
{Speck}, A.~K., {Barlow}, M.~J., {Sylvester}, R.~J., \& {Hofmeister}, A.~M.
  2000, \aaps, 146, 437

\bibitem[{{Stetson} {et~al.}(2014){Stetson}, {Fiorentino}, {Bono}, {Bernard},
  {Monelli}, {Iannicola}, {Gallart}, \& {Ferraro}}]{stetson_etal_2014}
{Stetson}, P.~B., {Fiorentino}, G., {Bono}, G., {et~al.} 2014, \pasp, 126, 616

\bibitem[{{Trabucchi} {et~al.}(2021){Trabucchi}, {Wood}, {Mowlavi},
  {Pastorelli}, {Marigo}, {Girardi}, \& {Lebzelter}}]{Trabucchi2021}
{Trabucchi}, M., {Wood}, P.~R., {Mowlavi}, N., {et~al.} 2021, \mnras, 500, 1575

\bibitem[{{Uttenthaler} {et~al.}(2011){Uttenthaler}, {van Stiphout}, {Voet},
  {van Winckel}, {van Eck}, {Jorissen}, {Kerschbaum}, {Raskin}, {Prins},
  {Pessemier}, {Waelkens}, {Fr{\'e}mat}, {Hensberge}, {Dumortier}, \&
  {Lehmann}}]{Uttenthaler_etal_2011}
{Uttenthaler}, S., {van Stiphout}, K., {Voet}, K., {et~al.} 2011, \aap, 531,
  A88

\bibitem[{{Van Eck} {et~al.}(2017){Van Eck}, {Neyskens}, {Jorissen}, {Plez},
  {Edvardsson}, {Eriksson}, {Gustafsson}, {J{\o}rgensen}, \&
  {Nordlund}}]{2017A&A...601A..10V}
{Van Eck}, S., {Neyskens}, P., {Jorissen}, A., {et~al.} 2017, \aap, 601, A10

\bibitem[{{VanderPlas}(2018)}]{vanderplas_2018}
{VanderPlas}, J.~T. 2018, \apjs, 236, 16

\bibitem[{{Westerlund} {et~al.}(1981){Westerlund}, {Olander}, \&
  {Hedin}}]{1981A&AS...43..267W}
{Westerlund}, B.~E., {Olander}, N., \& {Hedin}, B. 1981, \aaps, 43, 267

\bibitem[{{Westerlund} {et~al.}(1978){Westerlund}, {Olander}, {Richer}, \&
  {Crabtree}}]{1978A&AS...31...61W}
{Westerlund}, B.~E., {Olander}, N., {Richer}, H.~B., \& {Crabtree}, D.~R. 1978,
  \aaps, 31, 61

\bibitem[{{Wood}(2015)}]{Wood2015}
{Wood}, P.~R. 2015, \mnras, 448, 3829

\bibitem[{{Wood} {et~al.}(1999){Wood}, {Alcock}, {Allsman}, {Alves}, {Axelrod},
  {Becker}, {Bennett}, {Cook}, {Drake}, {Freeman}, {Griest}, {King}, {Lehner},
  {Marshall}, {Minniti}, {Peterson}, {Pratt}, {Quinn}, {Stubbs}, {Sutherland},
  {Tomaney}, {Vandehei}, \& {Welch}}]{wood_etal_1999}
{Wood}, P.~R., {Alcock}, C., {Allsman}, R.~A., {et~al.} 1999, in IAU Symposium,
  Vol. 191, Asymptotic Giant Branch Stars, ed. T.~{Le Bertre}, A.~{Lebre}, \&
  C.~{Waelkens}, 151

\bibitem[{{Wray} {et~al.}(2004){Wray}, {Eyer}, \&
  {Paczy{\'n}ski}}]{Wray_etal_2004}
{Wray}, J.~J., {Eyer}, L., \& {Paczy{\'n}ski}, B. 2004, \mnras, 349, 1059

\bibitem[{{Yin} {et~al.}(2010){Yin}, {Magrini}, {Matteucci}, {Lanfranchi},
  {Gon{\c{c}}alves}, \& {Costa}}]{2010A&A...520A..55Y}
{Yin}, J., {Magrini}, L., {Matteucci}, F., {et~al.} 2010, \aap, 520, A55

\bibitem[{{Yung} {et~al.}(2014){Yung}, {Nakashima}, \&
  {Henkel}}]{yung_etal_2014}
{Yung}, B. H.~K., {Nakashima}, J.-i., \& {Henkel}, C. 2014, \apj, 794, 81

\bibitem[{{Zijlstra} \& {Bedding}(2002)}]{ZijlstraBedding_2002}
{Zijlstra}, A.~A. \& {Bedding}, T.~R. 2002, JAAVSO, 31, 2

\end{thebibliography}

\begin{appendix}
\section{Acknowledgements}
\label{app:acknowledgements}
This work presents results from the European Space Agency (ESA) space mission \Gaia. \Gaia data are being processed by the \Gaia Data Processing and Analysis Consortium (DPAC). Funding for the DPAC is provided by national institutions, in particular the institutions participating in the \Gaia MultiLateral Agreement (MLA). The \Gaia mission website is \url{https://www.cosmos.esa.int/gaia}. The \Gaia archive website is \url{https://archives.esac.esa.int/gaia}.

The \Gaia mission and data processing have financially been supported by, in alphabetical order by country:
\begin{itemize}
\item the Algerian Centre de Recherche en Astronomie, Astrophysique et G\'{e}ophysique of Bouzareah Observatory;
\item the Austrian Fonds zur F\"{o}rderung der wissenschaftlichen Forschung (FWF) Hertha Firnberg Programme through grants T359, P20046, and P23737;
\item the BELgian federal Science Policy Office (BELSPO) through various PROgramme de D\'{e}veloppement d'Exp\'{e}riences scientifiques (PRODEX) grants and the Polish Academy of Sciences - Fonds Wetenschappelijk Onderzoek through grant VS.091.16N, and the Fonds de la Recherche Scientifique (FNRS), and the Research Council of Katholieke Universiteit (KU) Leuven through grant C16/18/005 (Pushing AsteRoseismology to the next level with TESS, GaiA, and the Sloan DIgital Sky SurvEy -- PARADISE);  
\item the Brazil-France exchange programmes Funda\c{c}\~{a}o de Amparo \`{a} Pesquisa do Estado de S\~{a}o Paulo (FAPESP) and Coordena\c{c}\~{a}o de Aperfeicoamento de Pessoal de N\'{\i}vel Superior (CAPES) - Comit\'{e} Fran\c{c}ais d'Evaluation de la Coop\'{e}ration Universitaire et Scientifique avec le Br\'{e}sil (COFECUB);
\item the Chilean Agencia Nacional de Investigaci\'{o}n y Desarrollo (ANID) through Fondo Nacional de Desarrollo Cient\'{\i}fico y Tecnol\'{o}gico (FONDECYT) Regular Project 1210992 (L.~Chemin);
\item the National Natural Science Foundation of China (NSFC) through grants 11573054, 11703065, and 12173069, the China Scholarship Council through grant 201806040200, and the Natural Science Foundation of Shanghai through grant 21ZR1474100;  
\item the Tenure Track Pilot Programme of the Croatian Science Foundation and the \'{E}cole Polytechnique F\'{e}d\'{e}rale de Lausanne and the project TTP-2018-07-1171 `Mining the Variable Sky', with the funds of the Croatian-Swiss Research Programme;
\item the Czech-Republic Ministry of Education, Youth, and Sports through grant LG 15010 and INTER-EXCELLENCE grant LTAUSA18093, and the Czech Space Office through ESA PECS contract 98058;
\item the Danish Ministry of Science;
\item the Estonian Ministry of Education and Research through grant IUT40-1;
\item the European Commission’s Sixth Framework Programme through the European Leadership in Space Astrometry (\href{https://www.cosmos.esa.int/web/gaia/elsa-rtn-programme}{ELSA}) Marie Curie Research Training Network (MRTN-CT-2006-033481), through Marie Curie project PIOF-GA-2009-255267 (Space AsteroSeismology \& RR Lyrae stars, SAS-RRL), and through a Marie Curie Transfer-of-Knowledge (ToK) fellowship (MTKD-CT-2004-014188); the European Commission's Seventh Framework Programme through grant FP7-606740 (FP7-SPACE-2013-1) for the \Gaia European Network for Improved data User Services (\href{https://gaia.ub.edu/twiki/do/view/GENIUS/}{GENIUS}) and through grant 264895 for the \Gaia Research for European Astronomy Training (\href{https://www.cosmos.esa.int/web/gaia/great-programme}{GREAT-ITN}) network;
\item the European Cooperation in Science and Technology (COST) through COST Action CA18104 `Revealing the Milky Way with \Gaia (MW-Gaia)';
\item the European Research Council (ERC) through grants 320360, 647208, and 834148 and through the European Union’s Horizon 2020 research and innovation and excellent science programmes through Marie Sk{\l}odowska-Curie grant 745617 (Our Galaxy at full HD -- Gal-HD) and 895174 (The build-up and fate of self-gravitating systems in the Universe) as well as grants 687378 (Small Bodies: Near and Far), 682115 (Using the Magellanic Clouds to Understand the Interaction of Galaxies), 695099 (A sub-percent distance scale from binaries and Cepheids -- CepBin), 716155 (Structured ACCREtion Disks -- SACCRED), 951549 (Sub-percent calibration of the extragalactic distance scale in the era of big surveys -- UniverScale), and 101004214 (Innovative Scientific Data Exploration and Exploitation Applications for Space Sciences -- EXPLORE);
\item the European Science Foundation (ESF), in the framework of the \Gaia Research for European Astronomy Training Research Network Programme (\href{https://www.cosmos.esa.int/web/gaia/great-programme}{GREAT-ESF});
\item the European Space Agency (ESA) in the framework of the \Gaia project, through the Plan for European Cooperating States (PECS) programme through contracts C98090 and 4000106398/12/NL/KML for Hungary, through contract 4000115263/15/NL/IB for Germany, and through PROgramme de D\'{e}veloppement d'Exp\'{e}riences scientifiques (PRODEX) grant 4000127986 for Slovenia;  
\item the Academy of Finland through grants 299543, 307157, 325805, 328654, 336546, and 345115 and the Magnus Ehrnrooth Foundation;
\item the French Centre National d’\'{E}tudes Spatiales (CNES), the Agence Nationale de la Recherche (ANR) through grant ANR-10-IDEX-0001-02 for the `Investissements d'avenir' programme, through grant ANR-15-CE31-0007 for project `Modelling the Milky Way in the \Gaia era’ (MOD4Gaia), through grant ANR-14-CE33-0014-01 for project `The Milky Way disc formation in the \Gaia era’ (ARCHEOGAL), through grant ANR-15-CE31-0012-01 for project `Unlocking the potential of Cepheids as primary distance calibrators’ (UnlockCepheids), through grant ANR-19-CE31-0017 for project `Secular evolution of galxies' (SEGAL), and through grant ANR-18-CE31-0006 for project `Galactic Dark Matter' (GaDaMa), the Centre National de la Recherche Scientifique (CNRS) and its SNO \Gaia of the Institut des Sciences de l’Univers (INSU), its Programmes Nationaux: Cosmologie et Galaxies (PNCG), Gravitation R\'{e}f\'{e}rences Astronomie M\'{e}trologie (PNGRAM), Plan\'{e}tologie (PNP), Physique et Chimie du Milieu Interstellaire (PCMI), and Physique Stellaire (PNPS), the `Action F\'{e}d\'{e}ratrice \Gaia' of the Observatoire de Paris, the R\'{e}gion de Franche-Comt\'{e}, the Institut National Polytechnique (INP) and the Institut National de Physique nucl\'{e}aire et de Physique des Particules (IN2P3) co-funded by CNES;
\item the German Aerospace Agency (Deutsches Zentrum f\"{u}r Luft- und Raumfahrt e.V., DLR) through grants 50QG0501, 50QG0601, 50QG0602, 50QG0701, 50QG0901, 50QG1001, 50QG1101, 50\-QG1401, 50QG1402, 50QG1403, 50QG1404, 50QG1904, 50QG2101, 50QG2102, and 50QG2202, and the Centre for Information Services and High Performance Computing (ZIH) at the Technische Universit\"{a}t Dresden for generous allocations of computer time;
\item the Hungarian Academy of Sciences through the Lend\"{u}let Programme grants LP2014-17 and LP2018-7 and the Hungarian National Research, Development, and Innovation Office (NKFIH) through grant KKP-137523 (`SeismoLab');
\item the Science Foundation Ireland (SFI) through a Royal Society - SFI University Research Fellowship (M.~Fraser);
\item the Israel Ministry of Science and Technology through grant 3-18143 and the Tel Aviv University Center for Artificial Intelligence and Data Science (TAD) through a grant;
\item the Agenzia Spaziale Italiana (ASI) through contracts I/037/08/0, I/058/10/0, 2014-025-R.0, 2014-025-R.1.2015, and 2018-24-HH.0 to the Italian Istituto Nazionale di Astrofisica (INAF), contract 2014-049-R.0/1/2 to INAF for the Space Science Data Centre (SSDC, formerly known as the ASI Science Data Center, ASDC), contracts I/008/10/0, 2013/030/I.0, 2013-030-I.0.1-2015, and 2016-17-I.0 to the Aerospace Logistics Technology Engineering Company (ALTEC S.p.A.), INAF, and the Italian Ministry of Education, University, and Research (Ministero dell'Istruzione, dell'Universit\`{a} e della Ricerca) through the Premiale project `MIning The Cosmos Big Data and Innovative Italian Technology for Frontier Astrophysics and Cosmology' (MITiC);
\item the Netherlands Organisation for Scientific Research (NWO) through grant NWO-M-614.061.414, through a VICI grant (A.~Helmi), and through a Spinoza prize (A.~Helmi), and the Netherlands Research School for Astronomy (NOVA);
\item the Polish National Science Centre through HARMONIA grant 2018/30/M/ST9/00311 and DAINA grant 2017/27/L/ST9/03221 and the Ministry of Science and Higher Education (MNiSW) through grant DIR/WK/2018/12;
\item the Portuguese Funda\c{c}\~{a}o para a Ci\^{e}ncia e a Tecnologia (FCT) through national funds, grants SFRH/\-BD/128840/2017 and PTDC/FIS-AST/30389/2017, and work contract DL 57/2016/CP1364/CT0006, the Fundo Europeu de Desenvolvimento Regional (FEDER) through grant POCI-01-0145-FEDER-030389 and its Programa Operacional Competitividade e Internacionaliza\c{c}\~{a}o (COMPETE2020) through grants UIDB/04434/2020 and UIDP/04434/2020, and the Strategic Programme UIDB/\-00099/2020 for the Centro de Astrof\'{\i}sica e Gravita\c{c}\~{a}o (CENTRA);  
\item the Slovenian Research Agency through grant P1-0188;
\item the Spanish Ministry of Economy (MINECO/FEDER, UE), the Spanish Ministry of Science and Innovation (MICIN), the Spanish Ministry of Education, Culture, and Sports, and the Spanish Government through grants BES-2016-078499, BES-2017-083126, BES-C-2017-0085, ESP2016-80079-C2-1-R, ESP2016-80079-C2-2-R, FPU16/03827, PDC2021-121059-C22, RTI2018-095076-B-C22, and TIN2015-65316-P (`Computaci\'{o}n de Altas Prestaciones VII'), the Juan de la Cierva Incorporaci\'{o}n Programme (FJCI-2015-2671 and IJC2019-04862-I for F.~Anders), the Severo Ochoa Centre of Excellence Programme (SEV2015-0493), and MICIN/AEI/10.13039/501100011033 (and the European Union through European Regional Development Fund `A way of making Europe') through grant RTI2018-095076-B-C21, the Institute of Cosmos Sciences University of Barcelona (ICCUB, Unidad de Excelencia `Mar\'{\i}a de Maeztu’) through grant CEX2019-000918-M, the University of Barcelona's official doctoral programme for the development of an R+D+i project through an Ajuts de Personal Investigador en Formaci\'{o} (APIF) grant, the Spanish Virtual Observatory through project AyA2017-84089, the Galician Regional Government, Xunta de Galicia, through grants ED431B-2021/36, ED481A-2019/155, and ED481A-2021/296, the Centro de Investigaci\'{o}n en Tecnolog\'{\i}as de la Informaci\'{o}n y las Comunicaciones (CITIC), funded by the Xunta de Galicia and the European Union (European Regional Development Fund -- Galicia 2014-2020 Programme), through grant ED431G-2019/01, the Red Espa\~{n}ola de Supercomputaci\'{o}n (RES) computer resources at MareNostrum, the Barcelona Supercomputing Centre - Centro Nacional de Supercomputaci\'{o}n (BSC-CNS) through activities AECT-2017-2-0002, AECT-2017-3-0006, AECT-2018-1-0017, AECT-2018-2-0013, AECT-2018-3-0011, AECT-2019-1-0010, AECT-2019-2-0014, AECT-2019-3-0003, AECT-2020-1-0004, and DATA-2020-1-0010, the Departament d'Innovaci\'{o}, Universitats i Empresa de la Generalitat de Catalunya through grant 2014-SGR-1051 for project `Models de Programaci\'{o} i Entorns d'Execuci\'{o} Parallels' (MPEXPAR), and Ramon y Cajal Fellowship RYC2018-025968-I funded by MICIN/AEI/10.13039/501100011033 and the European Science Foundation (`Investing in your future');
\item the Swedish National Space Agency (SNSA/Rymdstyrelsen);
\item the Swiss State Secretariat for Education, Research, and Innovation through the Swiss Activit\'{e}s Nationales Compl\'{e}mentaires and the Swiss National Science Foundation through an Eccellenza Professorial Fellowship (award PCEFP2$\_$194638 for R.~Anderson);
\item the United Kingdom Particle Physics and Astronomy Research Council (PPARC), the United Kingdom Science and Technology Facilities Council (STFC), and the United Kingdom Space Agency (UKSA) through the following grants to the University of Bristol, the University of Cambridge, the University of Edinburgh, the University of Leicester, the Mullard Space Sciences Laboratory of University College London, and the United Kingdom Rutherford Appleton Laboratory (RAL): PP/D006511/1, PP/D006546/1, PP/D006570/1, ST/I000852/1, ST/J005045/1, ST/K00056X/1, ST/\-K000209/1, ST/K000756/1, ST/L006561/1, ST/N000595/1, ST/N000641/1, ST/N000978/1, ST/\-N001117/1, ST/S000089/1, ST/S000976/1, ST/S000984/1, ST/S001123/1, ST/S001948/1, ST/\-S001980/1, ST/S002103/1, ST/V000969/1, ST/W002469/1, ST/W002493/1, ST/W002671/1, ST/W002809/1, and EP/V520342/1.
\end{itemize}

The GBOT programme  uses observations collected at (i) the European Organisation for Astronomical Research in the Southern Hemisphere (ESO) with the VLT Survey Telescope (VST), under ESO programmes
092.B-0165,
093.B-0236,
094.B-0181,
095.B-0046,
096.B-0162,
097.B-0304,
098.B-0030,
099.B-0034,
0100.B-0131,
0101.B-0156,
0102.B-0174, and
0103.B-0165;
%
%
and (ii) the Liverpool Telescope, which is operated on the island of La Palma by Liverpool John Moores University in the Spanish Observatorio del Roque de los Muchachos of the Instituto de Astrof\'{\i}sica de Canarias with financial support from the United Kingdom Science and Technology Facilities Council, and (iii) telescopes of the Las Cumbres Observatory Global Telescope Network.

\section{Comparison of the SOS and classification tables}
\label{app:ComparisonClassificationSOS}

Like in DR2, long-period variable candidates are published in DR3 in two tables, namely vari$\_$classifier$\_$result and vari$\_$long$\_$period$\_$variable.
The first of these two tables is the result of the classification process, which attributes variability types to all variable star candidates within the DR3 dataset.
Long-period variables in that table are identified as "LPV".
This dataset, consisting of more than 2.3 million LPV candidates (see Table \ref{tab:catalog_numbers}), was the starting point for the star list entering the second$^{}$ Gaia LPV catalogue. 
In this appendix we briefly summarise the differences between the two datasets.
Tab.\,\ref{tab:catalog_numbers} shows that the classification table includes approximately 600\,000 (26\%) more objects than the second$^{}$ Gaia LPV catalogue. 
One reason for exclusion is a low S/N value for the G measurements. This primarily corresponds to a loss of stars at the low brightness end.

\begin{figure}[h]
\centering
\includegraphics[width=\hsize]{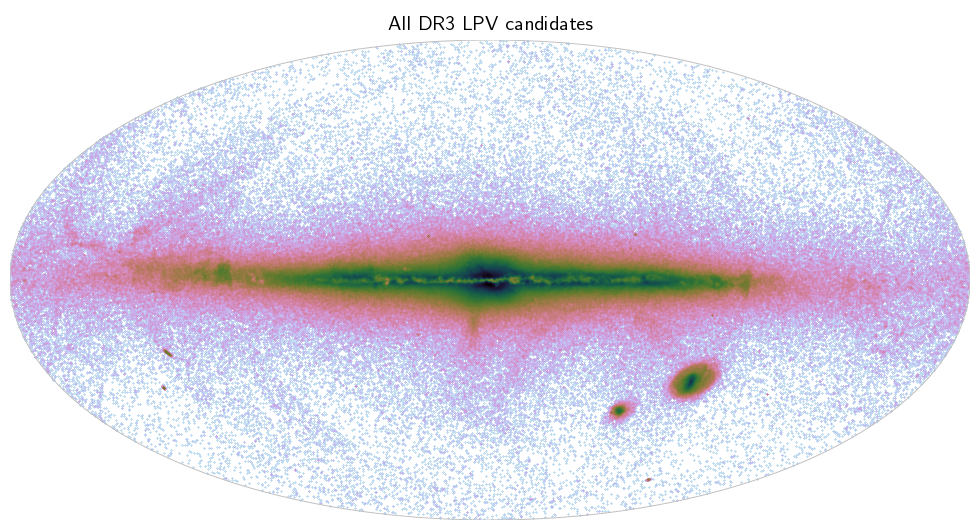}
\caption{Sky distribution of the entire set of \Gaia DR3 LPV candidates in the classification table.}
\label{Fig:sky_distribution_classif}
\end{figure}

Fig.\,\ref{Fig:sky_distribution_classif} shows the sky distribution of all LPV candidates in the classification table.
In comparison with the distribution of sources from the SOS table (Fig.\,\ref{Fig:sky_distribution}), the SOS selection filters caused the removal of a number of artefacts associated with the \Gaia scanning law, while preserving the galactic structures and external galaxies.

Fig.\,\ref{Fig:classif_SOS_qr5} compares the distribution of amplitudes for the samples in the classification table, in the SOS table, and the sample of stars with exported periods. 
The main difference between the classification table and the catalogue table is found at amplitudes around 1 mag. 
We suspect that the difference stems from red variables that look like LPVs due to their clear large-amplitude variability, but which do not comply with the stricter criteria applied for the catalogue table. 
Longer time series, expected for DR4, will likely allow classifying these stars unambiguously.
At larger amplitudes (QR$_{5}$(G)>2 mag), both tables include practically the same objects.
For the variables with the smallest amplitudes, we lack periods for a majority of the stars because they could not be identified properly.

\begin{figure}
\centering
\includegraphics[width=\hsize]{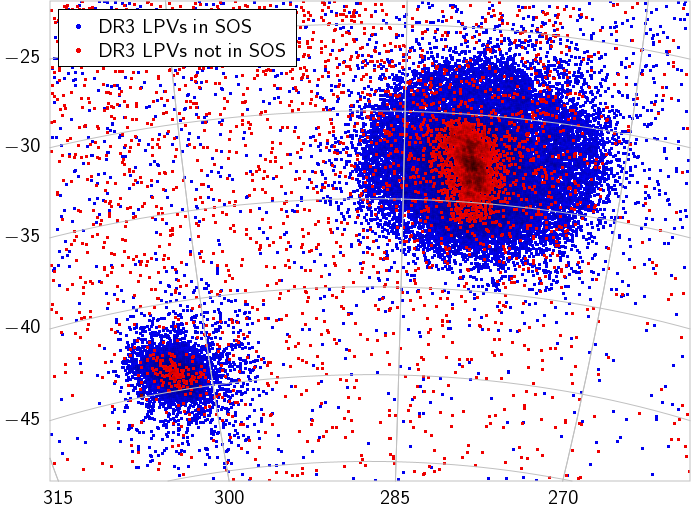}
\caption{Sky distribution (in Galactic coordinates) of \Gaia DR3 LPV candidates in the area of the Magellanic Clouds, showing sources that are part of the SOS table (blue) or only in the classification table (red). The latter are concentrated towards the central regions of the two galaxies, where the S/N is degraded by high levels of crowding.}
\label{Fig:nonSOS_sky}
\end{figure}

\begin{figure}
\centering
\includegraphics[width=\hsize]{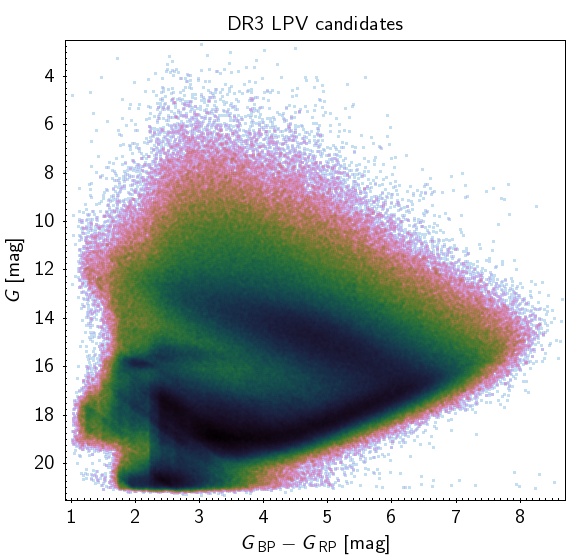}
\caption{Density-mapped colour-magnitude diagrams in the \Gaia passbands of the \Gaia DR3 LPV candidates in the classification table.}
\label{Fig:DR3_CM_classif}
\end{figure}

The average fraction of 26\% of objects missing from the SOS table compared to the classification table can show quite significant differences between various sky areas. 
In Fig.\,\ref{Fig:nonSOS_sky} we illustrate the situation for the sky area around the Magellanic Clouds.
Within this galaxy itself, the difference is only about 10\,\%, while in the field around the Galactic centre, this value can reach 74\,\%.
This supports the previous assumption that weak objects in crowded regions form the bulk of objects from the classification table that are absent from the catalogue table.
In a colour-magnitude diagram (Fig.\,\ref{Fig:DR3_CM_classif}) constructed from the classification table, these objects are located in the lower left corner.
They were cleaned from the SOS table (compare Fig.\,\ref{Fig:DR3_CM}).
No significant differences in the period distribution between the samples in the classification table and in the SOS table were found, except for a lack of very short-period objects in the SOS table.

\subsection{Completeness}
\label{app:ComparisonClassificationSOS:Completeness}

\begin{figure*}[h]
\centering
\includegraphics[width=.49\hsize]{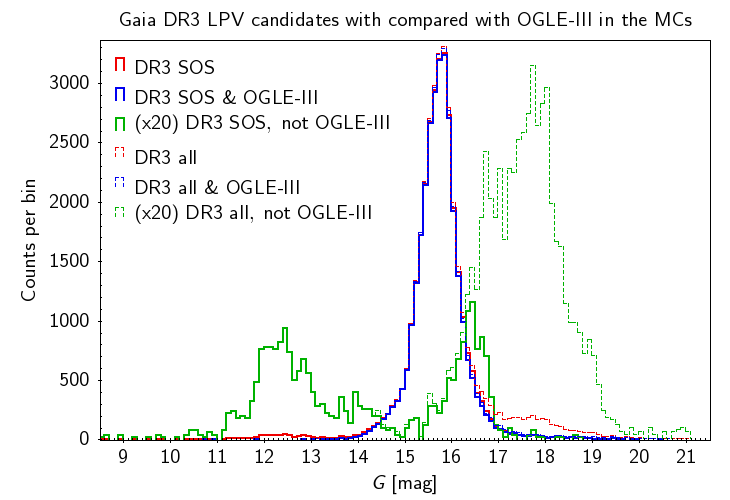}
\includegraphics[width=.49\hsize]{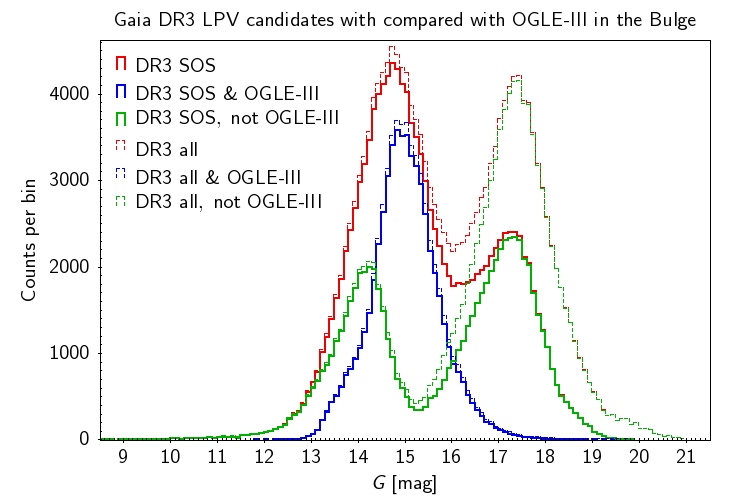}
\caption{Similar to Fig.~\ref{fig:hist_mediang_XMandNew_dr3sos_ogle3}, but showing the brightness distribution of the sources in the dataset obtained by combining the LPV candidates resulting from the \Gaia DR3 SOS and classification modules. The full set of sources in the direction of the MCs (left) or Galactic bulge (right) OGLE-III fields are shown with red lines. Blue lines correspond to sources matched with OGLE-III, while newly discovered candidates are represented by the green lines. Solid lines are limited to the sources resulting from the SOS module, and dashed lines include the output of the classification module as well. The green lines in the left panel show counts enhanced by a factor 20 for visual clarity.}
\label{fig:hist_mediang_XMandNew_dr3_ogle3}
\end{figure*}

In Sect.~\ref{sec:CatalogQuality:Completeness} we estimated the completeness of the second$^{}$ \Gaia catalogue of LPV candidates by quantifying its recovery rate with respect to OGLE and ASAS-SN. While in this case the \Gaia DR3 LPV candidates are limited to the table produced by the corresponding SOS module from the \Gaia DR3 pipeline, it interesting to extend this dataset by including the LPV candidates resulting from the classification module of the \Gaia DR3 pipeline. The corresponding recovery rates are given in Tables~\ref{tab:ogle3_recovery_rates_clas}, \ref{tab:ogle4_recovery_rates_clas}, and~\ref{tab:asassn_recovery_rates_clas}.

The inclusion of LPVs from the classification module typically raises the recovery rates by 1-2\% relative to OGLE (III or IV) and ASAS-SN. The only exception is represented by miras, whose recovery rate in creases by 5-8\%. In particular, the recovery rate of C-rich miras (in the Magellanic Clouds) relative to OGLE-III is increased by more than 10\%. In other words, many C-rich miras that did not pass the filtering criteria adopted for the construction of the second$^{}$ \Gaia catalogue of LPV candidates (mainly because they are relatively faint at optical wavelengths; see Sect.~\ref{sec:CatalogConstruction:FilteringCriteria}) can still be found in the \Gaia DR3 classification table.

\begin{table}
\caption{Similar to Table~\ref{tab:ogle3_recovery_rates}, but showing the recovery rates (relative to OGLE-III) of the dataset obtained by combining the LPV candidates resulting from the \Gaia DR3 SOS and classification modules.}
\label{tab:ogle3_recovery_rates_clas}
\centering
\begin{tabular}{c c c c}
\hline
Selection & OGLE-III & Matched $\leq2\arcsec$ & Recovery rate \\
\hline
\hline
All       &    84\,897 &    72\,655 &     85.6\% \\
Mira      &     5\,843 &     5\,155 &     88.2\% \\
O-Mira    &       494 &       477 &     96.6\% \\
C-Mira    &     1\,479 &     1\,182 &     79.9\% \\
SRV       &    32\,630 &    30\,854 &     94.6\% \\
O-SRV     &     6\,413 &     5\,982 &     93.3\% \\
C-SRV     &     6\,461 &     6\,208 &     96.1\% \\
OSARG     &    46\,424 &    36\,646 &     78.9\% \\
\hline
\end{tabular}
\end{table}

\begin{table}
\caption{Similar to Table~\ref{tab:ogle4_recovery_rates}, but showing the recovery rates (relative to OGLE-IV) of the dataset obtained by combining the LPV candidates resulting from the \Gaia DR3 SOS and classification modules.}
\label{tab:ogle4_recovery_rates_clas}
\centering
\begin{tabular}{c c c c}
\hline
Selection & OGLE-IV & Matched $\leq2\arcsec$ & Recovery rate \\
\hline
\hline
Mira     &    50\,311 &    46\,342 &     92.1\% \\
BLG-Mira &    33\,806 &    30\,683 &     90.8\% \\
GD-Mira  &    16\,505 &    15\,659 &     94.9\% \\
\hline
\end{tabular}
\end{table}

\begin{table}
\caption{Similar to Table~\ref{tab:asassn_recovery_rates}, but showing the recovery rates (relative to ASAS-SN) of the dataset obtained by combining the LPV candidates resulting from the \Gaia DR3 SOS and classification modules.}
\label{tab:asassn_recovery_rates_clas}
\centering
\begin{tabular}{c c c c}
\hline
Selection & ASAS-SN & Matched $\leq2\arcsec$ & Recovery rate \\
\hline
\hline
LPV        &   225\,726 &   180\,619 &     80.0\% \\
Mira       &    11\,249 &    10\,708 &     95.2\% \\
SR         &   139\,980 &   113\,081 &     80.8\% \\
L          &    74\,497 &    56\,830 &     76.3\% \\
\hline
\end{tabular}
\end{table}

\subsection{New LPV candidates}
\label{app:ComparisonClassificationSOS:NewLPVCandidates}

We extended the analysis performed in Sect.~\ref{sec:CatalogQuality:NewLPVCandidates} of the number of newly discovered LPV candidates to the dataset consisting of the output from both the SOS and the classification modules. This is summarised in Table~\ref{tab:newdiscoveries_clas}. As expected, the sources from the classification module lead to a further increase in the rate of new discoveries, in particular compared with the ASAS-SN catalogue (the new candidates are almost ten times as numerous as the known sources).

\begin{figure}[h]
\centering
\includegraphics[width=\hsize]{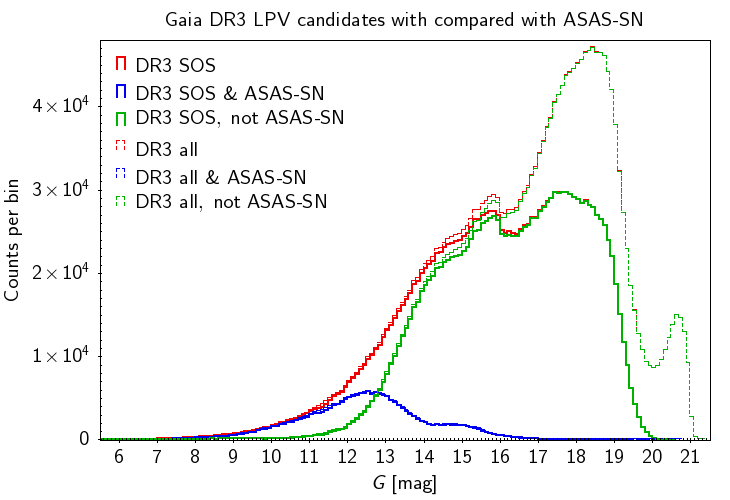}
\caption{Similar to Fig.~\ref{fig:hist_mediang_XMandNew_dr3_ogle3}, but concerning the cross match with the ASAS-SN catalogue of pulsating stars.}
\label{fig:hist_mediang_XMandNew_dr3_asassn}
\end{figure}

Compared with OGLE-III, we find a number of new candidates in the direction of the bulge that is twice the number of known LPVs, while the increase is still relatively small towards the Magellanic Clouds. Figures~\ref{fig:hist_mediang_XMandNew_dr3_ogle3} and~\ref{fig:hist_mediang_XMandNew_dr3_asassn} illustrate the brightness range of the new LPV candidates in \Gaia DR3 compared to the known candidates.

\begin{table}
\caption{Similar to Table~\ref{tab:newdiscoveries}, but showing the new discovery rates of the dataset obtained by combining the LPV candidates resulting from the \Gaia DR3 SOS and classification modules.}
\label{tab:newdiscoveries_clas}
\centering
\begin{tabular}{lrrrrr}
\hline\hline
Catalogue             &      Known &         New & Discovery rate \\
\hline
OGLE-III            &    84\,897 &    116\,664 &        137.4\% \\
\hspace{6mm}LMC     &    25\,015 &      3\,386 &         13.5\% \\
\hspace{6mm}SMC     &     4\,238 &         262 &          6.2\% \\
\hspace{6mm}BLG     &    55\,644 &    113\,016 &        203.1\% \\
ASAS-SN &   & &        \\
(LPVs only) & 225\,726 & 2\,122\,974 & 940.5\% \\
\hline
\end{tabular}
\end{table}

\section{Additional definitions}
\label{app:AdditionalDefinitions}

\subsection{Sky area common to \Gaia DR3 and OGLE-III}
\label{app:AdditionalDefinitions:CommonSkyOGLE3}
\begin{figure}
\centering
\includegraphics[width=\hsize]{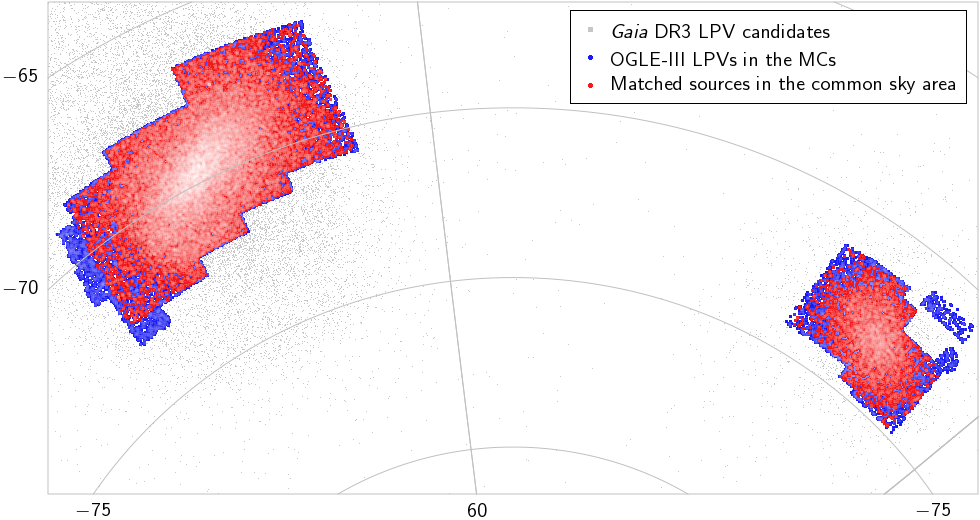}
\includegraphics[width=\hsize]{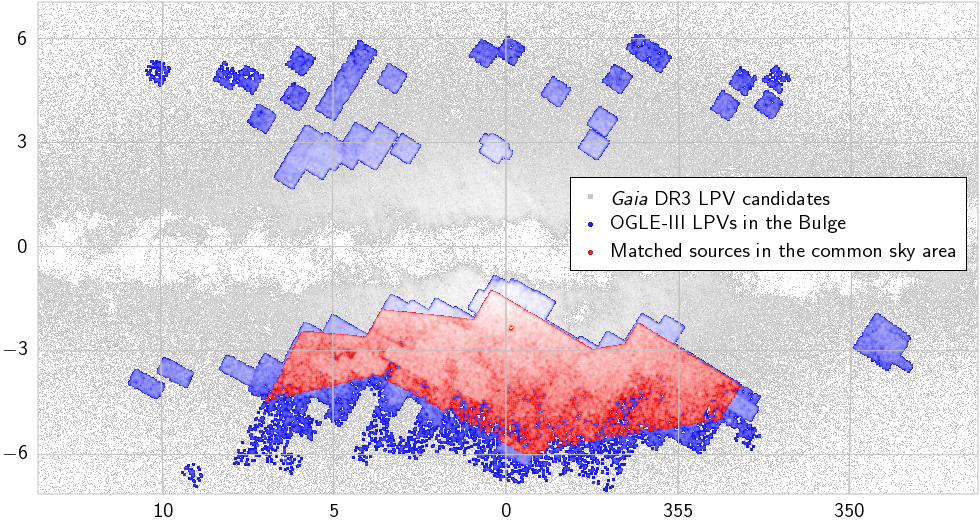}
\caption{Sky coverage of the second$^{}$ \Gaia catalogue of LPV candidates and the OGLE-III LPV catalogue in the region of the Magellanic Clouds (top, in equatorial coordinates) and towards the Galactic bulge (bottom, in galactic coordinates). Grey symbols show \Gaia DR3 LPV sources, blue symbols show OGLE-III sources, and red symbols show sources that matched within $2\arcsec$ in the two catalogues that lie within the region selected as the common sky area of the two surveys for comparison.}
\label{Fig:OGLE3GaiaCommonSkyArea}
\end{figure}

The definition of a common sky area is necessary for a comparison of \Gaia DR3 and OGLE-III. Because \Gaia is an all-sky survey, it would be enough to restrict it to the OGLE-III sky coverage. However, the latter exhibits a complex pattern of fields of view, especially towards the Galactic bulge, the exact replication of which is cumbersome and unnecessary. Instead, we selected a sub-region of the OGLE sky area with a simpler shape, as displayed in Fig.~\ref{Fig:OGLE3GaiaCommonSkyArea}.

\begin{figure*}
\centering
\includegraphics[width=\hsize]{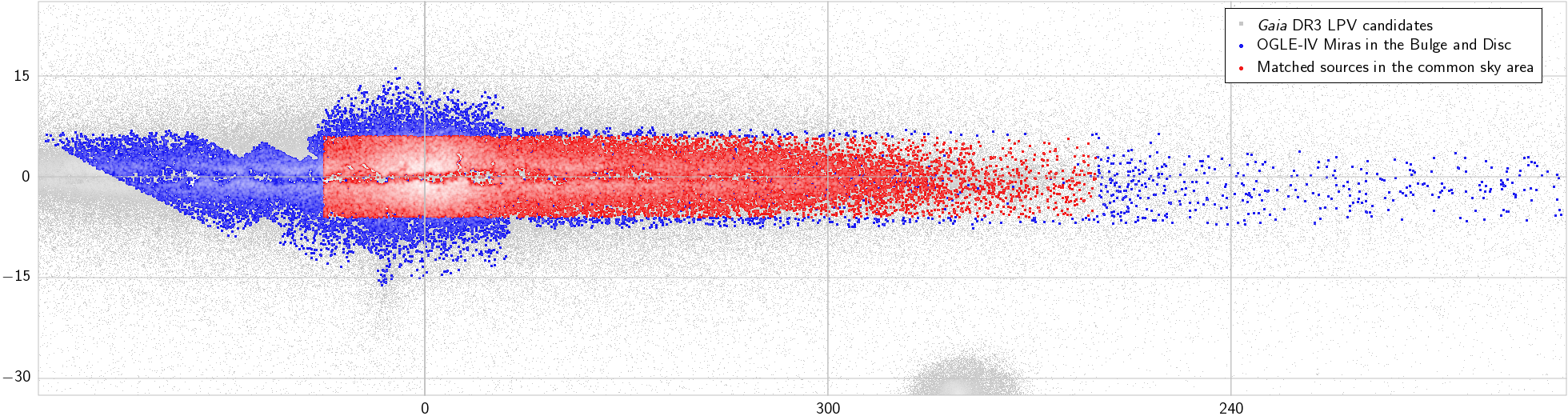}
\caption{Similar to Fig.~\ref{Fig:OGLE3GaiaCommonSkyArea}, but showing the common sky area (in Galactic coordinates) selected for comparing the second$^{}$ \Gaia catalogue of LPV candidates with the OGLE-IV catalogue of miras in the Galactic bulge and disc.}
\label{Fig:OGLE4GaiaCommonSkyArea}
\end{figure*}

This simplification allows for an easier comparison and reproducibility of our results. The conditions on equatorial sky coordinates ($\alpha$, $\delta$) for selecting sources within the common sky area are the following.

The LMC common sky area consists of the union of four regions that are defined by the conditions
\begin{equation}\label{eq:lmc1}
    \begin{array}{c}
    (\delta \leq -2.0\,\alpha + 114.5\degr) \mbox{ and } (\delta \geq 0.005\,\alpha - 72.9\degr) \\
     \mbox{ and } (\delta \leq 0.005\,\alpha - 68.85\degr) \mbox{ and } (\alpha \geq 84.3\degr) \,,
    \end{array}
\end{equation}
\begin{equation}\label{eq:lmc2}
    \begin{array}{c}
    (\alpha \leq 86.6\degr)  \mbox{ and } (\delta \geq -71.9\degr) \mbox{ and } \\
    (\delta \leq -67.8\degr) \mbox{ and } (\alpha \geq 78.95\degr) \,,
    \end{array}
\end{equation}
\begin{equation}\label{eq:lmc3}
    \begin{array}{c}
    (\alpha \leq 78.95\degr) \mbox{ and } (\delta \geq -71.3\degr) \mbox{ and } \\
    (\delta \leq -66.6\degr) \mbox{ and } (\alpha \geq 73.65\degr) \,,
    \end{array}
\end{equation}
\begin{equation}\label{eq:lmc4}
    \begin{array}{c}
    (\alpha \leq 73.65\degr) \mbox{ and } (\delta \geq -70.7\degr) \mbox{ and } \\
    (\delta \leq -66.6\degr) \mbox{ and } (\delta \leq 2.0\,\alpha - 204.5\degr) \,,
    \end{array}
\end{equation}
which are combined as (\ref{eq:lmc1}) or (\ref{eq:lmc2}) or (\ref{eq:lmc3}) or (\ref{eq:lmc4}).

Similarly, for the SMC common sky area,
\begin{equation}\label{eq:smc1}
    \begin{array}{c}
    (\delta \leq -2.0\,\alpha - 30.5\degr) \mbox{ and } (\delta \geq 0.005\,\alpha - 74.25\degr) \\
    \mbox{ and } (\delta \leq 0.005\,\alpha - 71.4\degr) \mbox{ and } (\alpha \geq 13.5\degr) \,,
    \end{array}
\end{equation}
\begin{equation}\label{eq:smc2}
    \begin{array}{c}
    (\alpha \leq 13.5\degr) \mbox{ and } (\delta \geq 0.005\,\alpha - 74.65\degr) \mbox{ and } \\
    (\delta \leq 0.005\,\alpha - 71.75\degr) \mbox{ and } (\alpha \geq 11.7\degr) \,,
    \end{array}
\end{equation}
\begin{equation}\label{eq:smc3}
    \begin{array}{c}
    (\alpha \leq 11.7\degr) \mbox{ and } (\delta \geq 0.005\,\alpha - 74.8\degr) \mbox{ and } \\
    (\delta \leq 0.005\,\alpha - 72.5\degr) \mbox{ and }  (\delta \leq 3.0 \alpha - 90.0) \,,
    \end{array}
\end{equation}
which are combined as (\ref{eq:smc1}) or (\ref{eq:smc2}) or (\ref{eq:smc3}).

The selected area towards the Galactic bulge is limited to sources below the Galactic plane (see Fig.~\ref{Fig:OGLE3GaiaCommonSkyArea}), and is defined by the conditions
\begin{equation}\label{eq:gb1}
    \begin{array}{c}
    (\delta \geq \alpha - 299.7\degr) \mbox{ or } ((\alpha \leq 272.1\degr) \mbox{ and } \\
    (\delta \geq \alpha - 304.5\degr) \mbox{ and } (\delta \geq -36.8\degr)) \mbox{ and } (\alpha \geq 266.3\degr) \,,
    \end{array}
\end{equation}
\begin{equation}\label{eq:gb2}
    \begin{array}{c}
    (\alpha \leq 267.9\degr) \mbox{ and } ((\delta \leq -33.35\degr) \\
    \mbox{ or } (\delta \leq \alpha - 300.5\degr)) \,,
    \end{array}
\end{equation}
\begin{equation}\label{eq:gb3}
    \begin{array}{c}
    (\alpha \geq 267.9\degr) \mbox{ and } ((\delta \leq -29.25\degr) \mbox{ or } \\
    ((\delta \leq 2.0\,\alpha - 567.3\degr) \mbox{ and } (\delta \leq -26.8\degr)) \mbox{ or } \\
    ((\delta \leq 2.0\,\alpha - 569.3\degr) \mbox{ and } (\delta \leq -25.1\degr))) \,,
    \end{array}
\end{equation}
which are combined as (\ref{eq:gb1}) or (\ref{eq:gb2}) or (\ref{eq:gb3}).

\subsection{Sky area common to \Gaia DR3 and OGLE-IV}
\label{app:AdditionalDefinitions:CommonSkyOGLE4}

The footprint of OGLE-IV observations of miras in the Galactic bulge and disc is substantially less complex than the case of OGLE-III LPVs and allows for a relatively simple definition of the common sky area (displayed in Fig.~\ref{Fig:OGLE4GaiaCommonSkyArea}):
\begin{equation}\label{eq:ogle3blggd}
    ((l < 15.0\degr) \mbox{ or } (l > 260.0\degr)) \mbox{ and } (|b|<6.0\degr) \,,
\end{equation}
where $l$ and $b$ are the Galactic coordinate longitude and latitude, respectively.

\subsection{Regions in the $\mediandeltawl$ vs. $\gbp-\grp$ plane}
\label{app:AdditionalDefinitions:RegionsMediandeltawlColor}

\begin{figure}
\centering
\includegraphics[width=\hsize]{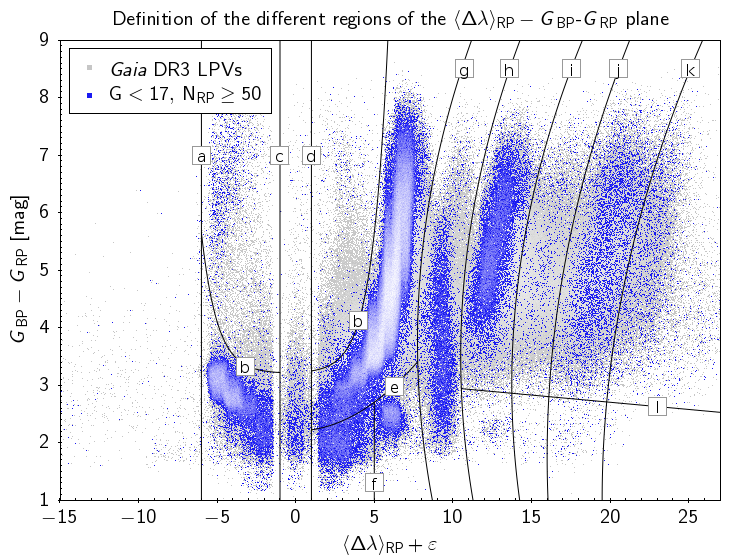}
\caption{Boundary lines in the $\mediandeltawl$ vs. $\gbp-\grp$ plane and corresponding labels (see Table~\ref{tab:mediandeltawl_color_plane_lines}).}
\label{Fig:lines_in_plane_mediandeltawlrp_vs_bprp}
\end{figure}

In order to characterise the C-star classification of the \Gaia DR3 LPV catalogue, we considered several regions in the diagram showing $\mediandeltawl$ as a function of the \Gaia colour $\gbp-\grp$. The regions are defined by 12 boundary lines (Fig.~\ref{Fig:lines_in_plane_mediandeltawlrp_vs_bprp}), labelled with lower-case letters $a$ through $l$, whose equations are given in Table~\ref{tab:mediandeltawl_color_plane_lines}.

\begin{table}
\caption{Equations of the boundary lines defining the regions in the $\mediandeltawl$ vs. $\gbp-\grp$ plane.}
\label{tab:mediandeltawl_color_plane_lines}
\centering
\begin{tabular}{c c}
Label & Equation \\
\hline
$a$ & $\mediandeltawl = -6.0$ \\
$b$ & $\gbp-\grp = 3.2 + 0.01\,e^{|\mediandeltawl+0.5|}$ \\
$c$ & $\mediandeltawl = -1.0$ \\
$d$ & $\mediandeltawl = 1.0$ \\
$e$ & $\gbp-\grp = 2.2 + 0.02\,\mediandeltawl^2$ \\
$f$ & $\mediandeltawl = 5.0$ \\
$g$ & $\mediandeltawl =  7.75 + 0.125\,(\gbp-\grp-3.75)^2$ \\
$h$ & $\mediandeltawl = 10.50 + 0.125\,(\gbp-\grp-3.50)^2$ \\
$i$ & $\mediandeltawl = 13.75 + 0.125\,(\gbp-\grp-3.00)^2$ \\
$j$ & $\mediandeltawl = 16.00 + 0.100\,(\gbp-\grp-1.75)^2$ \\
$k$ & $\mediandeltawl = 19.50 + 0.100\,(\gbp-\grp-1.00)^2$ \\
$l$ & $\gbp-\grp = 3.2 + 0.025\,\mediandeltawl$ \\
\end{tabular}
\end{table}

\subsection{Selection of sources in Local Group galaxies}
\label{app:AdditionalDefinitions:LocalGroupGalaxies}

\begin{table*}
\caption{Selection of \Gaia DR3 LPV sources in Local Group galaxies.}
\label{tab:LocalGroupGalaxiesSelection}
\centering
\begin{tabular}{c c}
Galaxy      & Selection \\
\hline
\multirow{2}{*}{Sgr dSph}    & $(\delta>-0.25\,\alpha+36.0)\mbox{ and }(\delta<-0.25\,\alpha+46.0)\mbox{ and }(\delta>2.5\,\alpha-780.0)\mbox{ and }(\delta<2.5\alpha-715.0)$ \\
                             & $\mbox{ \& }(\varpi<0.1)\mbox{ and }(-3.05<\mu_{\alpha^{\ast}}<-2.25)\mbox{ and }(-1.78<\mu_{\delta}<-0.98)$ \\
\multirow{2}{*}{M31$^{(a)}$} & $(\delta>\alpha+29.75)\mbox{ and }(\delta<\alpha+31.50)\mbox{ and }(\delta>-0.75\,\alpha+46.5)\mbox{ and }(\delta<-0.75\alpha+52.0)$ \\
                             & $\mbox{ and }(\varpi<0.5)\mbox{ and }(|\mu_{\alpha^{\ast}}|<1.0)\mbox{ and }(|\mu_{\delta}|<1.0)$ \\
M33$^{(a)}$                  & $(23.0<\alpha<23.9)\mbox{ and }(30.1<\delta<31.2)\mbox{ and }(\varpi<0.5)\mbox{ and }(|\mu_{\alpha^{\ast}}|<1.0)\mbox{ and }(|\mu_{\delta}|<1.0)$ \\
Fornax dSph$^{(a)}$          & $(39.3<\alpha<40.7)\mbox{ and }(-35.3<\delta<-33.7)\mbox{ and }(\varpi<0.5)\mbox{ and }(|\mu_{\alpha^{\ast}}|<1.0)\mbox{ and }(|\mu_{\delta}|<1.0)$ \\
NGC 6822$^{(a)}$             & $(296.05<\alpha<296.40)\mbox{ and }(-15.0<\delta<-14.6)\mbox{ and }(\varpi<0.5)\mbox{ and }(|\mu_{\alpha^{\ast}}|<1.0)\mbox{ and }(|\mu_{\delta}|<1.0)$ \\
IC 10$^{(a)}$                & $(4.9<\alpha<5.2)\mbox{ and }(59.2<\delta<59.4)\mbox{ and }(\varpi<0.5)\mbox{ and }(|\mu_{\alpha^{\ast}}|<1.0)\mbox{ and }(|\mu_{\delta}|<1.0)$ \\
Leo I dSph$^{(a)}$           & $(152.0<\alpha<152.2)\mbox{ and }(12.2<\delta<12.4)\mbox{ and }(\varpi<0.5)\mbox{ and }(|\mu_{\alpha^{\ast}}|<1.0)\mbox{ and }(|\mu_{\delta}|<1.0)$ \\
\end{tabular}
\tablefoot{$^{(a)}$: sources without a published value of the parallax or of the proper motions are assumed to be non-Galactic and thus are retained.}
\end{table*}

\begin{figure*}
\centering
\includegraphics[width=.33\hsize]{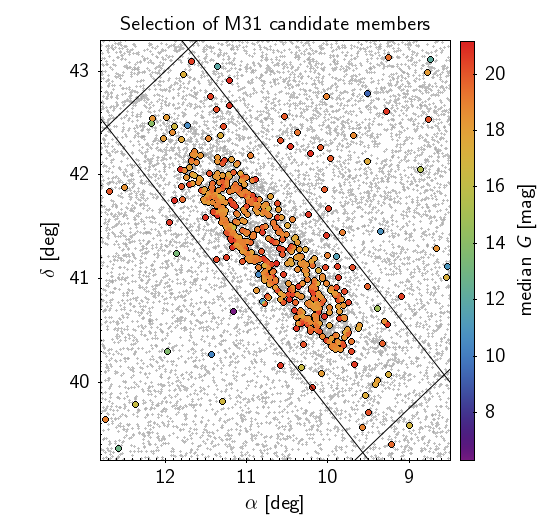}
\includegraphics[width=.33\hsize]{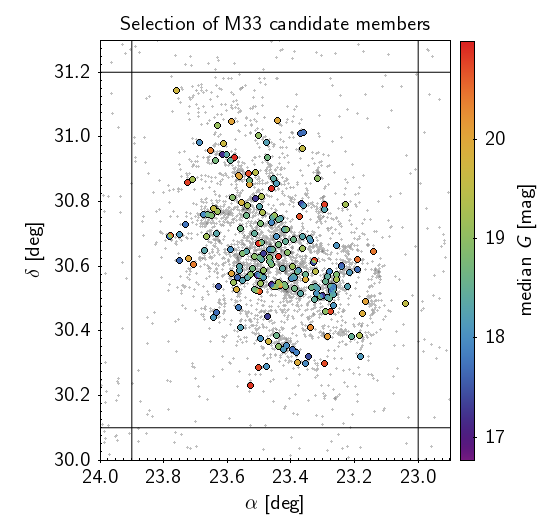}
\includegraphics[width=.33\hsize]{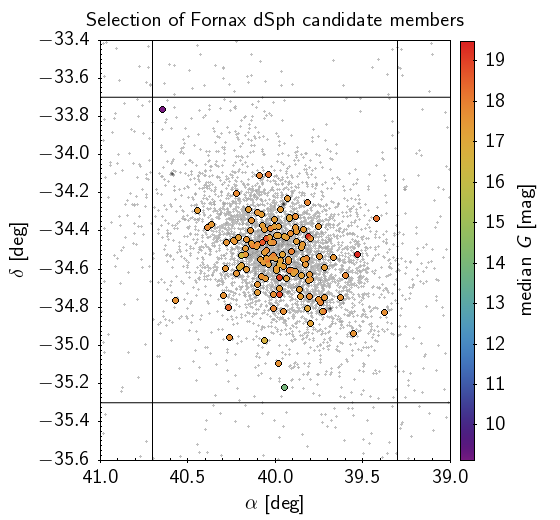}
\caption{Sky maps towards M31 (left panel), M33 (central panel), and Fornax dSph (right panel). \Gaia DR3 LPV candidates are colour-coded according to their median $G$-band magnitude, and a sample of \Gaia EDR3 sources is displayed as grey symbols in the background for reference. Solid lines represent the cuts in equatorial coordinates that we applied to select sources in each galaxy (see Table~\ref{tab:LocalGroupGalaxiesSelection}).}
\label{Fig:sky_M31_M33_Fornax}
\end{figure*}

\begin{figure*}
\centering
\includegraphics[width=.33\hsize]{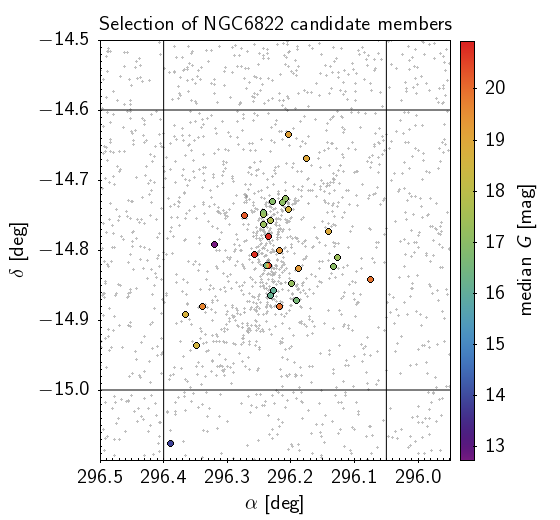}
\includegraphics[width=.33\hsize]{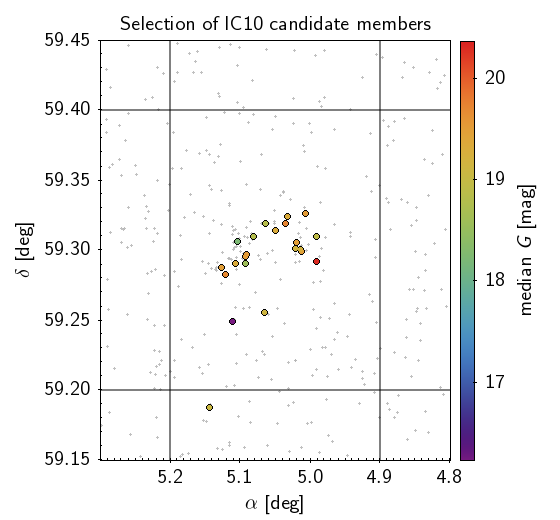}
\includegraphics[width=.33\hsize]{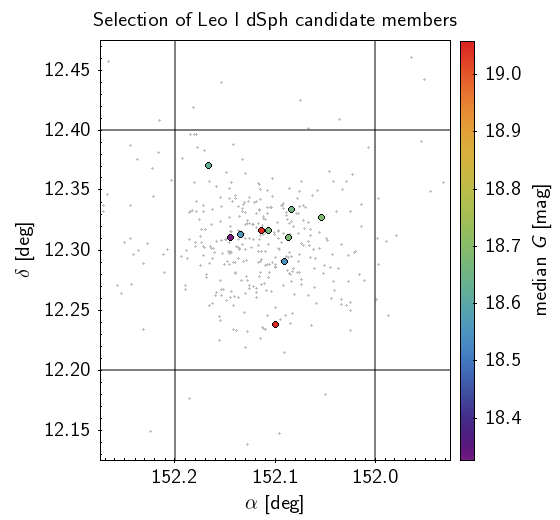}
\caption{Similar to Fig.~\ref{Fig:sky_M31_M33_Fornax}, but showing (from left to right) Fornax dSph, NGC 6822, IC 10, and Leo I dSph.}
\label{Fig:sky_Fornax_NGC6822_IC10_LeoI}
\end{figure*}

The selection of candidate members of Local Group galaxies in the \Gaia DR3 LPV catalogue is based on the equatorial coordinates ($\alpha$ and $\delta$, in degrees), the corresponding proper motions ($\mu_{\alpha^{\ast}}$ and $\mu_{\delta}$, in mas $yr^{-1}$), and the parallax ($\varpi$, in mas). The expressions for each galaxy are listed in Table~\ref{tab:LocalGroupGalaxiesSelection}, with the exception of the LMC and the SMC, for which the expression given by \citet{mowlavitrabucchilebzelter_2019} were adopted.

\section{Comparison sample of well-known LPVs}

Table \ref{tab:wellknownLPVs} lists the basic parameters of the 98 stars used in the comparison in Sect.\,\ref{sec:CatalogQuality:PeriodRecovery:WellObservedLPVs}.

\longtab[1]{
\begin{longtable}{lrccccccl}
\caption{Comparison sample of bright LPVs monitored from ground. Column 8 gives the {\scshape \large median\_delta\_wl\_rp} value for stars in the SOS table.}
\label{tab:wellknownLPVs}
\\
\hline
\hline
Name & \Gaia EDR3 ID & P$_{Lit.}^1$ & in SOS & P$_{SOS}$  & Spectrum$^1$ & isCstar & $\mediandeltawl$ & remark \\
     &               &     [days] &        &     [days] &          &         &                  &        \\
\hline
\endfirsthead
\caption{continued.}
\\
\hline
\hline
Name & \Gaia EDR3 ID & P$_{Lit.}$ & in SOS & P$_{SOS}$ & Spectrum & isCstar & $\mediandeltawl$ & remark \\
     &               &     [days] &        &    [days] &          &         &                  &        \\
\hline
\endhead
\hline
\endfoot
R And    &  379224622734044800 &  409 & y &  406.5    & S &  n & -4.922 &     \\
VX And   &  385742630743259008 &  375 & y &  371.01   & C &  y &  9.122 &     \\
EU And   & 1941516640193852800 &   -- & y &   55.17   & C &  y &  9.998 &     \\
V Aql    & 4206206750288121728 &  353 & n &      --   & C & -- &     -- & [1] \\
R Aqr    & 2419576358847950592 &  390 & y &  384.14   & M &  n &  6.891 &     \\ 
T Aqr    & 6913517223245165696 &  201 & y &  200.3    & M &  n &  5.152 &     \\
T Ari    &   83396552513705344 &  323 & y &  310.44* & M &  n &  6.146 &     \\
R Aur    &  266367210206027776 &  458 & y &  478.6    & M &  n &  6.632 &     \\
S Aur    &  182625449700126336 &  596 & y &  527.24   & C &  y &  9.472 &     \\
TV Aur   &  256155014926211968 &  183 & y &   12.74* & S &  n &  2.444 &     \\
UU Aur   &  944939847899350784 &  441 & y &  227.74* & C &  y &  9.665 &     \\
UV Aur   &  180919213811383680 &  393 & y &  389.39   & C &  y &  9.474 &     \\
T Cam    &  483143830359915648 &  374 & y &  376.71   & S &  n & -5.070 &     \\
U Cam    &  486859664266293376 & 2800 & y &  400      & C &  y &  9.474 &     \\
ST Cam   &  483958671558728576 &  372 & y &  195.62* & C &  y &  9.842 &     \\
TX Cam   &  279632268237135488 &  557 & n &      --   & M & -- &     -- & [2] \\
BD Cam   &  487305619316615680 &   -- & n &      --   & S & -- &     -- & [2] \\
RT Cap   & 6853966780133956096 &  393 & y &   11.52* & C &  y &  9.122 &     \\
R Car    & 5250535985379773184 &  309 & y &  307.52   & M &  n &  6.766 &     \\
S Car    & 5253294179046079872 &  149 & y &  148.66   & M &  n &  5.378 &     \\
R Cas    & 1944073004732961152 &  434 & y &  424.57   & M &  n &  6.632 &     \\
T Cas    &  421412815180931456 &  445 & y &  426.13   & M &  n &  6.890 &     \\
W Cas    &  425942081532426752 &  408 & y &  402.44   & C &  n &  9.472 &     \\
SV Cas   & 1993001306528852608 &  276 & y &  211.1    & M &  n &  6.086 &     \\
WZ Cas   &  429338816550168960 &  186 & y &  388.8    & C &  y &  9.840 &     \\
R Cen    & 5866952440466180736 &  546 & y &  268.51   & M &  n &  6.154 &     \\
T Cen    & 6165341341986059520 &   90 & n &      --   & M & -- &     -- & [1] \\
WW Cen   & 6055860804562317568 &  304 & y &  302.59   & M &  n &  6.000 &     \\
S Cep    & 2284711568256711040 &  487 & y &  498.36   & C &  y &  9.175 &     \\
T Cep    & 2270451142963759744 &  388 & y &  375.05   & M &  n &  6.181 &     \\
U Cet    & 5170512944979310208 &  235 & y &  242.98   & M &  n &  6.152 &     \\
W CMa    & 3045373881522901632 &   -- & n &      --   & C & -- &     -- & [1] \\
S CMi    & 3143124657116728448 &  332 & y &  339.02   & M &  n &  6.632 &     \\
R Cnc    &  649386277628656128 &  361 & y &  358.73   & M &  n &  6.766 &     \\
X Cnc    &  611797170530193408 &  193 & y &  191.4*  & C &  y &  9.84  &     \\
S CrB    & 1277100833181749760 &  360 & y &  359.97   & M &  n &  6.766 &     \\
V CrB    & 1376933877642712320 &  357 & y &  379.66   & C &  y &  9.473 &     \\
Y CVn    & 1542553623374596352 &  157 & y &  152.88* & C &  y & 10.000 &     \\
chi Cyg  & 2034702312280673792 &  407 & y &  422.93   & S &  n & -4.832 &     \\
R Cyg    & 2135109439204739200 &  426 & y &  448.75   & S &  n & -5.000 &     \\
RU Cyg   & 2174336062728231040 &  234 & y &  398.71   & M &  n &  6.210 &     \\
RV Cyg   & 1952830855365548416 &  263 & y &   20.64* & C &  y &  9.840 &     \\
RZ Cyg   & 2166885393983390208 &  276 & y &  608.01   & M &  n &  6.634 &     \\
U Cyg    & 2084221361016638208 &  465 & y &  501.56   & C &  y &  9.311 &     \\
V Cyg    & 2167591280437990656 &  421 & y &  394.18   & C &  y &  8.856 &     \\
CY Cyg   & 2166355704250883968 &   -- & y &  156.38   & S &  y &  8.527 &     \\
V460 Cyg & 1950794078794961536 &  180 & y &   83.28* & C &  y &  9.841 &     \\
R Dor    & 4677205714465503104 &  338 & n &      --   & M & -- &     -- & [2] \\
RY Dra   & 1678844308746410880 &  200 & y &  643.39   & C &  y &  9.314 &     \\
UX Dra   & 2289510386756671872 &  168 & y &   38.43* & C &  y &  9.664 &     \\
RT Eri   & 5111694364293223040 &  371 & y &  365.77   & M &  n &  6.487 &     \\
R For    & 5118511817421484544 &  389 & y &  393.41   & C &  y &  8.698 &     \\
TU Gem   &  675394812867813376 &  230 & y &  296.67   & C &  n &  4.530 &     \\
DY Gem   & 3356050581395042048 & 1145 & y &  144.24   & S &  n &  6.000 &     \\
g Her    & 1381119031215320576 &   89 & y &  834.4*  & M &  n &  5.906 &     \\
U Her    & 1200834239913483392 &  405 & y &  408.54   & M &  n &  6.211 &     \\
OP Her   & 1349637195812633984 &  121 & y &   20.37* & S &  n & -5.000 &     \\
RU Her   & 1303081537030212480 &  485 & n &      --   & M & -- &     -- & [2] \\
R Hor    & 4748477123328077696 &  408 & n &      --   & M & -- &     -- & [2] \\
R Hya    & 6195030801635544704 &  388 & y &  349      & M &  n &  6.042 &     \\
T Hya    & 5749870429386271488 &  291 & y &  287.2    & M &  n &  5.953 &     \\
U Hya    & 3751290548759011712 &  450 & n &      --   & C & -- &     -- & [1] \\ 
W Hya    & 6177092406867764352 &  361 & n &      --   & M & -- &     -- & [1] \\
R Leo    &  612958873284344448 &  313 & n &      --   & M & -- &     -- & [1] \\
R Lep    & 2987082722815713792 &  427 & y &  427.7    & C &  y &  9.310 &     \\
R LMi    &  794754943320201600 &  372 & y &  369.3    & M &  n &  6.632 &     \\
Y Lyn    &  973871911539040000 &  110 & y &  852.7    & S &  n &  6.000 &     \\
T Lyr    & 2096185937305282048 &   -- & y &   32.51* & C &  y &  9.311 &     \\
TT Mon   & 3055512615681140736 &  318 & y &  326.4    & M &  n &  6.634 &     \\
CL Mon   & 3129886983796551936 &  476 & y &  467.28   & C &  y &  9.474 &     \\
V613 Mon & 3132397473782935424 &   -- & n &      --   & S & -- &     -- & [2] \\
R Nor    & 5985676640941632384 &  507 & y &  505.54   & M &  n &  6.082 &     \\
W Nor    & 5933137130595750272 &  135 & y &   25.34* & M &  n &  6.084 &     \\
X Oct    & 5191703179748307456 &  200 & y &  203.79   & M &  n &  6.001 &     \\
o1 Ori   & 3296627028792695168 &   30 & y &   51.29* & S &  n &  0.528 &     \\
RR Ori   & 3349083629043128320 &  252 & y &  251.24   & M &  n &  6.632 &     \\
BL Ori   & 3356702213833773440 &   -- & y &  158.63* & C &  y &  9.752 &     \\
R Peg    & 2715274995932228352 &  378 & y &  383.32   & M &  n &  6.701 &     \\
W Peg    & 2844608899441322112 &  344 & y &  361.91   & M &  n &  6.767 &     \\
RZ Peg   & 1900047116043182848 &  439 & y &  444.29   & S &  y &  9.232 &     \\
GZ Peg   & 2713996126469641984 &   93 & n &      --   & S & -- &     -- & [2] \\
HR Peg   & 2829186324715284480 &   50 & y &   24.48* & S &  n & -4.070 &     \\
Z Psc    &  294734094804790400 &  144 & y &  169.68   & C &  y &  9.840 &     \\
TX Psc   & 2743004129429424000 &   -- & y &  142.07* & C &  y &  9.472 &     \\
DT Psc   &  308866701872125568 &   -- & y &   17.88* & S & -- &     -- &     \\
L2 Pup   & 5559704601966334848 &  141 & y &   14.73* & M &  n &  5.070 &     \\
NQ Pup   & 3037131942360132480 &   -- & y &   16.8*  & S &  n & -3.790 &     \\
R Scl    & 5016138145186249088 &  370 & y &  371.53   & C &  y &  9.314 &     \\
RR Sco   & 6029292686708214016 &  281 & y &  278.55   & M &  n &  6.002 &     \\
R Ser    & 1192855977386610688 &  357 & y &  359.09   & M &  n &  6.634 &     \\
BG Ser   & 4404055896203005952 &  143 & y &  393.91* & M &  n &  6.890 &     \\
RX Tau   & 3292447407136388480 &  337 & y &  324.29   & M &  n &  6.766 &     \\
IK Tau   & 3303343395568710016 &  450 & n &      --   & M & -- &     -- & [2] \\
R UMi    & 1653382471306715264 &  324 & y &   30.96* & M &  n &  6.632 &     \\
S UMi    & 1708259612045319808 &  326 & y &  324.4    & M &  n &  6.890 &     \\
VZ Vel   & 5357938205327223936 &  317 & n &      --   & M & -- &     -- & [1] \\
R Vir    & 3709971554622524800 &  146 & y &  144.64   & M &  n &  5.952 &     \\
RU Vir   & 3704116483406003072 &  434 & y &  428.28   & C &  y &  9.312 &     \\
\hline 
\end{longtable}
\noindent $^1$ The period P$_{Lit.}$ (third column) and the spectral type (fifth column) are taken from the General Catalogue of Variable Stars \citep[GCVS;][]{gcvs}.\\
* period not exported; 
[1] in classification table only; 
[2] neither in classification nor in SOS table. 
}

\section{Catalogue retrieval}
\label{app:catalogRetrieval}

Here are two examples on how to retrieve LPV-related data from the \Gaia archive.
To retrieve few parameters from the source and the LPV catalogue:

\begin{footnotesize}
\begin{verbatim}
SELECT gs.source_id,
  gs.ra, gs.dec,
  lpv.frequency, lpv.frequency_error,
  ...
FROM gaiadr3.gaia_source AS gs
LEFT JOIN gaiadr3.vari_classifier_result AS claslpv
  ON gs.source_id = claslpv.source_id
LEFT JOIN gaiadr3.vari_long_period_variable AS lpv
  ON gs.source_id = lpv.source_id
WHERE claslpv.best_class_name = 'LPV'
  OR lpv.source_id IS NOT NULL
\end{verbatim}
\end{footnotesize}

To retrieve all source data, statistics data, and data from both the LPV catalogue and classification tables:
\begin{footnotesize}
\begin{verbatim}
SELECT src.*,
  variStat.*,
  classif.*,
  lpv.*
FROM gaiadr3.vari_summary AS variStat
LEFT JOIN gaiadr3.gaia_source AS src
  ON src.source_id = variStat.source_id
LEFT OUTER JOIN gaiadr3.vari_long_period_variable AS lpv
  ON lpv.source_id = variStat.source_id
LEFT OUTER JOIN gaiadr3.vari_classifier_result AS classif
  ON classif.source_id = variStat.source_id
WHERE lpv.source_id = variStat.source_id
  OR classif.best_class_name = 'LPV'
\end{verbatim}
\end{footnotesize}

\end{appendix}

\end{document}